\documentclass[twocolumn]{aastex62}   
\usepackage{natbib}
\usepackage{xspace}

\newcommand{\be}{\begin{equation}}
\newcommand{\ee}{\end{equation}}

\newcommand{\Fermi}{\textit{Fermi}\xspace}
\newcommand{\Fermilat}{\textit{Fermi}-LAT\xspace}

\shorttitle{\Fermilat Fourth Catalog}
\shortauthors{\Fermilat collaboration}

\begin{document}

 

\title{Fermi Large Area Telescope Fourth Source Catalog}
    

\author{S.~Abdollahi}
\affiliation{Department of Physical Sciences, Hiroshima University, Higashi-Hiroshima, Hiroshima 739-8526, Japan}
\author{F.~Acero}
\affiliation{AIM, CEA, CNRS, Universit\'e Paris-Saclay, Universit\'e Paris Diderot, Sorbonne Paris Cit\'e, F-91191 Gif-sur-Yvette, France}
\author{M.~Ackermann}
\affiliation{Deutsches Elektronen Synchrotron DESY, D-15738 Zeuthen, Germany}
\author{M.~Ajello}
\affiliation{Department of Physics and Astronomy, Clemson University, Kinard Lab of Physics, Clemson, SC 29634-0978, USA}
\author{W.~B.~Atwood}
\affiliation{Santa Cruz Institute for Particle Physics, Department of Physics and Department of Astronomy and Astrophysics, University of California at Santa Cruz, Santa Cruz, CA 95064, USA}
\author{M.~Axelsson}
\affiliation{Department of Physics, Stockholm University, AlbaNova, SE-106 91 Stockholm, Sweden}
\affiliation{Department of Physics, KTH Royal Institute of Technology, AlbaNova, SE-106 91 Stockholm, Sweden}
\author{L.~Baldini}
\affiliation{Universit\`a di Pisa and Istituto Nazionale di Fisica Nucleare, Sezione di Pisa I-56127 Pisa, Italy}
\author{J.~Ballet}
\email{jean.ballet@cea.fr}
\affiliation{AIM, CEA, CNRS, Universit\'e Paris-Saclay, Universit\'e Paris Diderot, Sorbonne Paris Cit\'e, F-91191 Gif-sur-Yvette, France}
\author{G.~Barbiellini}
\affiliation{Istituto Nazionale di Fisica Nucleare, Sezione di Trieste, I-34127 Trieste, Italy}
\affiliation{Dipartimento di Fisica, Universit\`a di Trieste, I-34127 Trieste, Italy}
\author{D.~Bastieri}
\affiliation{Istituto Nazionale di Fisica Nucleare, Sezione di Padova, I-35131 Padova, Italy}
\affiliation{Dipartimento di Fisica e Astronomia ``G. Galilei'', Universit\`a di Padova, I-35131 Padova, Italy}
% ...

\begin{abstract}
We present the fourth \Fermi Large Area Telescope catalog (4FGL) of $\gamma$-ray sources.
Based on the first eight years of science data from the \Fermi \textit{Gamma-ray Space Telescope} mission in the energy range from 50~MeV to 1~TeV, it is the deepest yet in this energy range. Relative to the 3FGL catalog, the 4FGL catalog has twice as much exposure as well as a number of analysis improvements, including an updated model for the Galactic diffuse $\gamma$-ray emission, and two sets of light curves (1-year and 2-month intervals). The 4FGL catalog includes 5064 sources above $4\sigma$ significance, for which we provide localization and spectral properties. Seventy-five sources are modeled explicitly as spatially extended, and overall 358 sources are considered as identified based on angular extent, periodicity or correlated variability observed at other wavelengths. For 1336 sources we have not found plausible counterparts at other wavelengths. More than 3130 of the identified or associated sources are active galaxies of the blazar class, and 239 are pulsars.
\end{abstract}


\section{Introduction}
\label{introduction}

The \Fermi \textit{Gamma-ray Space Telescope} was launched in June 2008, and the Large Area Telescope (LAT) onboard has been continually surveying the sky in the GeV energy range since then. Integrating the data over many years, the \Fermilat collaboration produced several generations of high-energy $\gamma$-ray source catalogs (Table \ref{tab:LATcatalogs}). The previous all-purpose catalog \citep[3FGL,][]{LAT15_3FGL} contained 3033 sources, mostly active galactic nuclei (AGN) and pulsars, but also a variety of other types of extragalactic and Galactic sources.

This paper presents the fourth catalog of sources, abbreviated as 4FGL (for \Fermi Gamma-ray LAT) detected in the first eight years of the mission.
As in previous catalogs, sources are included based on the statistical significance of their detection considered over the entire time period of the analysis. For this reason the 4FGL catalog does not contain transient $\gamma$-ray sources which are detectable only over a short duration, including Gamma-ray Bursts \citep[GRBs, ][]{GRBCat2}, solar flares \citep{SolarFlares}, and most novae \citep{LAT14_novae}.

\begin{deluxetable*}{lllll}

\tablecaption{Previous \Fermilat catalogs
\label{tab:LATcatalogs}}
\tablehead{
\colhead{Acronym} & \colhead{IRFs/Diffuse model} & \colhead{Energy range/Duration} & \colhead{Sources} & \colhead{Analysis/Reference}
}

\startdata
1FGL & P6\_V3\_DIFFUSE & 0.1 -- 100~GeV & 1451 (P) & Unbinned, F/B \\
 & gll\_iem\_v02 & 11 months &  & \citet{LAT10_1FGL} \\
2FGL & P7SOURCE\_V6 & 0.1 -- 100~GeV & 1873 (P) & Binned, F/B \\
 & gal\_2yearp7v6\_v0 & 2 years &  & \citet{LAT12_2FGL} \\
3FGL & P7REP\_SOURCE\_V15 & 0.1 -- 300~GeV & 3033 (P) & Binned, F/B \\
 & gll\_iem\_v06 & 4 years &  & \citet{LAT15_3FGL} \\
FGES & P8R2\_SOURCE\_V6 & 10~GeV -- 2~TeV & 46 (E) & Binned, PSF, $|b| < 7\degr$ \\
 & gll\_iem\_v06 & 6 years &  & \citet{LAT17_10GeVES} \\
3FHL & P8R2\_SOURCE\_V6 & 10~GeV -- 2~TeV & 1556 (P) & Unbinned, PSF \\
 & gll\_iem\_v06 & 7 years &  & \citet{LAT17_3FHL} \\
FHES & P8R2\_SOURCE\_V6 & 1~GeV -- 1~TeV & 24 (E) & Binned, PSF, $|b| > 5\degr$ \\
 & gll\_iem\_v06 & 7.5 years &  & \citet{FHES_2018} \\
\hline
4FGL & P8R3\_SOURCE\_V2 & 0.05~GeV -- 1~TeV & 5064 (P) & Binned, PSF \\
 & gll\_iem\_v07 (\S~\ref{GalacticIEM}) & 8 years & & this work \\
\enddata
\tablecomments{In the Analysis column, F/B stands for $Front/Back$, and PSF for PSF event types\tablenotemark{$a$}.
In the Sources column, we write (P) when the catalog's objective is to look for point-like sources, (E) when it looks for extended sources.
}
\tablenotetext{a}{See \url{https://fermi.gsfc.nasa.gov/ssc/data/analysis/LAT\_essentials.html}.}
\end{deluxetable*}

The 4FGL catalog benefits from a number of improvements with respect to the 3FGL, besides the twice longer exposure:
\begin{enumerate}
\item We used Pass 8 data\footnote{See \url{http://fermi.gsfc.nasa.gov/ssc/data/analysis/documentation/Pass8\_usage.html}.} (\S~\ref{LATData}).  The principal difference relative to the P7REP data used for 3FGL is improved angular resolution above 3~GeV and about 20\% larger acceptance at all energies, reaching 2.5 m$^2$ sr between 2 and 300~GeV. The acceptance is defined here as the integral of the effective area over the field of view. It is the most relevant quantity for a survey mission such as \Fermilat.
\item We developed a new model of the underlying diffuse Galactic emission (\S~\ref{DiffuseModel}).
\item We introduced weights in the maximum likelihood analysis (\S~\ref{catalog_significance}) to mitigate the effect of systematic errors due to our imperfect knowledge of the Galactic diffuse emission.
\item We accounted for the effect of energy dispersion (reconstructed event energy not equal to the true energy of the incoming $\gamma$ ray). This is a small correction (\S~\ref{compare_stepbystep}) and was neglected in previous \Fermilat catalogs because the energy resolution (measured as the 68\% containment half width) is better than 15\% over most of the LAT energy range and the $\gamma$-ray spectra have no sharp features.
\item We tested all sources with three spectral models (power law, log normal and power law with subexponential cutoff, \S~\ref{catalog_spectral_shapes}).
\item We explicitly modeled 75 sources as extended emission regions (\S~\ref{catalog_extended}), up from 25 in 3FGL.
\item We built light curves and tested variability using two different time bins (one year and two months, \S~\ref{catalog_variability}).
\item To study the associations of LAT sources with counterparts at other wavelengths, we updated several of the counterpart catalogs, and correspondingly recalibrated the association procedure.
\end{enumerate}
A preliminary version of this catalog (FL8Y\footnote{See \url{https://fermi.gsfc.nasa.gov/ssc/data/access/lat/fl8y/}.}) was built from the same data and the same software, but using the previous interstellar emission model (gll\_iem\_v06) as background, starting at 100~MeV and switching to curved spectra at $TS_{\rm curv} > 16$ (see \S~\ref{catalog_spectral_shapes} for definition). We use it as a starting point for source detection and localization, and to estimate the impact of changing the underlying diffuse model.
The result of a dedicated effort for studying the AGN population in the 4FGL catalog is published in the accompanying fourth LAT AGN catalog \citep[4LAC,][]{LAT19_4LAC} paper.

Section \ref{lat_and_background} describes the LAT, the data, and the models for the diffuse  backgrounds, celestial and otherwise.  Section \ref{catalog_main} describes the construction of the catalog, with emphasis on what has changed since the analysis for the 3FGL catalog.
Section \ref{4fgl_description} describes the catalog itself, Section \ref{sec:associations} explains the association and identification procedure, and Section \ref{sec:assocsum} details the association results. We conclude in Section \ref{conclusions}.
We provide appendices with technical details of the analysis and of the format of the electronic version of the catalog.

    \section{Instrument \& Background}
\label{lat_and_background}

\subsection{The Large Area Telescope}
\label{LATDescription}

\begin{figure}[!ht]
\centering
\includegraphics[width=\linewidth]{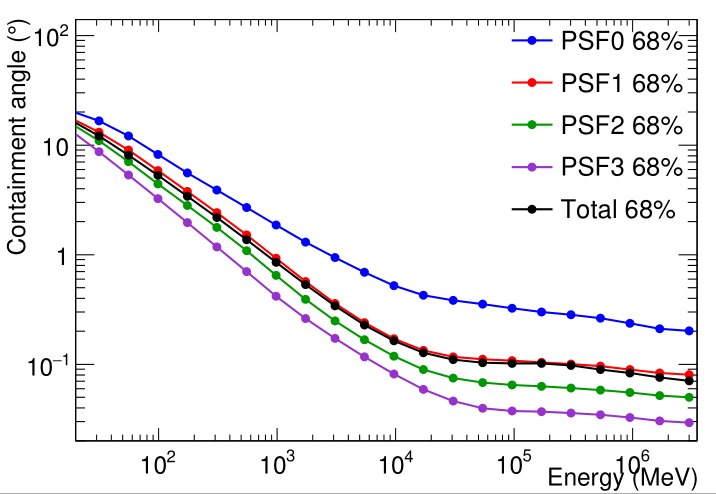}
\caption{Containment angle (68\%) of the \Fermilat PSF as a function of energy, averaged over off-axis angle. The black line is the average over all data, whereas the colored lines illustrate the difference between the four categories of events ranked by PSF quality from worst (PSF0) to best (PSF3).
}
\label{fig:psfwidth}
\end{figure}

The LAT detects $\gamma$ rays in the energy range from 20~MeV to more than 1~TeV, measuring their arrival times, energies, and directions.
The field of view of the LAT is $\sim$ 2.7~sr at 1~GeV and above.   The per-photon angular resolution (point-spread function, PSF, 68\% containment radius) is $\sim 5\degr$ at 100~MeV, improving to $0\fdg8$ at 1~GeV (averaged over the acceptance of the LAT), varying with energy approximately as $E^{-0.8}$ and asymptoting at $\sim 0\fdg1$ above 20~GeV (Figure~\ref{fig:psfwidth}). The tracking section of the LAT has 36 layers of silicon strip detectors interleaved with 16 layers of tungsten foil (12 thin layers, 0.03 radiation length, at the top or {\it Front} of the instrument, followed by 4 thick layers, 0.18 radiation lengths, in the {\it Back} section).  The silicon strips track charged particles, and the tungsten foils facilitate conversion of $\gamma$ rays to positron-electron pairs.  Beneath the tracker is a calorimeter composed of an 8-layer array of CsI crystals ($\sim$8.5 total radiation lengths) to determine the $\gamma$-ray energy.
More information about the LAT is provided in  \citet{LAT09_instrument}, and the in-flight calibration of the LAT is described in \citet{LAT09_calib}, \citet{LAT12_calib} and \citet{LAT_energyscale2012}.

The LAT is also an efficient detector of the intense background of charged particles from cosmic rays and trapped radiation at the orbit of the \Fermi satellite.
A segmented charged-particle anticoincidence detector (plastic scintillators 
read out by photomultiplier tubes) around the tracker is used to reject charged-particle background events.
Accounting for $\gamma$ rays lost in filtering charged particles from the data, the effective collecting area at normal incidence (for the P8R3\_SOURCE\_V2 event selection used here; see below)\footnote{See \url{http://www.slac.stanford.edu/exp/glast/groups/canda/lat_Performance.htm}.} exceeds 0.3~m$^2$ at 0.1~GeV, 0.8~m$^2$ at 1~GeV, and remains nearly constant at $\sim$ 0.9~m$^2$ from 2 to 500~GeV.
The live time is nearly 76\%, limited primarily by interruptions of data taking when \Fermi is passing through the South Atlantic Anomaly (SAA, $\sim$15\%) and readout dead-time fraction ($\sim$9\%).

\subsection{The LAT Data}
\label{LATData}

The data for the 4FGL catalog were taken during the period 2008 August 4 (15:43 UTC) to 2016 August 2 (05:44 UTC) covering eight years.
During most of this time, \Fermi was operated in sky-scanning survey mode (viewing direction rocking north and south of the zenith on alternate orbits).
As in 3FGL, intervals around solar flares and bright GRBs were excised.
Overall, about two days were excised due to solar flares, and 39 ks due to 30 GRBs.
The precise time intervals corresponding to selected events are recorded in the \texttt{GTI} extension of the FITS file (Appendix \ref{appendix_fits_format}).
The maximum exposure ($4.5 \times 10^{11}$ cm$^2$ s at 1~GeV) is reached at the North celestial pole. The minimum exposure ($2.7 \times 10^{11}$ cm$^2$ s at 1~GeV) is reached at the celestial equator.

The current version of the LAT data is Pass 8 P8R3 \citep{LAT13_P8, LAT18_P305}.
It offers 20\% more acceptance than P7REP \citep{LAT13_P7repro} and a narrower PSF at high energies.
Both aspects are very useful for source detection and localization \citep{LAT17_3FHL}.
We used the Source class event selection, with the Instrument Response Functions (IRFs) P8R3\_SOURCE\_V2.
Pass 8 introduced a new partition of the events, called PSF event types, based on the quality of the angular reconstruction (Figure~\ref{fig:psfwidth}), with approximately equal effective area in each event type at all energies. The angular resolution is critical to distinguish point sources from the background, so we split the data into those four categories to avoid diluting high-quality events (PSF3) with poorly localized ones (PSF0). We split the data further into 6 energy intervals (also used for the spectral energy distributions in \S~\ref{catalog_flux_determination}) because the extraction regions must extend further at low energy (broad PSF) than at high energy, but the pixel size can be larger. After applying the zenith angle selection (\S~\ref{zmax}), we were left with the 15 components described in Table~\ref{tab:components}. The log-likelihood is computed for each component separately, then they are summed for the SummedLikelihood maximization (\S~\ref{catalog_significance}).

The lower bound of the energy range was set to 50~MeV, down from 100~MeV in 3FGL, to constrain the spectra better at low energy. It does not help detecting or localizing sources because of the very broad PSF below 100~MeV.
The upper bound was raised from 300~GeV in 3FGL to 1~TeV. This is because as the source-to-background ratio decreases, the sensitivity curve \citep[Figure~18 of][1FGL]{LAT10_1FGL} shifts to higher energies. The 3FHL catalog \citep{LAT17_3FHL} went up to 2 TeV, but only 566 events exceed 1~TeV over 8 years (to be compared to 714,000 above 10~GeV).

\subsection{Zenith angle selection}
\label{zmax}

\begin{deluxetable*}{lrrrlllll}
\tablecaption{4FGL Summed Likelihood components
\label{tab:components}
}
\tablehead{
\colhead{Energy interval} & \colhead{NBins} & \colhead{ZMax} & \colhead{Ring width} & \multicolumn{5}{c}{Pixel size (deg)} \\
(GeV) &  & (deg) & (deg) & \colhead{PSF0} & \colhead{PSF1} & \colhead{PSF2} & \colhead{PSF3} & \colhead{All}
}
\startdata
0.05 -- 0.1 & 3 & 80 & 7 & \nodata & \nodata & \nodata & 0.6 & \nodata \\
0.1 -- 0.3 & 5 & 90 & 7 & \nodata & \nodata & 0.6 & 0.6 & \nodata \\
0.3 -- 1 & 6 & 100 & 5 & \nodata & 0.4 & 0.3 & 0.2 & \nodata \\
1 -- 3 & 5 & 105 & 4 & 0.4 & 0.15 & 0.1 & 0.1 & \nodata \\
3 -- 10 & 6 & 105 & 3 & 0.25 & 0.1 & 0.05 & 0.04 & \nodata \\
10 -- 1000 & 10 & 105 & 2 & \nodata & \nodata & \nodata & \nodata & 0.04 \\
\enddata
\tablecomments{We used 15 components (all in binned mode) in the 4FGL Summed Likelihood approach (\S~\ref{catalog_significance}). Components in a given energy interval share the same number of energy bins, the same zenith angle selection and the same RoI size, but have different pixel sizes in order to adapt to the PSF width (Figure~\ref{fig:psfwidth}). Each filled entry under Pixel size corresponds to one component of the summed log-likelihood. NBins is the number of energy bins in the interval, ZMax is the zenith angle cut, Ring width refers to the difference between the RoI core and the extraction region, as explained in item 5 of \S~\ref{catalog_significance}.
}
\end{deluxetable*}

\begin{figure}[!ht]
\centering
\includegraphics[width=\linewidth]{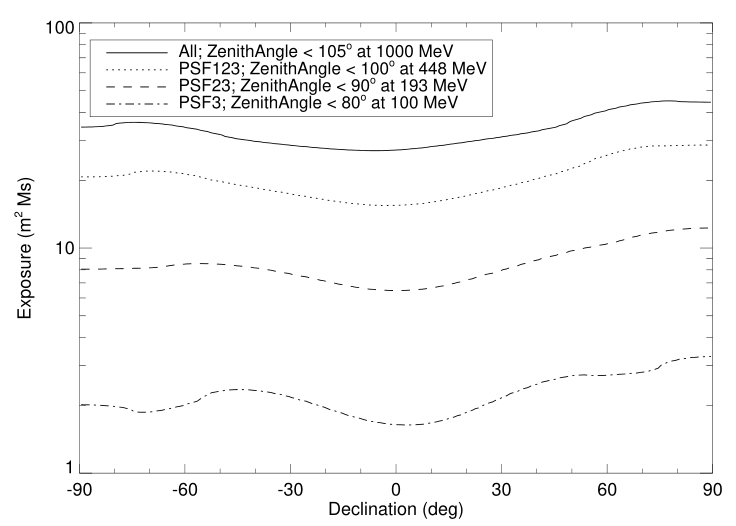}
\caption{Exposure as a function of declination and energy, averaged over right ascension, summed over all relevant event types as indicated in the figure legend.
}
\label{fig:exposure}
\end{figure}

The zenith angle cut was set such that the contribution of the Earth limb at that zenith angle was less than 10\% of the total (Galactic + isotropic) background. Integrated over all zenith angles, the residual Earth limb contamination is less than 1\%.
We kept PSF3 event types with zenith angles less than 80$\degr$ between 50 and 100~MeV, PSF2 and PSF3 event types with zenith angles less than 90$\degr$ between 100 and 300~MeV, and PSF1, PSF2 and PSF3 event types with zenith angles less than 100$\degr$ between 300~MeV and 1~GeV. Above 1~GeV we kept all events with zenith angles less than 105$\degr$ (Table~\ref{tab:components}).

The resulting integrated exposure over 8 years is shown in Figure~\ref{fig:exposure}. The dependence on declination is due to the combination of the inclination of the orbit ($25\fdg6$), the rocking angle, the zenith angle selection and the off-axis effective area. The north-south asymmetry is due to the SAA, over which no scientific data is taken. Because of the regular precession of the orbit every 53 days, the dependence on right ascension is small when averaged over long periods of time. The main dependence on energy is due to the increase of the effective area up to 1~GeV, and the addition of new event types at 100~MeV, 300~MeV and 1~GeV. The off-axis effective area depends somewhat on energy and event type. This, together with the different zenith angle selections, introduces a slight dependence of the shape of the curve on energy.

Selecting on zenith angle applies a kind of time selection (which depends on direction in the sky). This means that the effective time selection at low energy is not exactly the same as at high energy. The periods of time during which a source is at zenith angle $< 105\degr$ but (for example) $> 90\degr$ last typically a few minutes every orbit. This is shorter than the main variability time scales of astrophysical sources in 4FGL, and therefore not a concern.
There remains however the modulation due to the precession of the spacecraft orbit on longer time scales over which blazars can vary. This is not a problem for a catalog (it can at most appear as a spectral effect, and should average out when considering statistical properties) but it should be kept in mind when extracting spectral parameters of individual variable sources.
We used the same zenith angle cut for all event types in a given energy interval, to reduce systematics due to that time selection.

Because the data are limited by systematics at low energies everywhere in the sky (Appendix \ref{appendix_weights}) rejecting half of the events below 300~MeV and 75\% of them below 100~MeV does not impact the sensitivity (if we had kept these events, the weights would have been lower).

    \subsection{Model for the Diffuse Gamma-Ray Background}
\label{DiffuseModel}

\subsubsection{Diffuse emission of the Milky Way}
\label{GalacticIEM}
We extensively updated the model of the Galactic diffuse emission for the 4FGL analysis, using the same P8R3 data selections (PSF types, energy ranges, and zenith angle limits). The development of the model is described in more detail (including illustrations of the templates and residuals) online\footnote{\url{https://fermi.gsfc.nasa.gov/ssc/data/analysis/software/aux/4fgl/Galactic_Diffuse_Emission_Model_for_the_4FGL_Catalog_Analysis.pdf}}.  Here we summarize the primary differences from the model developed for the 3FGL catalog \citep{LAT16_DiffuseModel}.  In both cases, the model is based on linear combinations of templates representing components of the Galactic diffuse emission.  For 4FGL we updated all of the templates, and added a new one as described below.

We have adopted the new, all-sky high-resolution, 21-cm spectral line HI4PI survey \citep{HI4PI} as our tracer of H\,{\sc i}, and extensively refined the procedure for partitioning the H\,{\sc i} and H$_2$ (traced by the 2.6-mm CO line) into separate ranges of Galactocentric distance (`rings'), by decomposing the spectra into individual line profiles, so the broad velocity dispersion of massive interstellar clouds does not effectively distribute their emission very broadly along the line of sight.  We also updated the rotation curve, and adopted a new procedure for interpolating the rings across the Galactic center and anticenter, now incorporating a general model for the surface density distribution of the interstellar medium to inform the interpolation, and defining separate rings for the Central Molecular Zone (within $\sim$150~pc of the Galactic center and between 150~pc and 600~pc of the center).  With this approach, the Galaxy is divided into ten concentric rings.

The template for the inverse Compton emission is still based on a model interstellar radiation field and cosmic-ray electron distribution \citep[calculated in GALPROP v56, described in][]{GALPROP17}\footnote{\url{http://galprop.stanford.edu}} but now we formally subdivide the model into rings (with the same Galactocentric radius ranges as for the gas templates), which are fit separately in the analysis, to allow some spatial freedom relative to the static all-sky inverse-Compton model.

We have also updated the template of the `dark gas' component \citep{Grenier2005}, representing interstellar gas that is not traced by the H\,{\sc i} and CO line surveys, by comparison with the {\it Planck} dust optical depth map\footnote{\texttt{COM\_CompMap\_Dust-GNILC-Model-Opacity\_2048\_R2.01.fits}, \citet{Planck_dust}}.  The dark gas is inferred as the residual component after the best-fitting linear combination of total $N$(H\,{\sc i}) and $W_{\rm CO}$ (the integrated intensity of the CO line) is subtracted, i.e., as the component not correlated with the atomic and molecular gas spectral line tracers, in a procedure similar to that used in \citet{LAT16_DiffuseModel}.  In particular, as before we retained the negative residuals as a `column density correction map'.

New to the 4FGL model, we incorporated a template representing the contribution of unresolved Galactic sources. This was derived from the model spatial distribution and luminosity function developed based on the distribution of Galactic sources in \citet{LAT15_3FGL} and an analytical evaluation of the flux limit for source detection as a function of direction on the sky.

As for the 3FGL model, we iteratively determined and re-fit a model component that represents non-template diffuse $\gamma$-ray emission, primarily Loop~I and the {\it Fermi} bubbles.  To avoid overfitting the residuals, and possibly suppressing faint Galactic sources, we spectrally and spatially smoothed the residual template.

The model fitting was performed using Gardian \citep{Diffuse2}, as a summed log-likelihood analysis.  This procedure involves transforming the ring maps described above into spatial-spectral templates evaluated in GALPROP.  We used model $^S$L$^Z$6$^R$30$^T$150$^C$2 from \citet{Diffuse2}.
The model is a linear combination of these templates, with free scaling functions of various forms for the individual templates.  For components with the largest contributions, a piecewise continuous function, linear in the logarithm of energy, with nine degrees of freedom was used.  Other components had a similar scaling function with five degrees of freedom, or power-law scaling, or overall scale factors, chosen to give the model adequate freedom while reducing the overall number of free parameters.  The model also required a template for the point and small-extended sources in the sky.  We iterated the fitting using preliminary versions of the 4FGL catalog.  This template was also given spectral degrees of freedom.  Other diffuse templates, described below and not related to Galactic emission, were included in the model fitting.

\subsubsection{Isotropic background}
\label{isotropic}

The isotropic diffuse background was derived over 45 energy bins covering the energy range 30~MeV to 1~TeV, from the eight-year data set excluding the Galactic plane ($|b| > 15\degr$). To avoid the Earth limb emission (more conspicuous around the celestial poles), we applied a zenith angle cut at $80 \degr$ and also excluded declinations higher than $60\degr$ below 300~MeV. The isotropic background was obtained as the residual between the spatially-averaged data and the sum of the Galactic diffuse emission model described above, a preliminary version of the 4FGL catalog and the solar and lunar templates (\S~\ref{SunMoon}), so it includes charged particles misclassified as $\gamma$ rays.  We implicitly assume that the acceptance for these residual charged particles is the same as for $\gamma$ rays in treating these diffuse background components together. To obtain a continuous model, the final spectral template was obtained by fitting the residuals in the 45 energy bins to a multiply broken power law with 18 breaks. For the analysis we derived the contributions to the isotropic background separately for each event type.

\subsubsection{Solar and lunar template}
\label{SunMoon}

The quiescent Sun and the Moon are fairly bright $\gamma$-ray sources.  The Sun moves in the ecliptic but the solar $\gamma$-ray emission is extended because of cosmic-ray interactions 
with the solar radiation field; detectable emission from inverse Compton scattering of cosmic-ray electrons on the radiation field of the Sun extends several degrees from the Sun \citep{OrlandoStrong, LAT11_Sun}.  The Moon is not an extended source in this way but the lunar orbit is inclined somewhat relative 
to the ecliptic and the Moon moves through a larger fraction of the sky than the Sun.  Averaged over time, the $\gamma$-ray emission from the Sun and Moon trace a region around the ecliptic. Without any correction this can seriously affect the spectra and light curves, so starting with 3FGL we model that emission.

The Sun and Moon emission are modulated by the solar magnetic field which deflects cosmic rays more (and therefore reduces $\gamma$-ray emission) when the Sun is at maximum activity. For that reason the model used in 3FGL (based on the first 18 months of data when the Sun was near minimum) was not adequate for 8 years. We used the improved model of the lunar emission \citep{LAT16_Moon} and a data-based model of the solar disk and inverse Compton scattering on the solar light (S. Raino, private communication).

We combined those models with calculations of their motions and of the exposure of the observations by the LAT 
to make templates for the equivalent diffuse component over 8 years using $gtsuntemp$ \citep{SST_2013}.
For 4FGL we used two different templates: one for the inverse Compton emission on the solar light (pixel size $0\fdg25$) and one for the sum of the solar and lunar disks.
For the latter we reduced the pixel size to $0\fdg125$ to describe the disks accurately, and computed a specific template for each event type / maximum zenith angle combination of Table~\ref{tab:components} (because their exposure maps are not identical).
As in 3FGL those components have no free parameter.

\subsubsection{Residual Earth limb template}
\label{EarthLimb}

For 3FGL we reduced the low-energy Earth limb emission by selecting zenith angles less than 100$\degr$, and modeled the residual contamination approximately.
For 4FGL we chose to cut harder on zenith angle at low energies and select event types with the best PSF (\S~\ref{zmax}). That procedure eliminates the need for a specific Earth limb component in the model.

\section{Construction of the Catalog}
\label{catalog_main}

The procedure used to construct the 4FGL catalog has a number of improvements relative to that of the 3FGL catalog.  In this section we review the procedure, emphasizing what was done differently.
The significances (\S~\ref{catalog_significance}) and spectral parameters (\S~\ref{catalog_spectral_shapes}) of all catalog sources were obtained using the standard $pyLikelihood$ framework (Python analog of $gtlike$) in the LAT Science Tools\footnote{See \url{http://fermi.gsfc.nasa.gov/ssc/data/analysis/documentation/Cicerone/}.} (version v11r7p0). The localization procedure (\S~\ref{catalog_detection}), which relies on $pointlike$ \citep{Kerr2010}, provided the source positions, the starting point for the spectral fitting in \S~\ref{catalog_significance}, and a comparison for estimating the reliability of the results (\S~\ref{catalog_analysismethod}).

Throughout the text we denote as RoIs, for Regions of Interest, the regions in which we extract the data.
We use the Test Statistic $TS = 2 \, \log (\mathcal{L} / \mathcal{L}_0)$ \citep{mattox96} to quantify how significantly a source emerges from the background, comparing the maximum value of the likelihood function $\mathcal{L}$ over the RoI including the source in the model with $\mathcal{L}_0$, the value without the source. Here and everywhere else in the text $\log$ denotes the natural logarithm.
The names of executables and libraries of the Science Tools are written in italics.
    \subsection{Detection and Localization}
\label{catalog_detection}

This section describes the generation of a list of candidate sources, with locations and initial spectral fits.
This initial stage uses $pointlike$. Compared with the $gtlike$-based analysis described in \S~\ref{catalog_significance} to \ref{catalog_limitations}, it uses the same time range and IRFs, 
but the partitioning of the sky, the weights, the computation of the likelihood function and its optimization are independent. The zenith angle cut is set to $100\degr$. Energy dispersion is neglected for the sources (we show in \S~\ref{compare_stepbystep} that it is a small effect).
Events below 100~MeV are not useful for source detection and localization, and are ignored at this stage.

\subsubsection{Detection settings}

The process started with an initial set of sources, from the 8-year FL8Y analysis, including the 75 spatially extended sources listed in \S~\ref{catalog_extended}, and the three-component representation of the Crab (\S~\ref{catalog_spectral_shapes}). The same spectral models were considered for each source as in \S~\ref{catalog_spectral_shapes}, but the favored model (power law, curved, or pulsar-like) was not necessarily the same. The point-source locations were also re-optimized.

The generation of a candidate list of additional sources, with locations and initial spectral fits, is substantially the same as for 3FGL. The sky was partitioned using 
HEALPix\footnote{\url{http://healpix.sourceforge.net}.} \citep{Gorski2005} with 
$N_{\rm side} = 12$, resulting in 1728 tiles of $\sim$24 deg$^2$ area.  (Note: references to $N_{\rm side}$ in the following refer to HEALPix.)
The RoIs included events in cones of 5$\degr$ radius about the center of the tiles.
The data were binned according to energy, 16 energy bands from 100~MeV to 1~TeV (up from 14 bands to 316~GeV in 3FGL), $Front$ or $Back$ event types, and angular position using HEALPix, but with $N_{\rm side}$ varying from 64 to 4096 according to the PSF.
Only $Front$ events were used for the two bands below 316~MeV, to avoid the poor PSF and contribution of the Earth limb. Thus the log-likelihood calculation, for each RoI, is a sum over the contributions of 30 energy and event type bands. 

All point sources within the RoI and those nearby, such that the contribution to the RoI was at least 1\% (out to $11\degr$ for the lowest energy band), were included.
Only the spectral model parameters for sources within the central tile were allowed to vary to optimize the likelihood. To account for correlations with fixed nearby sources, and a factor of three overlap for the data (each photon contributes to $\sim$ 3 RoIs), the following iteration process was followed. All 1728 RoIs were optimized independently. Then the process was repeated, until convergence, for all RoIs for which the log-likelihood had changed by more than 10. Their nearest neighbors (presumably affected by the modified sources) were iterated as well.

Another difference from 3FGL was that the diffuse contributions were adjusted globally. We fixed the isotropic diffuse source to be actually constant over the sky, but globally refit its spectrum up to 10~GeV, since point-source fits are insensitive to diffuse energies above this. The Galactic diffuse emission component also was treated quite differently. Starting with a version of the Galactic diffuse model (\S~\ref{GalacticIEM}) without its non-template diffuse $\gamma$-ray emission, we derived an alternative adjustment by optimizing the Galactic diffuse normalization for each RoI and the eight bands below 10~GeV. These values were turned into an 8-layer map which was smoothed, then applied to the PSF-convolved diffuse model predictions for each band. Then the corrections were remeasured. This process converged after two iterations, such that no further corrections were needed. The advantage of the procedure, compared to fitting the diffuse spectral parameters in each RoI (\S~\ref{catalog_significance}), is that the effective predictions do not vary abruptly from an RoI to its neighbors, and are unique for each point. Also it does not constrain the spectral adjustment to be a power law.

After a set of iterations had converged, the localization procedure was applied, and source positions updated for a new set of iterations. At this stage, new sources were occasionally added using the residual $TS$ procedure described in \S~\ref{catalog_newseeds}.
The detection and localization process resulted in 7841 candidate point sources with $TS > 10$, of which 3179 were new. 
The fit validation and likelihood weighting were done as in 3FGL, except that, due to the improved representation of the Galactic diffuse, the effect of the weighting factor was less severe.

The $pointlike$ unweighting scheme is slightly different from that described in the 3FGL paper (\S~3.1.2). 
A measure of the sensitivity to the Galactic diffuse component is the average count density for the RoI divided by the peak value of the PSF, $N_{\rm diff}$, which represents a measure of the diffuse background under the point source.  
For the RoI at the Galactic center, and the lowest energy band, this is $4.15 \times 10^4$ counts. 
We unweight the likelihood for all energy bands by effectively limiting this implied precision to 2\%, corresponding to 2500 counts. 
As before, we divide the log-likelihood contribution from this energy band by $\max(1, N_{\rm diff}/2500)$. For the aforementioned case, this value is 16.6. 
A consequence is to increase the spectral fit uncertainty for the lowest energy bins for every source in the RoI.  
The value for this unweighting factor was determined by examining the distribution of the deviations between fluxes fitted in individual energy bins and the global spectral fit (similar to what is done in \S~\ref{catalog_flux_determination}). The 2\% precision was set such that the RMS for the distribution of positive deviations in the most sensitive lowest energy band was near the statistical expectation. (Negative deviations are distorted by the positivity constraint, resulting in an asymmetry of the distribution.)

\begin{figure}[ht]
\centering
\includegraphics[width=0.5\textwidth]{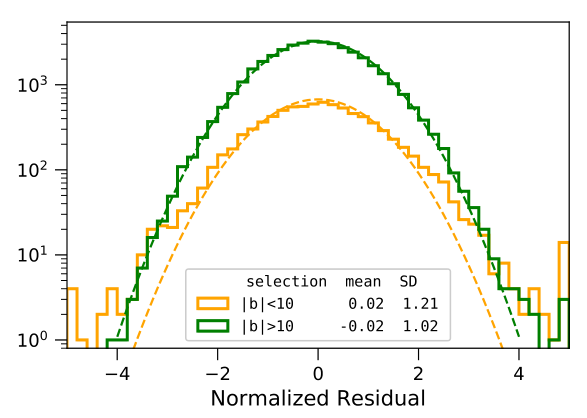}
\caption{Photon count residuals with respect to the model per $N_{\rm side}=64$ bin, for energies 100 -- 177 MeV,  normalized by the Poisson uncertainty, that is, $(N_{data}-N_{model})/\sqrt{ N_{model}}$. Histograms are shown for the values at high latitude ($|b|>10\degr$) and low latitude ($|b|<10\degr$) (capped at $\pm 5\sigma$). 
Dashed lines are the Gaussian expectations for the same number of sources. 
The legend shows the mean and standard deviation for the two subsets.}
\label{fig:pointlikepulls}
\end{figure}

An important validation criterion is the all-sky counts residual map. Since the source overlaps and diffuse uncertainties are most severe at the lowest energy, we present, in Figure~\ref{fig:pointlikepulls}, the distribution of normalized residuals per pixel, binned with $N_{\rm side}=64$, in the 100 -- 177 MeV $Front$ energy band.  There are 49,920 such pixels, with data counts varying from 92 to $1.7 \times 10^4$. For $|b|>10\degr$, the agreement with the expected Gaussian distribution is very good, while it is clear that there are issues along the plane. These are of two types.  First, around very strong sources, such as Vela, the discrepancies are perhaps a result of inadequacies of the simple spectral models used, but the (small) effect of energy dispersion and the limited accuracy of the IRFs may contribute.  Regions along the Galactic ridge are also evident, a result of the difficulty modeling the emission precisely, the reason we unweight contributions to the likelihood.

\subsubsection{Detection of additional sources}
\label{catalog_newseeds}

As in 3FGL, the same implementation of the likelihood used for optimizing source parameters was used to test for the presence of additional point sources. This is inherently iterative, in that the likelihood is valid to the extent that the model used to calculate it is a fair representation of the data. Thus, the detection of the faintest sources depends on accurate modeling of all nearby brighter sources and the diffuse contributions.

\begin{figure}[!ht]
\centering
\includegraphics[width=0.4\textwidth]{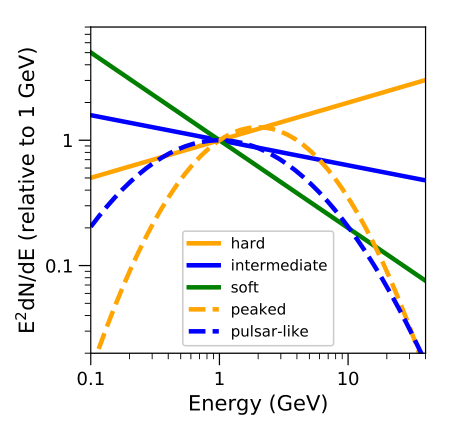}

\caption{Spectral shape templates used in source finding. }
\label{fig:sourcetemplates}
\end{figure}

The FL8Y source list from which this started represented several such additions from the 4-year 3FGL.  As before, an iteration starts with choosing a HEALPix $N_{\rm side}$ = 512 grid, 3.1~M points with average separation ~0.15 degrees.  But now, instead of testing a single power-law spectrum, we try five spectral shapes; three are power laws with different indices, two with significant curvature. Table~\ref{tab:sourcesearch} lists the spectral shapes used for the templates. 
They are shown in Figure~\ref{fig:sourcetemplates}.

\begin{deluxetable}{rrrlrr}

\tablecaption{Spectral shapes for source search
\label{tab:sourcesearch}}
\tablehead{
\colhead{$\alpha$} & \colhead{$\beta$} & \colhead{$E_0$ (GeV)} & \colhead{Template} & \colhead{Generated} & \colhead{Accepted}
}

\startdata
1.7 & 0.0 & 50.00 & Hard         & 471 & 101 \\
2.2 & 0.0 & 1.00  & Intermediate & 889 & 177 \\
2.7 & 0.0 & 0.25  & Soft         & 476 & 84 \\
2.0 & 0.5 & 2.00  & Peaked       & 686 & 151 \\
2.0 & 0.3 & 1.00  & Pulsar-like  & 476 & 84 \\
\enddata
\tablecomments{The spectral parameters $\alpha$, $\beta$ and $E_0$ refer to the LogParabola spectral shape (Eq.~\ref{eq:logparabola}).
The last two columns show the number, for each shape, that were successfully added to the $pointlike$ model, and the number accepted for the final 4FGL list.
}
\end{deluxetable}

For each trial position, and each of the five templates, the normalizations were optimized, and the resulting $TS$ associated with the pixel. Then, as before, but independently for each template, a cluster analysis selected groups of pixels with $TS > 16$, as compared to $TS > 10$ for 3FGL.  Each cluster defined a seed, with a position determined by weighting the $TS$ values. Finally, the five sets of potential seeds were compared and, for those within $1\degr$, the seed with the largest $TS$ was selected for inclusion. 

Each candidate was added to its respective RoI, then fully optimized, including localization, during a full likelihood optimization including all RoIs. 
 The combined results of two iterations of this procedure, starting from a $pointlike$ model including only sources imported from the FL8Y source list, are summarized in Table~\ref{tab:sourcesearch}, which shows the number for each template that was successfully added to the $pointlike$ model, and the number finally included in 4FGL.
The reduction is mostly due to  the $TS > 25$ requirement in 4FGL, as applied to the $gtlike$ calculation (\S~\ref{catalog_significance}), which uses different data and smaller weights. The selection is even stricter ($TS > 34$, \S~\ref{catalog_spectral_shapes}) for sources with curved spectra. Several candidates at high significance were not accepted because they were too close to even brighter sources, or inside extended sources, and thus unlikely to be independent point sources.

\subsubsection{Localization}
\label{catalog_localization}

The position of each source was determined by maximizing the likelihood with 
respect to its position only. That is, all other parameters are kept fixed. The possibility that 
a shifted position would affect the spectral models or positions of nearby sources is 
accounted for by iteration.
In the ideal limit of large statistics the log-likelihood is a 
quadratic form in any pair of orthogonal angular variables, assuming small angular offsets. 
We define LTS, for Localization Test Statistic, to be twice the log of the 
likelihood ratio of any position with respect to the maximum; the LTS evaluated 
for a grid of positions is called an LTS map. We fit the distribution of LTS  
to a quadratic form to determine the uncertainty ellipse (position, major and 
minor axes, and orientation).
The fitting procedure starts with a prediction of the LTS distribution from the current elliptical parameters. From this, it evaluates the LTS for eight positions in a circle of a radius corresponding to twice the geometric mean of the two Gaussian sigmas. We define a measure, the localization quality (LQ), of how well the actual LTS distribution matches this expectation as the sum of squares of differences at those eight positions. The fitting procedure determines a new set of elliptical parameters from the eight values. In the ideal case, this is a linear problem and one iteration is sufficient from any starting point. To account for finite statistics or distortions due to inadequacies of the model, we iterate until changes are small. The procedure effectively minimizes LQ.

We flagged apparently significant sources that do not have good localization fits (LQ $> 8$) with Flag 9  
(\S~\ref{catalog_analysis_flags}) and for them estimated the position and uncertainty by performing 
a moment analysis of an LTS map instead of fitting a quadratic form. 
Some sources that did not have a well-defined peak in the likelihood were discarded by hand, on the consideration that they were most likely related to residual diffuse emission. Another possibility 
is that two adjacent sources produce a dumbbell-like shape; 
for a few of these cases we added a new source by hand.

As in 3FGL, we checked the sources spatially associated with 984 AGN counterparts, comparing their locations with the well-measured positions of the counterparts. 
Better statistics allowed examination of the distributions of the differences separately for bright, dim, and moderate-brightness sources. From this we estimate the absolute precision $\Delta_{\rm abs}$ (at the 95\% confidence level) more accurately at $\sim 0\fdg0068$, up from $\sim 0\fdg005$ in 3FGL.
The systematic factor $f_{\rm rel}$ was 1.06, slightly up from 1.05 in 3FGL.
Eq.~\ref{eq:locsysts} shows how the statistical errors $\Delta_{\rm stat}$ are transformed into total errors $\Delta_{\rm tot}$:
\begin{equation}
\Delta_{\rm tot}^2 = (f_{\rm rel} \, \Delta_{\rm stat})^2 + \Delta_{\rm abs}^2
\label{eq:locsysts}
\end{equation}
which is applied to both ellipse axes.

    \subsection{Significance and Thresholding}
\label{catalog_significance}

The framework for this stage of the analysis is inherited from the 3FGL catalog. It splits the sky into RoIs, varying typically half a dozen sources near the center of the RoI at the same time. 
Each source is entered into the fit with the spectral shape and parameters obtained by $pointlike$ (\S~\ref{catalog_detection}), the brightest sources first. Soft sources from $pointlike$ within $0\fdg2$ of bright ones were intentionally deleted. They appear because the simple spectral models we use are not sufficient to account for the spectra of bright sources, but including them would bias the spectral parameters.
There are 1748 RoIs for 4FGL, listed in the \texttt{ROIs} extension of the catalog (Appendix \ref{appendix_fits_format}). The global best fit is reached iteratively, injecting the spectra of sources in the outer parts of the RoI from the previous step or iteration.
In this approach, the diffuse emission model (\S~\ref{DiffuseModel}) is taken from the global templates (including the spectrum, unlike what is done with $pointlike$ in \S~\ref{catalog_detection}) but it is modulated in each RoI by three parameters: normalization (at 1~GeV) and small corrective slope of the Galactic component, and normalization of the isotropic component. 

Among the more than 8,000 seeds coming from the localization stage, we keep only sources with $TS > 25$, corresponding to a significance of just over $4\sigma$ evaluated from the $\chi^2$ distribution with 4 degrees of freedom \citep[position and spectral parameters of a power-law source,][]{mattox96}. The model for the current RoI is readjusted after removing each seed below threshold. The low-energy flux of the seeds below threshold (a fraction of which are real sources) can be absorbed by neighboring sources closer than the PSF radius.
As in 3FGL, we manually added known LAT pulsars that could not be localized by the automatic procedure without phase selection. However none of those reached $TS > 25$ in 4FGL.

We introduced a number of improvements with respect to 3FGL (by decreasing order of importance):
\begin{enumerate}
\item In 3FGL we had already noted that systematic errors due to an imperfect modeling of diffuse emission were larger than statistical errors in the Galactic plane, and at the same level over the entire sky. With twice as much exposure and an improved effective area at low energy with Pass 8, the effect now dominates. The approach adopted in 3FGL (comparing runs with different diffuse models) allowed characterizing the effect globally and flagging the worst offenders, but left purely statistical errors on source parameters. In 4FGL we introduce weights in the maximum likelihood approach (Appendix \ref{appendix_weights}). This allows obtaining directly (although in an approximate way) smaller $TS$ and larger parameter errors, reflecting the level of systematic uncertainties. We estimated the relative spatial and spectral residuals in the Galactic plane where the diffuse emission is strongest. The resulting systematic level $\epsilon \sim$ 3\% was used to compute the weights. This is by far the most important improvement, which avoids reporting many dubious soft sources.
\item The automatic iteration procedure at the next-to-last step of the process was improved. There are now two iteration levels. In a standard iteration the sources and source models are fixed and only the parameters are free. An RoI and all its neighbors are run again until $\log \mathcal{L}$ does not change by more than 10 from the previous iteration. Around that we introduce another iteration level (superiterations). At the first iteration of a given superiteration we reenter all seeds and remove (one by one) those with $TS < 16$. We also systematically check a curved spectral shape versus a power-law fit to each source at this first iteration, and keep the curved spectral shape if the fit is significantly better (\S~\ref{catalog_spectral_shapes}). At the end of a superiteration an RoI (and its neighbors) enters the next superiteration until $\log \mathcal{L}$ does not change by more than 10 from the last iteration of the previous superiteration. This procedure stabilizes the spectral shapes, particularly in the Galactic plane. Seven superiterations were required to reach full convergence. 
\item The fits are now performed from 50~MeV to 1~TeV, and the overall significances (\texttt{Signif\_Avg}) as well as the spectral parameters refer to the full band. The total energy flux, on the other hand, is still reported between 100~MeV and 100~GeV. For hard sources with photon index less than 2 integrating up to 1~TeV would result in much larger uncertainties. The same is true for soft sources with photon index larger than 2.5 when integrating down to 50~MeV.
\item We considered the effect of energy dispersion in the approximate way implemented in the Science Tools. The effect of energy dispersion is calculated globally for each source, and applied to the whole 3D model of that source, rather than accounting for energy dispersion separately in each pixel. This approximate rescaling captures the main effect (which is only a small correction, see \S~\ref{compare_stepbystep}) at a very minor computational cost. In evaluating the likelihood function, the effects of energy dispersion were not applied to the isotropic background and the Sun/Moon components whose spectra were obtained from the data without considering energy dispersion.
\item We used smaller RoIs at higher energy because we are interested in the core region only, which contains the sources whose parameters come from that RoI (sources in the outer parts of the RoI are entered only as background). The core region is the same for all energy intervals, and the RoI is obtained by adding a ring to that core region, whose width adapts to the PSF and therefore decreases with energy (Table~\ref{tab:components}). This does not significantly affect the result because the outer parts of the RoI would not have been correlated to the inner sources at high energy anyway, but saves memory and CPU time.
\item At the last step of the fitting procedure we tested all spectral shapes described in \S~\ref{catalog_spectral_shapes} (including log-normal for pulsars and cutoff power law for other sources), readjusting the parameters (but not the spectral shapes) of neighboring sources.
\end{enumerate}

We used only binned likelihood analysis in 4FGL because unbinned mode is much more CPU intensive, and does not support weights or energy dispersion.
We split the data into fifteen components, selected according to PSF event type and described in Table~\ref{tab:components}. As explained in \S~\ref{EarthLimb} at low energy we kept only the event types with the best PSF.
 Each event type selection has its own isotropic diffuse template (because it includes residual charged-particle background, which depends on event type).
A single component is used above 10~GeV to save memory and CPU time: at high energy the background under the PSF is small, so keeping the event types separate does not markedly improve significance; it would help for localization, but this is done separately (\S~\ref{catalog_localization}).

A known inconsistency in acceptance exists between Pass 8 PSF event types. It is easy to see on bright sources or the entire RoI spectrum and peaks at the level of 10\% between PSF0 (positive residuals, underestimated effective area) and PSF3 (negative residuals, overestimated effective area) at a few GeV. In that range all event types were considered so the effect on source spectra average out. Below 1 GeV the PSF0 event type was discarded but the discrepancy is lower at low energy. We checked by comparing with preliminary corrected IRFs that the energy fluxes indeed tend to be underestimated, but by only 3\%. The bias on power-law index is less than 0.01.

    \subsection{Spectral Shapes}
\label{catalog_spectral_shapes}

The spectral representation of sources largely follows what was done in 3FGL, considering three spectral models (power law, power law with subexponential cutoff, and log-normal).
We changed two important aspects of how we parametrize the cutoff power law:
\begin{itemize}
\item The cutoff energy was replaced by an exponential factor ($a$ in Eq.~\ref{eq:expcutoff}) which is allowed to be positive. This makes the simple power law a special case of the cutoff power law and allows fitting that model to all sources, even those with negligible curvature.
\item We set the exponential index ($b$ in Eq.~\ref{eq:expcutoff}) to 2/3 (instead of 1) for all pulsars that are too faint for it to be left free. This recognizes the fact that $b < 1$ (subexponential) in all six bright pulsars that have $b$ free in 4FGL. Three have $b \sim 0.55$ and three have $b \sim 0.75$. We chose 2/3 as a simple intermediate value.
\end{itemize}

For all three spectral representations in 4FGL, the normalization (flux density $K$) is defined at a reference energy $E_0$ chosen such that the error on $K$ is minimal. $E_0$ appears as \texttt{Pivot\_Energy} in the FITS table version of the catalog (Appendix \ref{appendix_fits_format}). The 4FGL spectral forms are thus:
\begin{itemize}
\item a log-normal representation (\texttt{LogParabola} under \texttt{SpectrumType} in the FITS table) for all significantly curved spectra except pulsars, 3C 454.3 and the Small Magellanic Cloud (SMC):
\begin{equation}
\frac{{\rm d}N}{{\rm d}E} = K \left (\frac{E}{E_0}\right )^{-\alpha -
\beta\log(E/E_0)}.
\label{eq:logparabola}
\end{equation}
The parameters $K$, $\alpha$ (spectral slope at $E_0$) and the curvature $\beta$ appear as \texttt{LP\_Flux\_Density}, \texttt{LP\_Index} and \texttt{LP\_beta} in the FITS table, respectively. No significantly negative $\beta$ (spectrum curved upwards) was found. The maximum allowed $\beta$ was set to 1 as in 3FGL. Those parameters were used for fitting because they allow minimizing the correlation between $K$ and the other parameters, but a more natural representation would use the peak energy $E_{\rm peak}$ at which the spectrum is maximum (in $\nu F_\nu$ representation)
\begin{equation}
E_{\rm peak} = E_0 \; \exp \left( \frac{2 - \alpha}{2 \, \beta} \right ).
\label{eq:EPeak}
\end{equation}
\item a subexponentially cutoff power law for all significantly curved pulsars (\texttt{PLSuperExpCutoff} under \texttt{SpectrumType} in the FITS table):
\begin{equation}
\frac{{\rm d}N}{{\rm d}E} = K \left (\frac{E}{E_0}\right )^{-\Gamma} \exp \left(a \, (E_0^b - E^b) \right)
\label{eq:expcutoff}
\end{equation}
where $E_0$ and $E$ in the exponential are expressed in MeV. The parameters $K$, $\Gamma$ (low-energy spectral slope), $a$ (exponential factor in MeV$^{-b}$) and $b$ (exponential index) appear as \texttt{PLEC\_Flux\_Density}, \texttt{PLEC\_Index}, \texttt{PLEC\_Expfactor} and \texttt{PLEC\_Exp\_Index} in the FITS table, respectively. Note that in the Science Tools that spectral shape is called \texttt{PLSuperExpCutoff2} and no $E_0^b$ term appears in the exponential, so the error on $K$ (\texttt{Unc\_PLEC\_Flux\_Density} in the FITS table) was obtained from the covariance matrix. The minimum $\Gamma$ was set to 0 (in 3FGL it was set to 0.5, but a smaller $b$ results in a smaller $\Gamma$). No significantly negative $a$ (spectrum curved upwards) was found.
\item a simple power-law form (Eq.~\ref{eq:expcutoff} without the exponential term) for all sources not significantly curved. For those parameters $K$ and $\Gamma$ appear as \texttt{PL\_Flux\_Density} and \texttt{PL\_Index} in the FITS table.
\end{itemize}

The power law is a mathematical model that is rarely sustained by astrophysical sources over as broad a band as 50~MeV to 1~TeV. All bright sources in 4FGL are actually significantly curved downwards. Another drawback of the power-law model is that it tends to exceed the data at both ends of the spectrum, where constraints are weak. It is not a worry at high energy, but at low energy (broad PSF) the collection of faint sources modeled as power laws generates an effectively diffuse excess in the model, which will make the curved sources more curved than they should be. Using a \texttt{LogParabola} spectral shape for all sources would be physically reasonable, but the very large correlation between sources at low energy due to the broad PSF makes that unstable. 

We use the curved representation in the global model (used to fit neighboring sources) if $TS_{\rm curv} > 9$ ($3 \, \sigma$ significance) where $TS_{\rm curv} = 2 \log (\mathcal{L}$(curved spectrum)$/\mathcal{L}$(power-law)). This is a step down from 3FGL or FL8Y, where the threshold was at 16, or $4 \, \sigma$, while preserving stability.
The curvature significance is reported as \texttt{LP\_SigCurv} or \texttt{PLEC\_SigCurv}, replacing the former unique \texttt{Signif\_Curve} column of 3FGL. Both values were derived from $TS_{\rm curv}$ and corrected for systematic uncertainties on the effective area following Eq.~3 of 3FGL. As a result, 51 \texttt{LogParabola} sources (with $TS_{\rm curv} > 9$) have \texttt{LP\_SigCurv} less than 3.

Sources with curved spectra are considered significant whenever $TS > 25 + 9 = 34$. This is similar to the 3FGL criterion, which requested $TS > 25$ in the power-law representation, but accepts a few more strongly curved faint sources (pulsar-like).

One more pulsar (PSR J1057$-$5226) was fit with a free exponential index, besides the six sources modeled in this way in 3FGL.
The Crab was modeled with three spectral components as in 3FGL, but the inverse Compton emission of the nebula (now an extended source, \S~\ref{catalog_extended}) was represented as a log-normal instead of a simple power law. The parameters of that component were fixed to $\alpha=1.75$, $\beta=0.08$, $K=5.5 \times 10^{-13}$ ph cm$^{-2}$ MeV$^{-1}$ s$^{-1}$ at 10~GeV, mimicking the broken power-law fit by \citet{LAT12_Crab}. They were unstable (too much correlation with the pulsar) without phase selection.
Four extended sources had fixed parameters in 3FGL. The parameters in these sources (Vela X, MSH 15$-$52, $\gamma$ Cygni and the Cygnus X cocoon) were freed in 4FGL.

Overall in 4FGL seven sources (the six brightest pulsars and 3C 454.3) were fit as \texttt{PLSuperExpCutoff} with free $b$ (Eq.~\ref{eq:expcutoff}), 214 pulsars were fit as \texttt{PLSuperExpCutoff} with $b=2/3$, the SMC was fit as \texttt{PLSuperExpCutoff} with $b=1$, 1302 sources were fit as \texttt{LogParabola} (including the fixed inverse Compton component of the Crab and 38 other extended sources) and the rest were represented as power laws. The larger fraction of curved spectra compared to 3FGL is due to the lower $TS_{\rm curv}$ threshold.

\begin{figure*}
   \centering
   \begin{tabular}{cc}
   \includegraphics[width=0.5\textwidth]{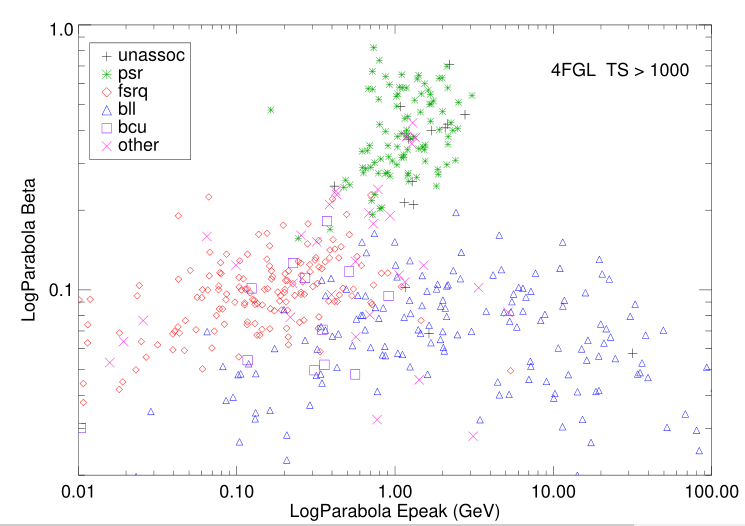} & 
   \includegraphics[width=0.5\textwidth]{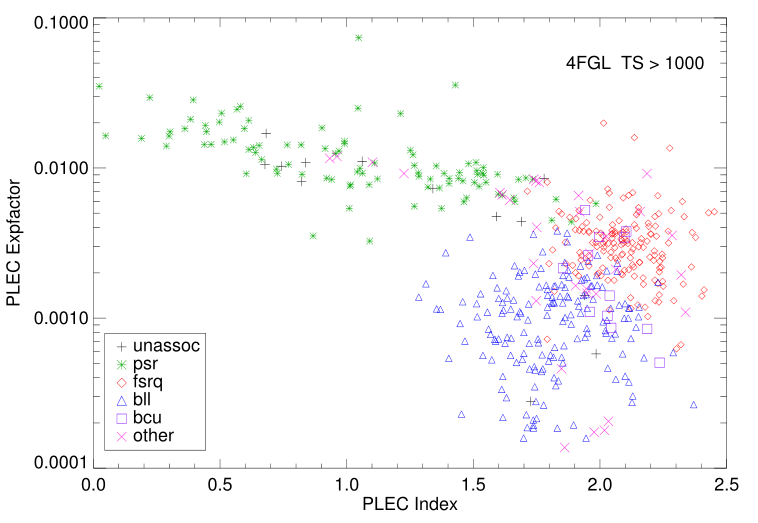}
   \end{tabular}
\caption{Spectral parameters of all bright sources ($TS > 1000$). The different source classes (\S~\ref{sec:assocsum}) are depicted by different symbols and colors. Left: log-normal shape parameters $E_{\rm peak}$ (Eq.~\ref{eq:EPeak}) and $\beta$. Right: subexponentially cutoff power-law shape parameters $\Gamma$ and $a$ (Eq.~\ref{eq:expcutoff}).}
\label{fig:spectralparams}
\end{figure*}

The way the parameters are reported has changed as well:
\begin{itemize}
\item The spectral shape parameters are now explicitly associated to the spectral model they come from. They are reported as Shape\_Param where Shape is one of PL (\texttt{PowerLaw}), PLEC (\texttt{PLSuperExpCutoff}) or LP (\texttt{LogParabola}) and Param is the parameter name. Columns Shape\_\texttt{Index} replace \texttt{Spectral\_Index} which was ambiguous.
\item All sources were fit with the three spectral shapes, so all fields are filled. The curvature significance is calculated twice by comparing power law with both log-normal and exponentially cutoff power law (although only one is actually used to switch to the curved shape in the global model, depending on whether the source is a pulsar or not). There are also three Shape\_\texttt{Flux\_Density} columns referring to the same \texttt{Pivot\_Energy}. The preferred spectral shape (reported as \texttt{SpectrumType}) remains what is used in the global model, when the source is part of the background (i.e., when fitting the other sources). It is also what is used to derive the fluxes, their uncertainties and the significance.
\end{itemize}
This additional information allows comparing unassociated sources with either pulsars or blazars using the same spectral shape. This is illustrated on Figure~\ref{fig:spectralparams}. Pulsar spectra are more curved than AGN, and among AGN flat-spectrum radio quasars (FSRQ) peak at lower energy than BL Lacs (BLL). It is clear that when the error bars are small (bright sources) any of those plots is very discriminant for classifying sources. They complement the variability versus curvature plot (Figure 8 of the 1FGL paper). We expect most of the (few) bright remaining unassociated sources (black plus signs) to be pulsars, from their location on those plots. The same reasoning implies that most of the unclassified blazars (bcu) should be flat-spectrum radio quasars, although the distinction with BL Lacs is less clear-cut than with pulsars. Unfortunately most unassociated sources are faint ($TS < 100$) and for those the same plots are very confused, because the error bars become comparable to the ranges of parameters.

    \subsection{Extended Sources}
\label{catalog_extended}

As in the 3FGL catalog, we explicitly model as spatially extended those LAT sources that have been shown in dedicated analyses to be spatially resolved by the LAT. 
The catalog process does not involve looking for new extended sources, testing possible extension of sources detected as point-like, nor refitting the spatial shapes of known extended sources.

Most templates are geometrical, so they are not perfect matches to the data and the source detection often finds residuals on top of extended sources, which are then converted into additional point sources.
As in 3FGL those additional point sources were intentionally deleted from the model, except if they met two of the following criteria: associated with a plausible counterpart known at other wavelengths, much harder than the extended source (\texttt{Pivot\_Energy} larger by a factor $e$ or more), or very significant ($TS > 100$). Contrary to 3FGL, that procedure was applied inside the Cygnus X cocoon as well.

The latest compilation of extended \Fermilat sources prior to this work consists of the 55 extended sources entered in the 3FHL catalog of sources above 10~GeV \citep{LAT17_3FHL}. This includes the result of the systematic search for new extended sources in the Galactic plane ($|b| < 7\degr$) above 10~GeV \citep[FGES,][]{LAT17_10GeVES}. Two of those were not propagated to 4FGL:
\begin{itemize}
\item FGES J1800.5$-$2343 was replaced by the W~28 template from 3FGL, and the nearby excesses \citep{W28_2014} were left to be modeled as point sources.
\item FGES J0537.6+2751 was replaced by the radio template of S~147 used in 3FGL, which fits better than the disk used in the FGES paper (S~147 is a soft source, so it was barely detected above 10~GeV).
\end{itemize}

The supernova remnant (SNR) MSH~15-56 was replaced by two morphologically distinct components, following \citet{MSH1556_2018}: one for the SNR (SNR mask in the paper), the other one for the pulsar wind nebula (PWN) inside it (radio template). We added back the W~30 SNR on top of FGES J1804.7$-$2144 (coincident with HESS J1804$-$216). The two overlap but the best localization clearly moves with energy from W~30 to HESS J1804$-$216.

Eighteen sources were added, resulting in 75 extended sources in 4FGL:
\begin{itemize}
\item The Rosette nebula and Monoceros SNR (too soft to be detected above 10~GeV) were characterized by \citet{LAT16_Monoceros}. We used the same templates.
\item The systematic search for extended sources outside the Galactic plane above 1~GeV \citep[FHES,][]{FHES_2018} found sixteen reliable extended sources. Three of them were already known as extended sources. Two were extensions of the Cen A lobes, which appear larger in $\gamma$ rays than the WMAP template that we use following \citet{LAT10_CenAlobes}. We did not consider them, waiting for a new morphological analysis of the full lobes. We ignored two others: M~31 \citep[extension only marginally significant, both in FHES and][]{LAT17_M31} and CTA 1 (SNR G119.5+10.2) around PSR J0007+7303 (not significant without phase gating). We introduced the nine remaining FHES sources, including the inverse Compton component of the Crab nebula and the $\rho$ Oph star-forming region (= FHES J1626.9$-$2431). One of them (FHES J1741.6$-$3917) was reported by \citet{FHESJ1741_2018} as well, with similar extension.
\item Four HESS sources were found to be extended sources in the \Fermilat range as well: HESS J1534$-$571 \citep{HESSJ1534_2017}, HESS J1808$-$204 \citep{HESSJ1808_2016}, HESS J1809$-$193 and HESS J1813$-$178 \citep{HESSJ1809_2018}.
\item Three extended sources were discovered in the search for GeV emission from magnetars \citep{Magnetars_2017}. They contain SNRs (Kes 73, Kes 79 and G42.8+0.6) but are much bigger than the radio SNRs. One of them (around Kes 73) was also noted by \citet{Kes73_2017}.
\end{itemize}

Table~\ref{tbl:extended} lists the source name, origin, spatial template and the reference for the dedicated analysis. These sources are tabulated with the point sources, with the only distinction being that no position uncertainties are reported and their names end in \texttt{e} (see Appendix \ref{appendix_fits_format}).  Unidentified point sources inside extended ones are indicated as ``xxx field'' in the \texttt{ASSOC2} column of the catalog.

\startlongtable
\begin{deluxetable*}{llllcl}
\tabletypesize{\scriptsize}
\tablecaption{Extended Sources Modeled in the 4FGL Analysis
\label{tbl:extended}}
\tablewidth{0pt}
\tablehead{
\colhead{4FGL Name}&
\colhead{Extended Source}&
\colhead{Origin}&
\colhead{Spatial Form}&
\colhead{Extent [deg]}&
\colhead{Reference}
}

\startdata
J0058.0$-$7245e & SMC Galaxy & Updated & Map & 1.5 & \citet{SMC_DM2016} \\
J0221.4+6241e & HB 3 & New & Disk & 0.8 & \citet{HB3_2016} \\
J0222.4+6156e & W 3 & New & Map & 0.6 & \citet{HB3_2016} \\
J0322.6$-$3712e & Fornax A & 3FHL & Map & 0.35 & \citet{Fornax2016} \\
J0427.2+5533e & SNR G150.3+4.5 & 3FHL & Disk & 1.515 & \citet{LAT17_10GeVES} \\
J0500.3+4639e & HB 9 & New & Map & 1.0 & \citet{Araya2014_HB9} \\
J0500.9$-$6945e & LMC FarWest & 3FHL & Map\tablenotemark{$a$} & 0.9 & \citet{LMC2016} \\
J0519.9$-$6845e & LMC Galaxy & New & Map\tablenotemark{$a$} & 3.0 & \citet{LMC2016} \\
J0530.0$-$6900e & LMC 30DorWest & 3FHL & Map\tablenotemark{$a$} & 0.9 & \citet{LMC2016} \\
J0531.8$-$6639e & LMC North & 3FHL & Map\tablenotemark{$a$} & 0.6 & \citet{LMC2016} \\
J0534.5+2201e & Crab nebula IC & New & Gaussian & 0.03 & \citet{FHES_2018} \\
J0540.3+2756e & S 147 & 3FGL & Disk & 1.5 & \citet{LAT12_S147} \\
J0617.2+2234e & IC 443 & 2FGL & Gaussian & 0.27 & \citet{LAT10_IC443} \\
J0634.2+0436e & Rosette & New & Map & (1.5, 0.875) & \citet{LAT16_Monoceros} \\
J0639.4+0655e & Monoceros & New & Gaussian & 3.47 & \citet{LAT16_Monoceros} \\
J0822.1$-$4253e & Puppis A & 3FHL & Disk & 0.443 & \citet{LAT17_10GeVES} \\
J0833.1$-$4511e & Vela X & 2FGL & Disk & 0.91 & \citet{LAT10_VelaX} \\
J0851.9$-$4620e & Vela Junior & 3FHL & Disk & 0.978 & \citet{LAT17_10GeVES} \\
J1023.3$-$5747e & Westerlund 2 & 3FHL & Disk & 0.278 & \citet{LAT17_10GeVES} \\
J1036.3$-$5833e & FGES J1036.3$-$5833 & 3FHL & Disk & 2.465 & \citet{LAT17_10GeVES} \\
J1109.4$-$6115e & FGES J1109.4$-$6115 & 3FHL & Disk & 1.267 & \citet{LAT17_10GeVES} \\
J1208.5$-$5243e & SNR G296.5+10.0 & 3FHL & Disk & 0.76 & \citet{LAT2016_SNRCat} \\
J1213.3$-$6240e & FGES J1213.3$-$6240 & 3FHL & Disk & 0.332 & \citet{LAT17_10GeVES} \\
J1303.0$-$6312e & HESS J1303$-$631 & 3FGL & Gaussian & 0.24 & \citet{HESS05_J1303} \\
J1324.0$-$4330e & Centaurus A (lobes) & 2FGL & Map & (2.5, 1.0) & \citet{LAT10_CenAlobes} \\
J1355.1$-$6420e & HESS J1356$-$645 & 3FHL & Disk & 0.405 & \citet{LAT17_10GeVES} \\
J1409.1$-$6121e & FGES J1409.1$-$6121 & 3FHL & Disk & 0.733 & \citet{LAT17_10GeVES} \\
J1420.3$-$6046e & HESS J1420$-$607 & 3FHL & Disk & 0.123 & \citet{LAT17_10GeVES} \\
J1443.0$-$6227e & RCW 86 & 3FHL & Map & 0.3 & \citet{RCW86_2016} \\
J1501.0$-$6310e & FHES J1501.0$-$6310 & New & Gaussian & 1.29 & \citet{FHES_2018} \\
J1507.9$-$6228e & HESS J1507$-$622 & 3FHL & Disk & 0.362 & \citet{LAT17_10GeVES} \\
J1514.2$-$5909e & MSH 15$-$52 & 3FHL & Disk & 0.243 & \citet{LAT17_10GeVES} \\
J1533.9$-$5712e & HESS J1534$-$571 & New & Disk & 0.4 & \citet{HESSJ1534_2017} \\
J1552.4$-$5612e & MSH 15$-$56 PWN & New & Map & 0.08 & \citet{MSH1556_2018} \\
J1552.9$-$5607e & MSH 15$-$56 SNR & New & Map & 0.3 & \citet{MSH1556_2018} \\
J1553.8$-$5325e & FGES J1553.8$-$5325 & 3FHL & Disk & 0.523 & \citet{LAT17_10GeVES} \\
J1615.3$-$5146e & HESS J1614$-$518 & 3FGL & Disk & 0.42 & \citet{LAT12_extended} \\
J1616.2$-$5054e & HESS J1616$-$508 & 3FGL & Disk & 0.32 & \citet{LAT12_extended} \\
J1626.9$-$2431e & FHES J1626.9$-$2431 & New & Gaussian & 0.29 & \citet{FHES_2018} \\
J1631.6$-$4756e & FGES J1631.6$-$4756 & 3FHL & Disk & 0.256 & \citet{LAT17_10GeVES} \\
J1633.0$-$4746e & FGES J1633.0$-$4746 & 3FHL & Disk & 0.61 & \citet{LAT17_10GeVES} \\
J1636.3$-$4731e & SNR G337.0$-$0.1 & 3FHL & Disk & 0.139 & \citet{LAT17_10GeVES} \\
J1642.1$-$5428e & FHES J1642.1$-$5428 & New & Disk & 0.696 & \citet{FHES_2018} \\
J1652.2$-$4633e & FGES J1652.2$-$4633 & 3FHL & Disk & 0.718 & \citet{LAT17_10GeVES} \\
J1655.5$-$4737e & FGES J1655.5$-$4737 & 3FHL & Disk & 0.334 & \citet{LAT17_10GeVES} \\
J1713.5$-$3945e & RX J1713.7$-$3946 & 3FHL & Map & 0.56 & \citet{RXJ1713_2018} \\
J1723.5$-$0501e & FHES J1723.5$-$0501 & New & Gaussian & 0.73 & \citet{FHES_2018} \\
J1741.6$-$3917e & FHES J1741.6$-$3917 & New & Disk & 1.65 & \citet{FHES_2018} \\
J1745.8$-$3028e & FGES J1745.8$-$3028 & 3FHL & Disk & 0.528 & \citet{LAT17_10GeVES} \\
J1801.3$-$2326e & W 28 & 2FGL & Disk & 0.39 & \citet{LAT10_W28} \\
J1804.7$-$2144e & HESS J1804$-$216 & 3FHL & Disk & 0.378 & \citet{LAT17_10GeVES} \\
J1805.6$-$2136e & W 30 & 2FGL & Disk & 0.37 & \citet{LAT12_W30} \\
J1808.2$-$2028e & HESS J1808$-$204 & New & Disk & 0.65 & \citet{HESSJ1808_2016} \\
J1810.3$-$1925e & HESS J1809$-$193 & New & Disk & 0.5 & \citet{HESSJ1809_2018} \\
J1813.1$-$1737e & HESS J1813$-$178 & New & Disk & 0.6 & \citet{HESSJ1809_2018} \\
J1824.5$-$1351e & HESS J1825$-$137 & 2FGL & Gaussian & 0.75 & \citet{LAT11_J1825} \\
J1834.1$-$0706e & SNR G24.7+0.6 & 3FHL & Disk & 0.214 & \citet{LAT17_10GeVES} \\
J1834.5$-$0846e & W 41 & 3FHL & Gaussian & 0.23 & \citet{W41_2015} \\
J1836.5$-$0651e & FGES J1836.5$-$0651 & 3FHL & Disk & 0.535 & \citet{LAT17_10GeVES} \\
J1838.9$-$0704e & FGES J1838.9$-$0704 & 3FHL & Disk & 0.523 & \citet{LAT17_10GeVES} \\
J1840.8$-$0453e & Kes 73 & New & Disk & 0.32 & \citet{Magnetars_2017} \\
J1840.9$-$0532e & HESS J1841$-$055 & 3FGL & 2D Gaussian  & (0.62, 0.38) & \citet{HESS08_UNID} \\
J1852.4+0037e & Kes 79 & New & Disk & 0.63 & \citet{Magnetars_2017} \\
J1855.9+0121e & W 44 & 2FGL & 2D Ring & (0.30, 0.19) & \citet{LAT10_W44} \\
J1857.7+0246e & HESS J1857+026 & 3FHL & Disk & 0.613 & \citet{LAT17_10GeVES} \\
J1908.6+0915e & SNR G42.8+0.6 & New & Disk & 0.6 & \citet{Magnetars_2017} \\
J1923.2+1408e & W 51C & 2FGL & 2D Disk & (0.375, 0.26) & \citet{LAT09_W51C} \\
J2021.0+4031e & $\gamma$ Cygni & 3FGL & Disk & 0.63 & \citet{LAT12_extended} \\
J2028.6+4110e & Cygnus X cocoon & 3FGL & Gaussian & 3.0 & \citet{LAT11_CygCocoon} \\
J2045.2+5026e & HB 21 & 3FGL & Disk & 1.19 & \citet{LAT13_HB21} \\
J2051.0+3040e & Cygnus Loop & 2FGL & Ring & 1.65 & \citet{LAT11_CygnusLoop} \\
J2129.9+5833e & FHES J2129.9+5833 & New & Gaussian & 1.09 & \citet{FHES_2018} \\
J2208.4+6443e & FHES J2208.4+6443 & New & Gaussian & 0.93 & \citet{FHES_2018} \\
J2301.9+5855e & CTB 109 & 3FHL & Disk & 0.249 & \citet{LAT17_10GeVES} \\
J2304.0+5406e & FHES J2304.0+5406 & New & Gaussian & 1.58 & \citet{FHES_2018} \\
\enddata

\tablenotetext{a}{Emissivity model.}

\tablecomments{~List of all sources that have been modeled as spatially extended. The Origin column gives the name of the \Fermilat catalog in which that spatial template was introduced. The Extent column indicates the radius for Disk (flat disk) sources, the 68\% containment radius for Gaussian sources, the outer radius for Ring (flat annulus) sources, and an approximate radius for Map (external template) sources. The 2D shapes are elliptical; each pair of parameters $(a, b)$ represents the semi-major $(a)$ and semi-minor $(b)$ axes.}

\end{deluxetable*}

    \subsection{Flux Determination}
\label{catalog_flux_determination}

\begin{figure*}
   \centering
   \begin{tabular}{cc}
   \includegraphics[viewport=20 20 700 540,clip,width=0.5\textwidth]{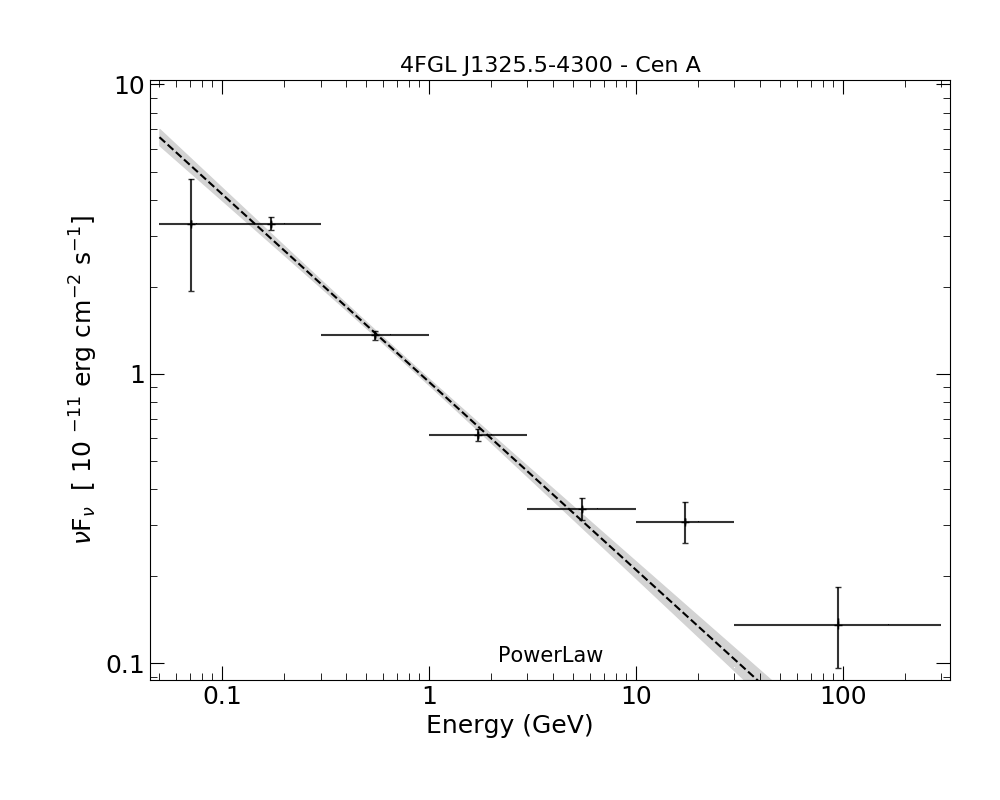} & 
   \includegraphics[viewport=20 20 700 540,clip,width=0.5\textwidth]{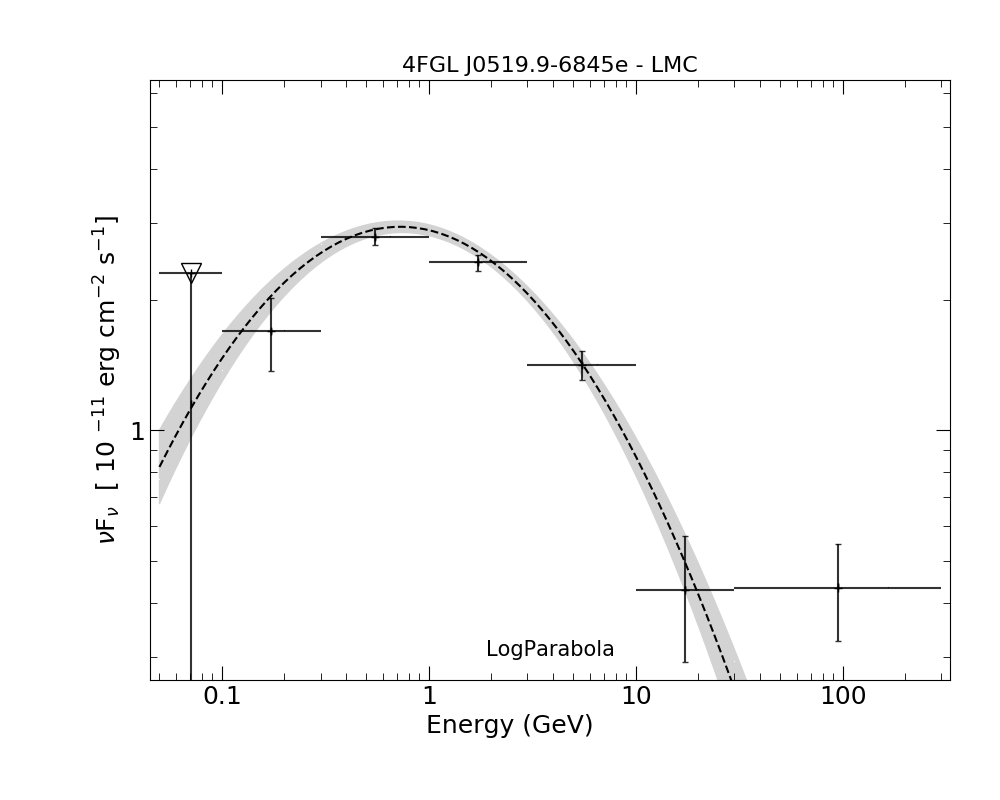} \\
   \includegraphics[viewport=20 20 700 540,clip,width=0.5\textwidth]{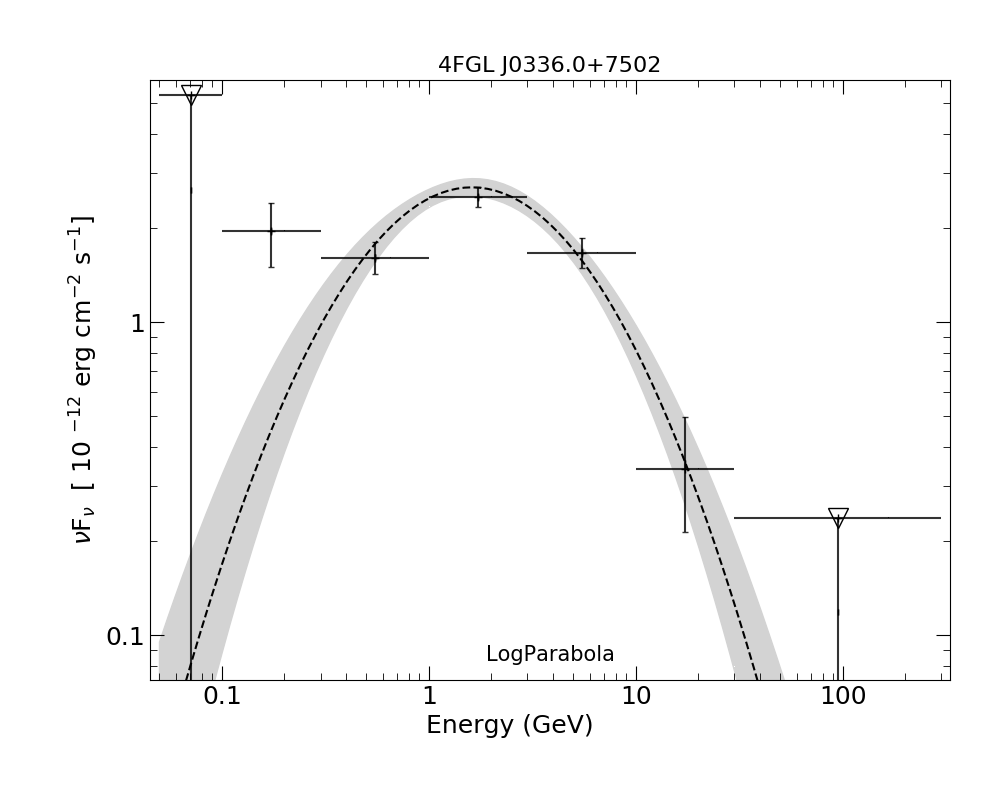} & 
   \includegraphics[viewport=20 20 700 540,clip,width=0.5\textwidth]{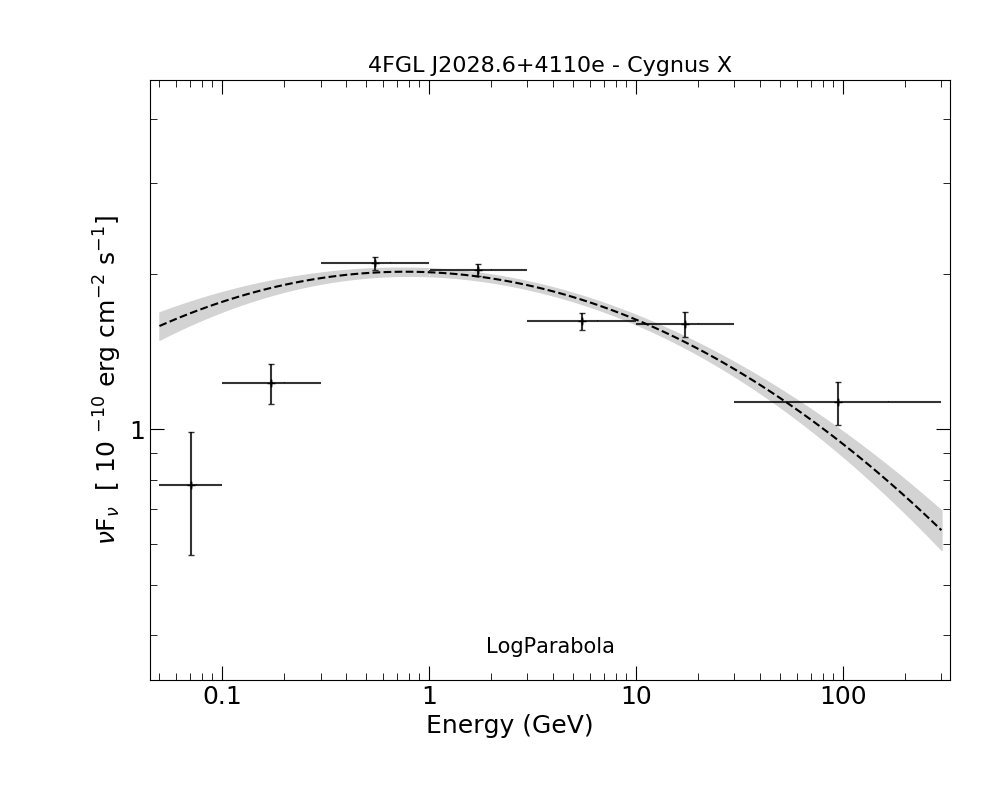} \\
   \end{tabular}
\caption{Spectral energy distributions of four sources flagged with bad spectral fit quality (Flag 10 in Table~\ref{tab:flags}). On all plots the dashed line is the best fit from the analysis over the full energy range, and the gray shaded area shows the uncertainty obtained from the covariance matrix on the spectral parameters. Downward triangles indicate upper limits at 95\% confidence level. The vertical scale is not the same in all plots.
Top left, the Cen A radio galaxy (4FGL J1325.5$-$4300) fit by a power law with $\Gamma = 2.65$: it is a good representation up to 10~GeV, but the last two points deviate from the power-law fit.
Top right, the Large Magellanic Cloud (4FGL J0519.9$-$6845e): the fitted LogParabola spectrum appears to drop too fast at high energy.
Bottom left, the unassociated source 4FGL J0336.0+7502: the low-energy points deviate from the LogParabola fit.
Bottom right, the Cygnus X cocoon (4FGL J2028.6+4110e): the deviation from the LogParabola fit at the first two points is probably spurious, due to source confusion.}
\label{fig:fourSEDs}
\end{figure*}



Thanks to the improved statistics, the source photon fluxes in 4FGL are reported in seven energy bands (1: 50 to 100~MeV; 2: 100 to 300~MeV; 3: 300~MeV to 1~GeV; 4: 1 to 3~GeV; 5: 3 to 10~GeV; 6: 10 to 30~GeV; 7: 30 to 300~GeV) extending both below and above the range (100~MeV to 100~GeV) covered in 3FGL. 
Up to 10~GeV, the data files were exactly the same as in the global fit (Table~\ref{tab:components}). To get the best sensitivity in band 6 (10 to 30~GeV), we split the data into 4 components per event type, using pixel size $0\fdg04$ for PSF3, $0\fdg05$ for PSF2, $0\fdg1$ for PSF1 and $0\fdg2$ for PSF0. Above 30~GeV (band 7) we used unbinned likelihood, which is as precise while using much smaller files. It does not allow correcting for energy dispersion, but this is not an important issue in that band.
The fluxes were obtained by freezing the power-law index to that obtained in the fit over the full range and adjusting the normalization in each spectral band. For the curved spectra (\S~\ref{catalog_spectral_shapes}) the photon index in a band was set to the local spectral slope at the logarithmic mid-point of the band $\sqrt{E_n E_{n+1}}$, restricted to be in the interval [0,5].

In each band, the analysis was conducted in the same way as for the 3FGL catalog. To adapt more easily to new band definitions, the results (photon fluxes and uncertainties, $\nu F_{\nu}$ differential fluxes, and significances) are reported in a set of four vector columns (Appendix \ref{appendix_fits_format}: \texttt{Flux\_Band}, \texttt{Unc\_Flux\_Band}, \texttt{nuFnu\_Band}, \texttt{Sqrt\_TS\_Band}) instead of a set of four columns per band as in previous FGL catalogs.

The spectral fit quality is computed in a more precise way than in 3FGL from twice the sum of log-likelihood differences, as we did for the variability index (Sect.~3.6 of the 2FGL paper). The contribution from each band $S_i^2$ also accounts for systematic uncertainties on effective area via
\begin{equation}
\label{eq:tsspec}
S_i^2 = \frac{2 \sigma_i^2}{\sigma_i^2 + (f_i^{\rm rel} F_i^{\rm fit})^2} \; \log \left[ \mathcal{L}_i(F_i^{\rm best}) / \mathcal{L}_i(F_i^{\rm fit}) \right]
\end{equation}
where $i$ runs over all bands, $F_i^{\rm fit}$ is the flux predicted by the global model, $F_i^{\rm best}$ is the flux fitted to band $i$ alone, $\sigma_i$ is the statistical error (upper error if $F_i^{\rm best} \le F_i^{\rm fit}$, lower error if $F_i^{\rm best} > F_i^{\rm fit}$) and the spectral fit quality is simply $\sum_i S_i^2$. The systematic uncertainties\footnote{See \url{https://fermi.gsfc.nasa.gov/ssc/data/analysis/LAT_caveats.html}.} $f_i^{\rm rel}$ are set to 0.15 in the first band, 0.1 in the second and the last bands, and 0.05 in bands 3 to 6. The uncertainty is larger in the first band because only PSF3 events are used.

Too large values of spectral fit quality are flagged (Flag 10 in Table~\ref{tab:flags}). Since there are 7 bands and (for most sources, which are fit with the power-law model) 2 free parameters, the flag is set when $\sum_i S_i^2 > 20.5$ (probability $10^{-3}$ for a $\chi^2$ distribution with 5 degrees of freedom). Only 6 sources trigger this. We also set the same flag whenever any individual band is off by more than $3 \sigma$ ($S_i^2 > 9$). This occurs in 26 sources. Among the 27 sources flagged with Flag 10 (examples in Figure~\ref{fig:fourSEDs}), the Vela and Geminga pulsars are very bright sources for which our spectral representation is not good enough. A few show signs of a real second component in the spectrum, such as Cen A \citep{CenA_HESS}. Several would be better fit by a different spectral model: the Large Magellanic Cloud (LMC) probably decreases at high energy as a power law like our own Galaxy, and 4FGL J0336.0+7502 is better fit by a PLSuperExpCutoff model. The latter is an unassociated source at $15\degr$ latitude, which has a strongly curved spectrum and is not variable: it is a good candidate for a millisecond pulsar. Other sources show deviations at low energy and are in confused regions or close to a brighter neighbor, such as the Cygnus X cocoon. This extended source contains many point sources inside it and the PSF below 300~MeV is too broad to provide a reliable separation.

The fluxes in the 50 to 100~MeV band are very hard to estimate because of the enormous confusion. The average distance between sources ($1\fdg7$) is about equal to the half width at half maximum of PSF3 events in that band, so it is nearly always possible to set a source to 0 and compensate by a suitable combination of flux adjustments in its neighbors. This is why only 34 sources have $TS > 25$ in that band (all are bright sources with global $TS > 700$). This is far fewer than the 198 low-energy (30 - 100~MeV) \Fermilat sources reported by \citet[][1FLE]{LAT18_1FLE}. The reason is that in 4FGL we consider that even faint sources in the catalog can have strong low-energy emission, so the total source flux is distributed over 5000 sources, whereas 1FLE focused on finding individual peaks.

At the other extreme, 618 sources have $TS > 25$ in the 30 to 300~GeV band, which is entirely limited by photon counting ($TS > 25$ in that band corresponds to about 5 events). Only 13 of those are not associated to a 3FHL or FHES source. The brightest of them (at $TS = 54$ in that band) is a hard source associated with 1RXS J224123.5+294244, mostly significant in the last year, after the 3FHL time range.

As in past FGL catalogs, the photon fluxes between 1 and 100~GeV as well as the energy fluxes between 100~MeV and 100~GeV 
were derived from the full-band analysis assuming the best spectral shape, and their uncertainties come from the covariance matrix. Even though the full analysis is carried out down to 50~MeV and up to 1~TeV in 4FGL, we have not changed the energy range over which we quote fluxes so that they can be easily compared with fluxes in past catalogs. The photon fluxes above 100~GeV are negligible except in the very hardest power-law sources, and the energy fluxes below 100~MeV and above 100~GeV are not precisely measured (even for soft and hard sources, respectively).


    \subsection{Variability}
\label{catalog_variability}

\subsubsection{One-year intervals}
\label{lconeyear}

\begin{figure}
\centering
\includegraphics[width=\linewidth]{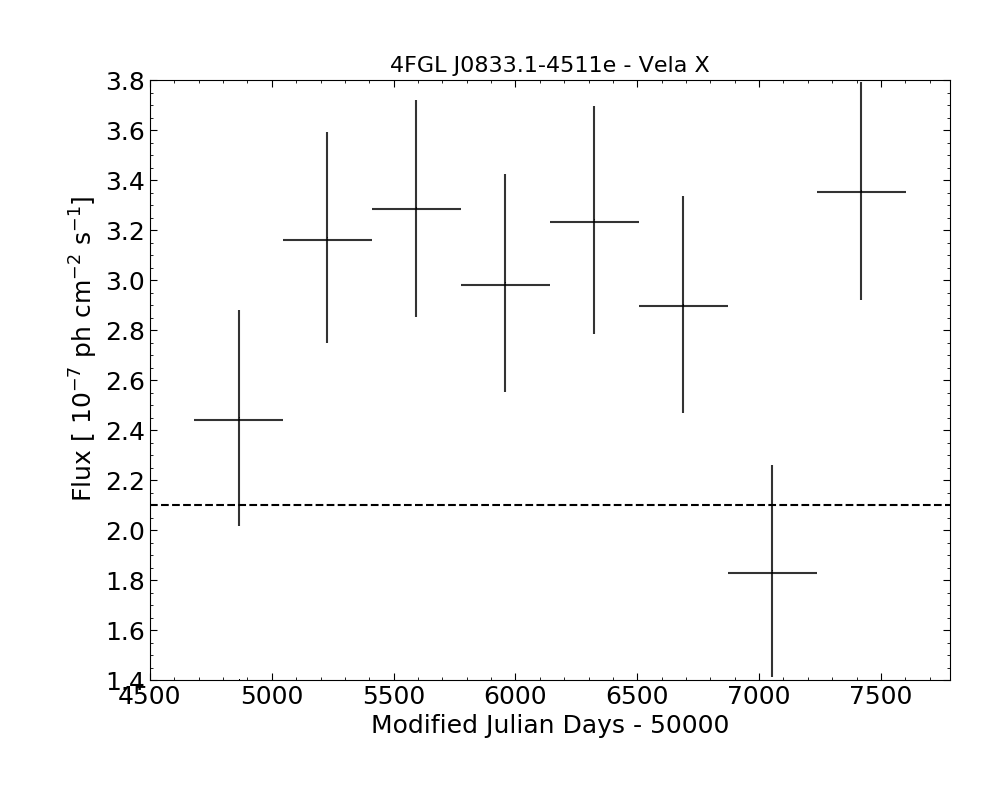}
\caption{Light curve of Vela X (4FGL J0833.1$-$4511e) in the 0.1 to 100~GeV band. It is an extended source that should not be variable. Indeed the yearly fluxes are compatible with a constant (the average flux is $2.9 \times 10^{-7}$ ph cm$^{-2}$ s$^{-1}$), but not with the flux extracted over the full eight years (dashed line, too low). That inconsistency is due to differences in the data analysis settings between the global fit and the fits per year (the weights in particular). Vela X is very close to the very bright Vela pulsar, so it is strongly attenuated by the weights. For most sources the average flux is much closer to the global flux.}
\label{fig:LCVelaX}
\end{figure}

\begin{figure}
\centering
\includegraphics[width=\linewidth]{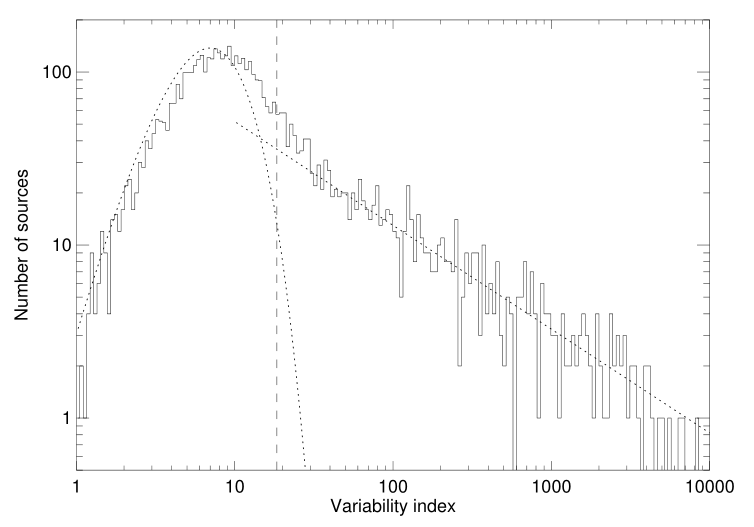}
\caption{Distribution of the variability index (Eq.~\ref{eq::VarIndex}) over one-year intervals. The dotted line at left is the $\chi^2$ distribution for 7 degrees of freedom, expected for a set of non-variable sources. The dotted line at right is a power-law decreasing as $TS_{\rm var}^{-0.6}$. The vertical dashed line is the threshold above which we consider that a source is likely variable.}
\label{fig:VarIndex}
\end{figure}

\begin{figure*}
   \centering
   \begin{tabular}{cc}
   \includegraphics[width=0.5\textwidth]{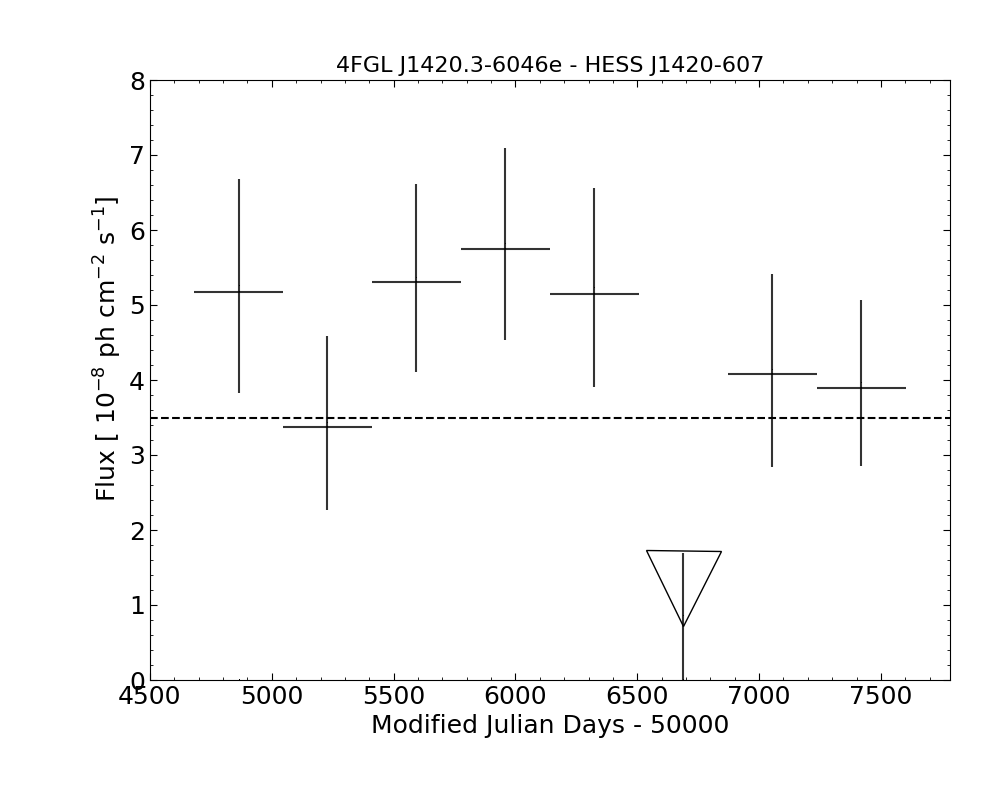} & 
   \includegraphics[width=0.5\textwidth]{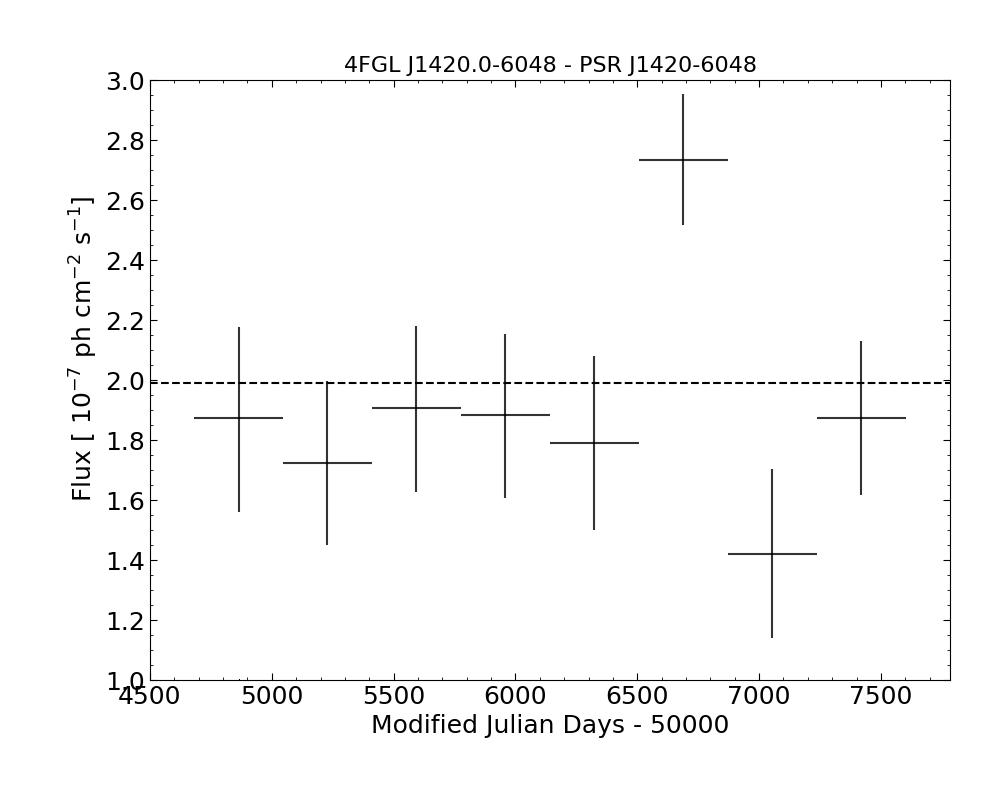}
   \end{tabular}
\caption{Light curves of the pulsar wind nebula HESS J1420$-$607 (4FGL J1420.3$-$6046e) at $TS_{\rm var} = 23.4$ over one-year intervals and its parent pulsar PSR J1420$-$6048 (4FGL J1420.0$-$6048). The apparent variability of HESS J1420$-$607 is due to the low point in the 6$^{\rm th}$ year (the downward triangle is an upper limit at 95\% confidence level), which corresponds to a high point in the light curve of PSR J1420$-$6048. This is clearly a case of incorrect flux transfer due to the strong spatial confusion (the nebula is only $0\fdg12$ in radius), despite the spectral difference between the two sources. The perturbation of the pulsar (brighter than the nebula) is not enough to exceed the variability threshold.}
\label{fig:HESSJ1420}
\end{figure*}


\begin{figure}
\centering
\includegraphics[width=\linewidth]{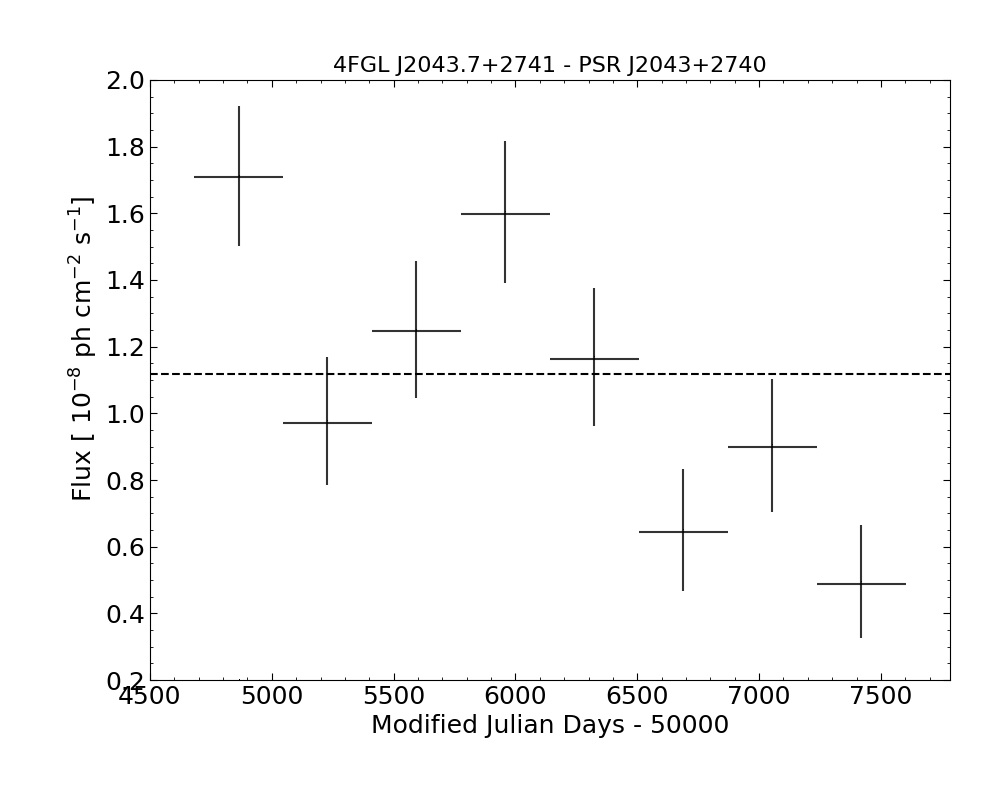}
\caption{Light curve of the pulsar PSR J2043+2740 (4FGL J2043.7+2741), at $TS_{\rm var} = 33$ over one-year intervals. The flux of this pulsar appears to be decreasing secularly.
}
\label{fig:LCPSRJ2043}
\end{figure}

We started by computing light curves over 1-year intervals. This is much faster and more stable than fitting smaller time intervals, and provides a good variability assessment already.
We used binned likelihood and the same data as in the main run up to 10~GeV (Table~\ref{tab:components}), but to save disk space and CPU time we merged event types together. Above 10 GeV we used unbinned likelihood (more efficient when there are few events). We ignored events above 100~GeV (unimportant for variability).

As in 3FGL the fluxes in each interval were obtained by freezing the spectral parameters to those obtained in the fit over the full range and adjusting the normalization. As in previous FGL catalogs, the fluxes in each interval are reported as photon fluxes between 0.1 and 100~GeV.

The weights appropriate for one year were computed using the procedure explained in Appendix~\ref{appendix_weights}, entering the same data cube divided by 8 (we use the same weights in each year), and ignoring the last steps specific to splitting event types. The weights are of course much larger than those for 8 years, but remain a significant correction (the weights are less than 0.2 in the Galactic Ridge up to 300~MeV).
We used the same Sun/Moon model for each year. This amounts to neglecting the modulation of their intrinsic flux along the 11-year solar cycle.

Because of the different weights between the full analysis and that in 1-year intervals, the average flux from the light curve $F_{\rm av}$ can differ somewhat from the flux in the total analysis $F_{\rm glob}$ (low energies are less attenuated in the analysis over 1-year intervals). This is illustrated in Figure~\ref{fig:LCVelaX}. 
In 4FGL we compute the variability index $TS_{\rm var}$ (reported as \texttt{Variability\_Index} in the FITS file) as
\begin{eqnarray}
TS_{\rm var} & = & 2\sum_i \log \left[ \frac{\mathcal{L}_i(F_i)}{\mathcal{L}_i(F_{\rm glob})} \right] - \max \left( \chi^2(F_{\rm glob}) - \chi^2(F_{\rm av}), 0 \right)
\label{eq::VarIndex} \\
\chi^2(F) & = & \sum_i \frac{(F_i - F)^2}{\sigma_i^2}
\end{eqnarray}
where $F_i$ are the individual flux values, $\mathcal{L}_i(F)$ the likelihood in interval $i$ assuming flux $F$ and $\sigma_i$ the errors on $F_i$ (upper error if $F_i \le F$, lower error if $F_i > F$).
The first term in Eq.~\ref{eq::VarIndex} is the same as Eq.~4 of 2FGL.
The second term corrects (in the Gaussian limit) for the difference between $F_{\rm glob}$ and $F_{\rm av}$
(since the average flux is known only at the very end, it could not be entered when computing $\mathcal{L}_i(F)$). We subtract the second term only when it is positive (it is not necessarily positive because the best $\chi^2$ is reached at the average weighted by $\sigma_i^{-2}$, not the straight average).
On the other hand, we did not correct the variability index for the relative systematic error, which is already accounted for in the weighting procedure.

The distribution of observed $TS_{\rm var}$ is shown in Figure~\ref{fig:VarIndex}. It looks like a composite of a power-law distribution and a $\chi^2(7)$ distribution with $N_{\rm int} - 1 = 7$ degrees of freedom, where $N_{\rm int}$ is the number of intervals. The left branch corresponds both to constant sources (such as most pulsars) and sources too faint to have measurable variability. There are many blazars among them, which are most likely just as variable as brighter blazars. This contribution of real variability to $TS_{\rm var}$ is the reason why the histogram is a little offset to the right of the $\chi^2(7)$ distribution (that offset is absent in the Galactic plane, and stronger off the plane).

Variability is considered probable when $TS_{\rm var} > 18.48$, corresponding to 99\% confidence in a $\chi^2(7)$ distribution. We find 1327 variable sources with that criterion. After the $\chi^2$-based correction of Eq.~\ref{eq::VarIndex}, Vela X remains below that threshold.
One extended source still exceeds the variability threshold. This is HESS J1420$-$607 (Figure~\ref{fig:HESSJ1420}), confused with its parent pulsar PSR J1420$-$6048. A similar flux transfer occured in the third year between the Crab pulsar and the Crab Nebula. This can be understood because the synchrotron emission of the nebula becomes much harder during flares, while our pipeline assumes the soft power-law fit over the full interval applies throughout. None of those variabilities are real.

Besides the Crab and the known variable pulsars PSR J1227$-$4853 \citep{PSRJ1227_2015} and PSR J2021+4026 \citep{LAT13_variablePSR}, three other pulsars are above the variability threshold. Two are just above it and can be chance occurrences (there are more than 200 pulsars, so we expect two above the 1\% threshold). The last one is PSR J2043+2740 (Figure~\ref{fig:LCPSRJ2043}), which looks like a case of real variability (secular flux decrease by a factor of 3).

\begin{figure}
\centering
\includegraphics[width=\linewidth]{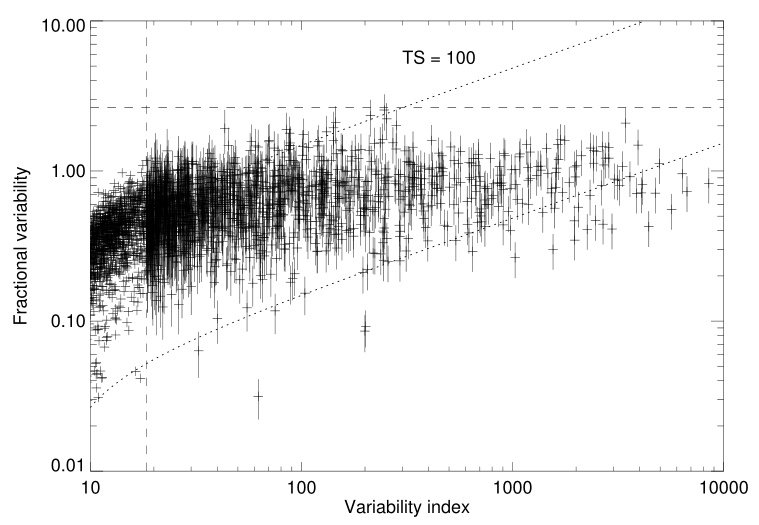}
\caption{Fractional variability of all sources plotted as a function of variability index, over one-year intervals. The vertical dashed line (below which the points have no error bar) is the variability threshold. The horizontal dashed line is the maximum fractional variability that can be reached ($\sqrt{N_{\rm int}-1}$). The dotted lines show how the variability index depends on $\delta F/F$ at $TS = 100$ and at $TS = 10,000$. At a given $TS$ threshold, the lower right part of the diagram is not accessible. The error bars are omitted below the variability threshold for clarity.
}
\label{fig:FracVar}
\end{figure}

In 4FGL we report the fractional variability of the sources in the FITS file as \texttt{Frac\_Variability}.
It is defined for each source from the excess variance on top of the statistical and systematic fluctuations:
\begin{eqnarray}
Var & = & \frac{1}{N_{\rm int}-1} \; \sum_i (F_i - F_{\rm av})^2
\label{eq:samplevar} \\
\delta F & = & \sqrt{\max \left( Var
              - \frac{\sum_i \sigma_i^2}{N_{\rm int}}, 0 \right)}
\label{eq:relvar} \\
\frac{\sigma_F}{F} & = & \max \left( \frac{1}{\sqrt{2(N_{\rm int}-1)}}
              \frac{V_i}{F_{\rm av} \; \delta F }, 10 \right)
\label{eq:uncrelvar}
\end{eqnarray}
where the fractional variability itself is simply $\delta F/F_{\rm av}$.
This is similar to Eq.~3 of 1FGL, except we omit the systematic error term because it is now incorporated in the $\sigma_i^2$ via the weights.
The error $\sigma_F/F$ is estimated from the expected scatter on the sample variance $Var$, which is the dominant source of uncertainty. We cap it at 10 to avoid reporting meaningless high uncertainties.
Figure~\ref{fig:FracVar} can be compared to Figure~8 of \citet{LAT09_BSL}, which was based on 1-week intervals (and contained many fewer sources, of course).
The fractional variability is similar in the two figures, going up to 1, reflecting the absence of a preferred variability time scale in blazars.
The criterion we use is not sensitive to relative variations smaller
than 50\% at $TS$ = 100, so only bright sources can populate the lower part of the plot.
There is no indication that fainter sources are less variable than brighter
ones, but we simply cannot measure their variability.

\subsubsection{Two-month intervals}
\label{lctwomonths}

To characterize variability, it is of course useful to have information on shorter time scales than one year.
Rather than use monthly bins as in 3FGL (which would have resulted in many upper limits), we have chosen to keep the same number of intervals and build light curves over 48 two-month bins.
Because the analysis is not limited by systematics at low energy over two months, we tried to optimize the data selection differently.
We used binned likelihood up to 3~GeV and the same zenith angle cuts as in Table~\ref{tab:components}, but included PSF2 events between 50 and 100~MeV (not only PSF3), and added PSF1 events between 100 and 300~MeV to our standard PSF2+3 selection. This improves the average source significance over one bin, and the Earth limb contamination remains minor.
Similarly to the one-year analyses, to save disk space and CPU time we merged event types together in the binned data sets. We used unbinned likelihood above 3~GeV and again ignored events above 100~GeV (unimportant for variability).

The weights appropriate for two months were computed using the same procedure (Appendix~\ref{appendix_weights}), entering the total data cube divided by 48 (same weights in each interval). The weights are of course larger than those for one year, but remain a significant correction in the Galactic plane. Up to 100~MeV the weights range from 0.2 in the Galactic Ridge to 0.85 at high latitude. At 300~MeV they increase to 0.55 in the Galactic Ridge and 0.99 at high latitude.
We used a different Sun/Moon model for each interval (the Sun averages out only over one year), but again assuming constant flux.

\begin{figure}
\centering
\includegraphics[viewport=0 0 399 359,clip,width=\linewidth]{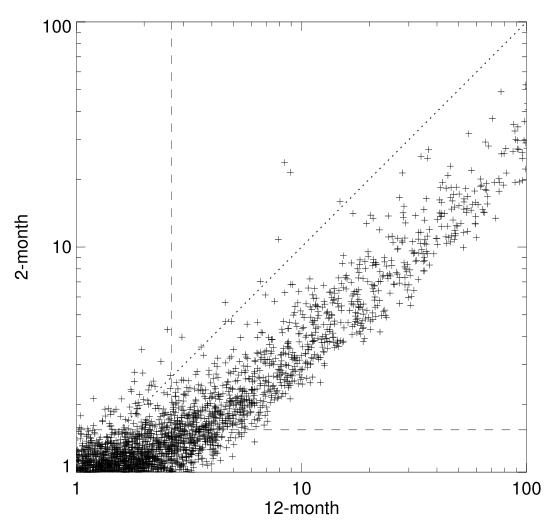}
\caption{Comparison of the reduced variability index (divided by $N_{\rm int}-1$) from two-month intervals with that for one-year intervals.
This illustrates that, for the majority of sources (AGN characterized by red noise) using longer intervals detects variability better.
The dotted line is the diagonal (expected for white noise).
The dashed lines show the two variability thresholds.}
\label{fig:VarIndex2}
\end{figure}

\begin{figure*}
   \centering
   \begin{tabular}{cc}
   \includegraphics[width=0.5\textwidth]{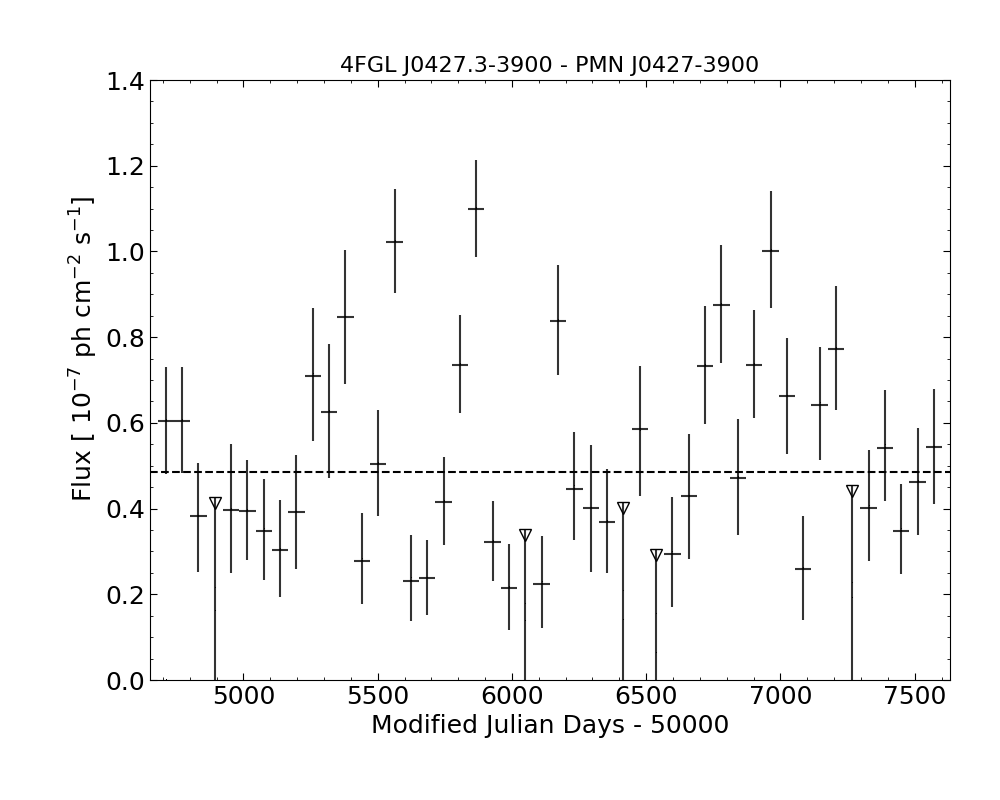} & 
   \includegraphics[width=0.5\textwidth]{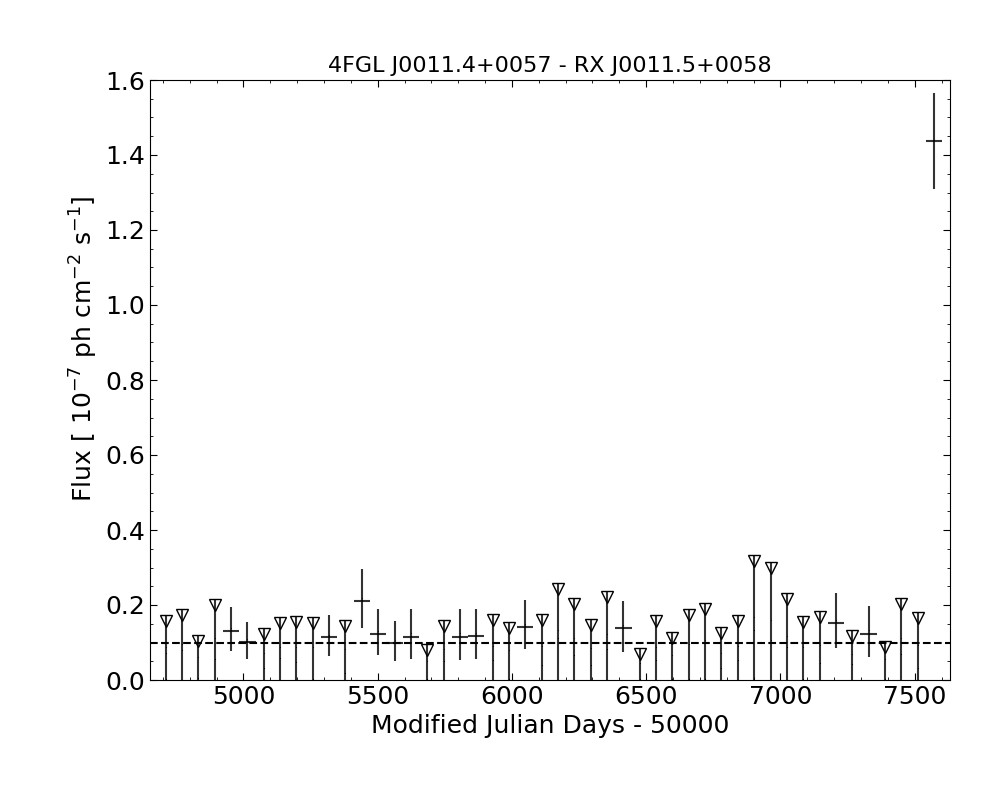}
   \end{tabular}
\caption{Light curves over two-month intervals of two blazars showing fast variability. Downward triangles indicate upper limits at 95\% confidence level. Left: unclassified blazar PMN J0427$-$3900 (4FGL J0427.3$-$3900) at $TS_{\rm var} = 202$. This is the highest $TS_{\rm var}$ among sources considered non-variable over one-year intervals ($TS_{\rm var} = 17.8$). Its variability is very fast (more like white noise than red noise) and averages out over one-year intervals. Right: flat-spectrum radio quasar RX J0011.5+0058 (4FGL J0011.4+0057) at $TS_{\rm var} = 278$, showing a single flare in the last 2-month bin. This source was detected as variable with one-year intervals ($TS_{\rm var} = 79$).}
\label{fig:lctwomonths}
\end{figure*}

Variability is considered probable when $TS_{\rm var} > 72.44$, corresponding to 99\% confidence in a $\chi^2$ distribution with $N_{\rm int} - 1 = 47$ degrees of freedom. We find 1173 variable sources with that criterion, 1057 of which were also considered variable with one-year intervals. Among the 116 sources considered variable only with 2-month light curves, 37 (1\% of 3738) would be expected by chance, so more than two thirds must be really variable. Similarly, 270 sources are considered variable only with one-year intervals (39 expected by chance).

Two extended sources exceed the two-month variability threshold. They are the Monoceros SNR and the Cen A lobes. Both are very extended (several degrees). It is likely that their variability is due to a flaring background source that was missed by the global source detection over eight years. Indeed the peak in the light curve of the Monoceros SNR is in June - July 2012, at the time of Nova V959 Mon 2012 \citep{LAT14_novae}. Another unexpected variable source is the Geminga pulsar. We think that its variability is not real but due to the direct pointings triggered toward the Crab when it was flaring (Geminga is 15$\degr$ away), combined with details of the effective area or PSF dependence on off-axis angle, that normally average out in scanning mode.

Because the source fluxes are not allowed to be negative, the distribution of fluxes for a given source is truncated at 0. For faint sources, this results in a slight overestimate of the average flux (of no consequence) but also an underestimate of the sample variance (Eq.~\ref{eq:samplevar}). As a result, the fractional variability (Eq.~\ref{eq:relvar}) is underestimated for faint sources and is often zero for weakly variable sources (below threshold). This even happens for two sources considered variable (just above threshold).

More sources are found to be variable using one-year intervals than using two-month intervals. The reason is illustrated in Figure~\ref{fig:VarIndex2}, which shows the variability indices divided by $N_{\rm int}-1$ (so that they become directly comparable). If the sources behaved like white noise (as the statistical errors) then the correlation would be expected to follow the diagonal. But blazars behave as red noise (more variability on longer time scales) so the correlation is shifted to the right and it is more advantageous to use longer intervals to detect variability with that criterion, because statistical errors decrease more than intrinsic variability.

Extending this relation to even shorter intervals, the 2FAV catalog of \Fermilat flaring sources \citep{LAT17_FAVA}, which used 1-week intervals, found 518 significantly varying sources. The methodology was completely different (it didn't start from a catalog over many years) and the duration a little shorter (7.4 years) but the same trend remains to find fewer variable sources on shorter intervals.
Not all sources are dominated by red noise however, and a fraction are above the diagonal in Figure~\ref{fig:VarIndex2}. An example is provided in Figure~\ref{fig:lctwomonths} (left). In all cases, the variability is of course much better characterized with smaller intervals. An extreme example is provided in Figure~\ref{fig:lctwomonths} (right).

    \subsection{Limitations and Systematic Uncertainties}
\label{catalog_limitations}

\subsubsection{Diffuse emission model}
\label{catalog_diffusemodel}

The model of diffuse emission is the main source of uncertainties for faint sources. Contrary to the effective area, it does not affect all sources equally: its effects are smaller outside the Galactic plane where the diffuse emission is fainter and varying on larger angular scales.
It is also less of a concern at high energy ($>$ 3~GeV) where the core of the PSF is narrow enough that the sources dominate the background under the PSF.
But it is a serious concern inside the Galactic plane at low energy ($<$ 1~GeV) and particularly inside the Galactic ridge ($|l| < 60\degr$) where the diffuse emission is strongest and very structured, following the molecular cloud distribution.
It is not easy to assess precisely how large the uncertainties are, because they relate to uncertainties in the distributions of interstellar gas, the interstellar radiation field, and cosmic rays, which depend in detail on position on the sky.

We estimate, from the residuals over the entire Galactic plane, that the systematics are at the 3\% level. This is already an achievement, but the statistical Poisson errors corresponding to the diffuse emission integrated over the PSF (as described in Appendix~\ref{appendix_weights}) are much smaller than this. Integrating energies up to twice the current one in the Galactic ridge, the statistical precision is 0.2, 0.4, 1, 2, 5\% above 100, 200, 500~MeV, 1, 2~GeV respectively.

The weights are able to mitigate the systematic effects globally, but cannot correct the model locally. In particular, underestimating the mass of an interstellar cloud will always tend to create spurious sources on top of it, and overestimating diffuse emission at a particular place tends to make the sources on top of it harder than they should be (because the model creates negative residuals there, and those are felt mostly at low energy). For an approximate local assessment, we have compared the 4FGL catalog with a version of the FL8Y source list (which used the 3FGL Galactic diffuse model \texttt{gll\_iem\_v06}) obtained with the same setup as 4FGL (see \S~\ref{compare_stepbystep}). Flags 1, 2 and 3 in Table~\ref{tab:flags} reflect that.

As we did for the 2FGL and 3FGL catalogs, we have checked which unidentified, non-variable sources with detection $TS < 150$ can be biased by large uncertainties in the modeling of the underlying Galactic interstellar emission. As described in more detail in the 2FGL paper, we have flagged sources that are potentially confused with complex small-scale structures in the interstellar emission. Their positions, fluxes, and spectral characteristics may not be reliable because of the uncertain contributions of the different gas components in their direction. Most flagged sources have $TS < 100$, but a large $TS$ value does not guarantee their reliability since a deficit in the bright interstellar background is necessarily compensated by one bright, statistically significant, point source (or several of them). Most of the flagged sources have power-law indices above 2.2, but nine of them are harder. This is possible if the interstellar deficit is at sub-degree angular scales. The diffuse model can adapt spectrally up to the energy at which the PSF is at the same angular scale as the interstellar deficit, leaving only a high-energy excess.
Those sources are assigned Flag 6 in the catalog (Table~\ref{tab:flags}). We also append \texttt{c} to the source names (except the extended ones). Most (64, $\sim$70\%) of those suspect sources have no association with a counterpart at other wavelengths, 10 have class UNK and 7 have class SPP (\S~\ref{sec:associations}).

\subsubsection{Analysis method}
\label{catalog_analysismethod}

As in 3FGL, we use the $pointlike$-based method described in \S~\ref{catalog_detection} to estimate systematic errors due to the way the main $gtlike$-based method (\S~\ref{catalog_significance}) is set up in detail. Many aspects differ between the two methods: the code, the weights implementation, the RoIs, and the diffuse model adjustments. The $pointlike$-based method does not remove faint sources (with $TS < 25$) from the model. Even the data differ, since the $pointlike$-based method uses $Front$ and $Back$ event types whereas the $gtlike$-based method uses PSF event types with a different zenith angle cut. Both methods reject a fraction of the events below 1~GeV, but not the same one.

Because of all those differences, we expect that comparing the results of the two methods source by source can provide an estimate of the sensitivity of the source list to details of the analysis. In particular we use it to flag sources whose spectral characterization differs strongly with the two methods (Flags 1 and 3 in Table~\ref{tab:flags}).

\subsubsection{Analysis Flags}
\label{catalog_analysis_flags}

\begin{deluxetable*}{crl}

\tablecaption{Definitions of the Analysis Flags
\label{tab:flags}}
\tablehead{
\colhead{Flag\tablenotemark{a}} & \colhead{$N_{\rm sources}$} & \colhead{Meaning}
}

\startdata
  1  & 215 & Source with $TS > 35$ which went to $TS < 25$ when changing the diffuse model \\
  &   & (\S~\ref{catalog_diffusemodel}) or the analysis method (\S~\ref{catalog_analysismethod}). Sources with $TS \le 35$ are not flagged \\
  &   & with this bit because  normal statistical fluctuations can push them to $TS < 25$. \\
  2  & 215 & Moved beyond its 95\% error ellipse when changing the diffuse model. \\
  3  & 343 & Flux ($>$ 1~GeV) or energy flux ($>$ 100~MeV) changed by more than $3\sigma$ when \\
  &   & changing the diffuse model or the analysis method. Requires also that the flux \\ 
  &   & change by more than 35\% (to not flag strong sources). \\
  4  & 212 & Source-to-background ratio less than 10\% in highest band in which $TS > 25$. \\
  &   & Background is integrated over $\pi r_{68}^2$ or 1 square degree, whichever is smaller.\\
  5  & 399 & Closer than $\theta_{\rm ref}$\tablenotemark{b} from a brighter neighbor. \\
  6  & 92 & On top of an interstellar gas clump or small-scale defect in the model of \\
     &  & diffuse emission; equivalent to the \texttt{c} designator in the source name (\S~\ref{catalog_diffusemodel}). \\
  7  & \nodata & Not used. \\
  8  & \nodata & Not used. \\
  9  & 136 & Localization Quality $>$ 8 in {\it pointlike} (\S~\ref{catalog_detection}) or long axis of 95\% ellipse $> 0\fdg25$. \\
 10  &  27 & $\sum_i S_i^2 > 20.5$ or $S_i^2 > 9$ in any band (Eq.~\ref{eq:tsspec}). \\
 11  & \nodata & Not used. \\
 12  & 102 & Highly curved spectrum; \texttt{LP\_beta} fixed to 1 or \texttt{PLEC\_Index} fixed to 0 (see \S~\ref{catalog_spectral_shapes}). \\
 \enddata
 
 \tablenotetext{a}{In the FITS version (see Appendix \ref{appendix_fits_format}) the values are encoded as individual bits in a single column, with Flag $n$ having value $2^{(n-1)}$.}
 \tablenotetext{b}{$\theta_{\rm ref}$ is defined in the highest band in which source $TS > 25$, or the band with highest $TS$ if all are $< 25$. $\theta_{\rm ref}$ is set to $3\fdg77$ below 100~MeV, $1\fdg68$ between 100 and 300~MeV (FWHM), $1\fdg03$ between 300~MeV and 1~GeV, $0\fdg76$ between 1 and 3~GeV (in-between FWHM and $2 \, r_{68}$), $0\fdg49$ between 3 and 10~GeV and $0\fdg25$ above 10~GeV ($2 \, r_{68}$).}

\end{deluxetable*}

As in 3FGL we identified a number of conditions
that should be considered cautionary regarding the reality of a source or the magnitude of the systematic uncertainties of its measured properties. 
They are described in Table~\ref{tab:flags}, together with the number of sources flagged for each reason.
Flags 1, 2 and 3 alert to a different result with $pointlike$ or the previous diffuse model. Flag 4 indicates a low source-to-background ratio. Flag 5 alerts to confusion, Flag 6 to a possible contamination by diffuse emission, Flag 9 to a bad localization, Flag 10 to a bad spectral representation and Flag 12 to a very highly curved spectrum.
We have changed slightly the definition of Flag 5 on the conservative side. For any source, we define its best band $k_0$ as before ({\it i.e.}, the highest-energy band in which it has $TS > 25$, or the band with highest $TS$ if none reaches 25). Defining $TS_0$ as the $TS$ of the source in that band, we now consider that a neighbor is brighter whenever it has $TS > TS_0$ in band $k_0$ or in any higher-energy band. This catches soft sources close to a harder neighbor only somewhat more significant.
The localization check with $gtfindsrc$ (Flag 7 in 3FGL) was not done because unbinned likelihood is very slow and does not support energy dispersion nor weights. The Sun check (Flag 11 in 3FGL) is no longer necessary since we now have a good model of the solar emission.

In total 1163 sources are flagged in 4FGL (about 23\%, similar to 3FGL).
Only 15\% of the sources with power-law index $\Gamma < 2.5$ are flagged, but 47\% of the soft sources with $\Gamma \ge 2.5$. This attests to the exacerbated sensitivity of soft sources to the underlying background emission and nearby sources. For the same reason, and also because of more confusion, 52\% of sources close to the Galactic plane (latitude less than $10\degr$) are flagged while only 12\% outside that region are. Only 15\% of associated sources are flagged but 45\% of the non-associated ones are flagged. This is in part because the associated sources tend to be brighter, therefore more robust, and also because many flagged sources are close to the Galactic plane where the association rate is low.

\section{The 4FGL Catalog}
\label{4fgl_description}

\begin{figure*}
   \centering
      \includegraphics[width=0.99\textwidth]{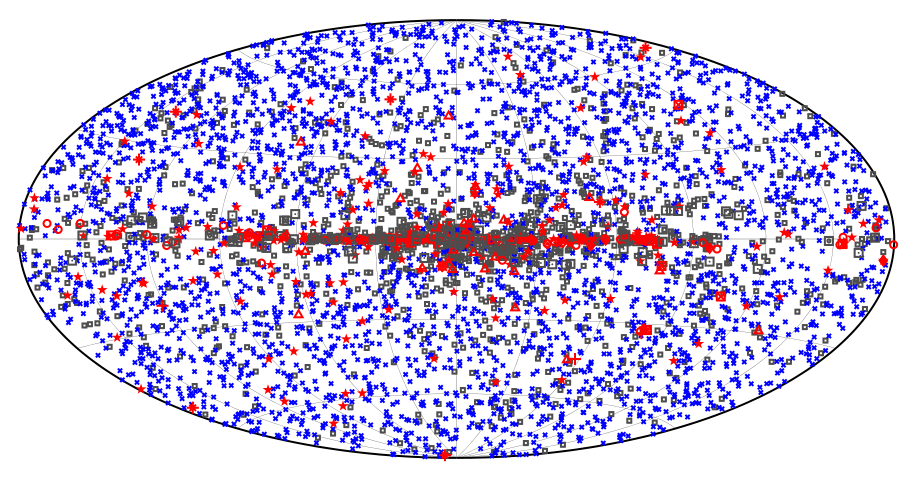}
      \includegraphics[width=0.80\textwidth]{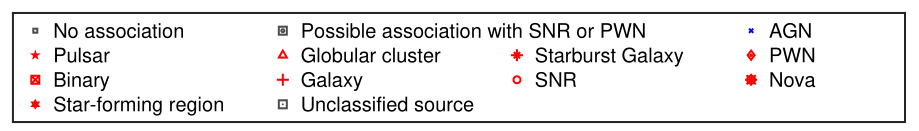}
      \includegraphics[width=0.99\textwidth]{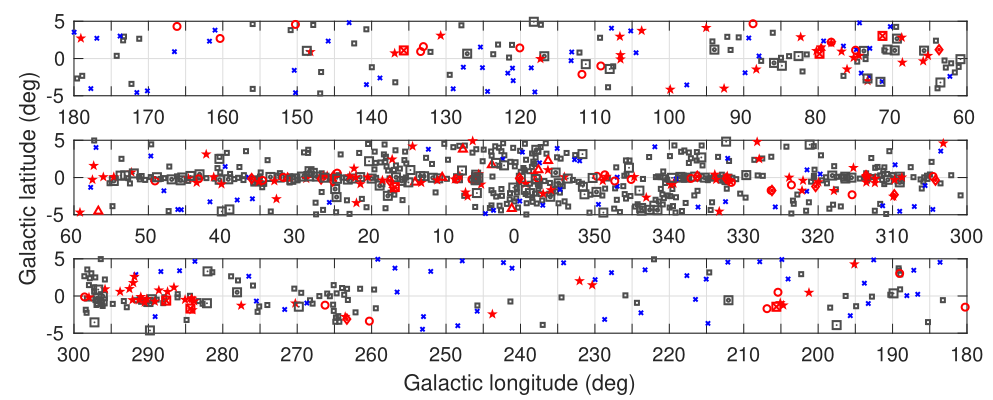}
   \caption{Full sky map (top) and blow-up of the Galactic plane split
   into three longitude bands (bottom) showing sources by source class (see \S~\ref{sec:assocsum}, no distinction is made between associations and identifications). All AGN classes are plotted with the same blue symbol for simplicity. Other associations to a well-defined class are plotted in red. Unassociated sources and sources associated to counterparts of unknown nature are plotted in black.}
   \label{fig:map_id_assoc}
\end{figure*}

\subsection{Catalog Description}
\label{catalog_description}

The catalog is available online\footnote{See \url{https://fermi.gsfc.nasa.gov/ssc/data/access/lat/8yr_catalog/}.}, together with associated products.
It contains 5064 sources\footnote{The file has 5065 entries because the Crab PWN is represented by two components (\S~\ref{catalog_spectral_shapes}).}. The source designation is \texttt{4FGL JHHMM.m+DDMM} where the \texttt{4} indicates that this is the fourth LAT catalog, \texttt{FGL} represents {\it Fermi} Gamma-ray LAT.
Sources confused with interstellar cloud complexes are singled out by a \texttt{c} appended to their names, where the \texttt{c} indicates that caution should be used in interpreting or analyzing these sources. 
The 75 sources that were modeled as extended for 4FGL (\S~\ref{catalog_extended}) are singled out by an \texttt{e} appended to their names.
The catalog columns are described in Appendix \ref{appendix_fits_format}.
Figure \ref{fig:map_id_assoc} illustrates the distribution of the 4FGL sources over the sky, separately for AGN (blue) and other (red) classes.

   \subsection{Comparison with 3FGL and earlier}


\subsubsection{General comparison}
\label{fgl_comparison}


\begin{figure}[!ht]
\includegraphics[width=\linewidth]{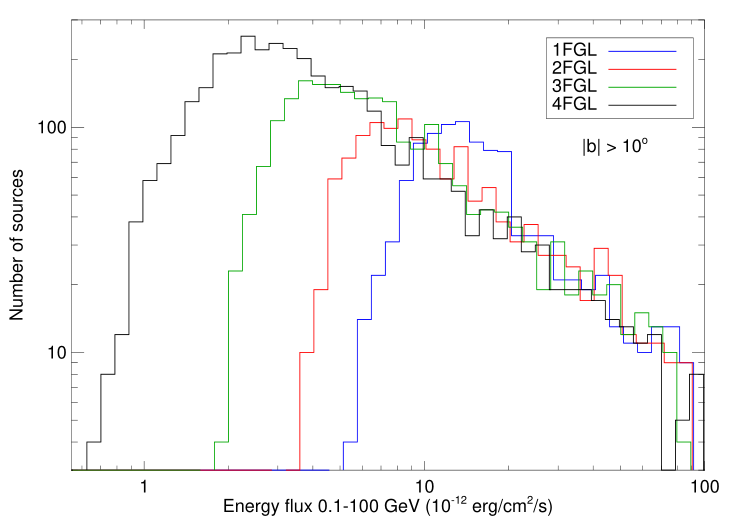}
\caption{Distributions of the energy flux for the high-latitude sources ($|b|>10\degr$) in the 1FGL (1043 sources, blue), 2FGL (1319 sources, red), 3FGL (2193 sources, green) and 4FGL (3646 sources, black) catalogs, illustrating the approximate detection threshold.}
\label{fig:enflux_HL}
\end{figure}

\begin{figure}[!ht]
\includegraphics[width=\linewidth]{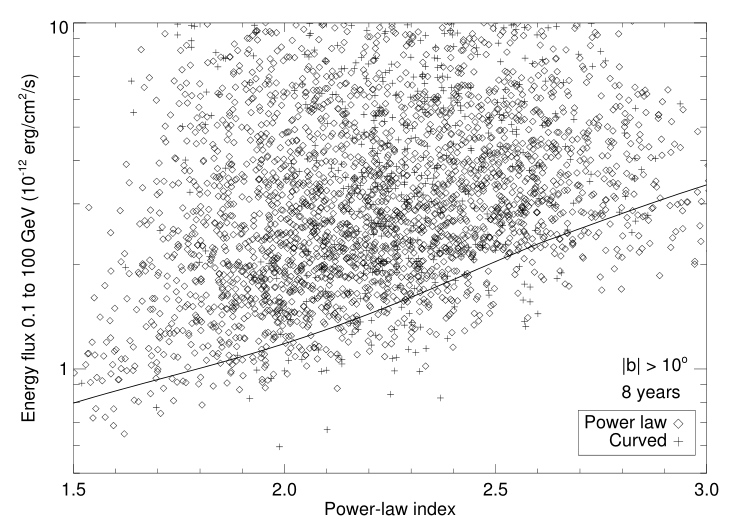}
\caption{Energy flux and power-law index of all sources outside the Galactic plane ($|b|>10\degr$). The solid line shows the expected detection threshold for a power-law spectrum. It is consistent with the fluxes of detected power-law sources (diamonds). The four sources furthest below the line are all curved (+ signs). Indeed the detection threshold (in terms of energy flux from 0.1 to 100~GeV) is lower for curved sources.}
\label{fig:detthresh_HL}
\end{figure}

Figure~\ref{fig:enflux_HL} shows the energy flux distribution in 1FGL, 2FGL, 3FGL and 4FGL outside the Galactic plane. Comparing the current flux threshold with those published in previous LAT Catalog papers we see that in 4FGL the threshold is down to $\simeq 2 \times 10^{-12}$ erg cm$^{-2}$ s$^{-1}$.
This is about a factor of two better than 3FGL. In the background-limited regime (up to a few GeV) doubling the exposure time would lead only to a factor $\sqrt{2}$. The remaining factor is due to the increased acceptance, the better PSF, and splitting the data into the PSF event types (\S~\ref{LATData}). The weights (Appendix \ref{appendix_weights}) do not limit the general detection at high latitudes.
Above $10^{-11}$ erg cm$^{-2}$ s$^{-1}$ the 2FGL and 3FGL distributions are entirely compatible with 4FGL. The 1FGL distribution shows a distinct bump between 1 and $2 \times 10^{-11}$ erg cm$^{-2}$ s$^{-1}$. That accumulation of fluxes was clearly incorrect. We attribute it primarily to overestimating significances and fluxes due to the unbinned likelihood bias in the 1FGL analysis, and also to the less accurate procedure then used to extract source flux  (see discussion in the 2FGL paper).

The threshold at low flux is less sharp in 4FGL than it was in 2FGL or 3FGL. This reflects a larger dependence of the detection threshold on the power-law index (Figure~\ref{fig:detthresh_HL}). The expected detection threshold is computed from Eq.~A1 of \citet{LAT10_1FGL}. The systematic limitation $\epsilon$ (entered in the weighted log-likelihood as described in Appendix \ref{appendix_weights}) is accounted for approximately by limiting the integral over angles to $\theta_{max}(E)$ such that $g(\theta_{max},E) = \epsilon$, since $g(\theta_{max},E)$ in that equation is exactly the source to background ratio. The detection threshold for soft sources decreases only slowly with exposure due to that.
On the other hand, the detection threshold improves nearly inversely proportional to exposure for hard sources because energies above 10~GeV are still photon-limited (not background-limited).


\begin{figure}[!ht]
\includegraphics[width=\linewidth]{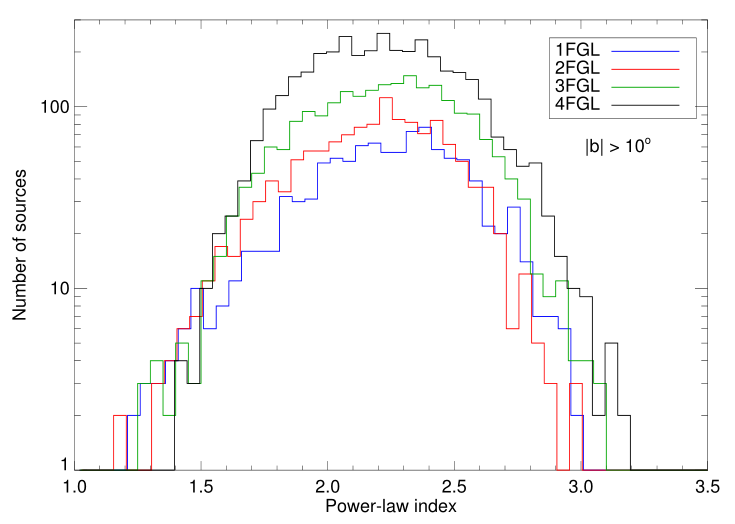}
\caption{Distributions of the power-law index for the high-latitude sources in the 1FGL (blue), 2FGL (red), 3FGL (green) and 4FGL (black) catalogs. The sources are the same as in Fig~\ref{fig:enflux_HL}.}
\label{fig:index_HL}
\end{figure}

The power-law index $\Gamma$ is a way to compare all sources over all catalog generations, ignoring the complexities of the curved models. Figure~\ref{fig:index_HL} shows the four distributions of the power-law indices of the sources at high Galactic latitude are very similar. Their averages and widths are $\Gamma_{\rm 1FGL} = 2.22 \pm 0.33$, $\Gamma_{\rm 2FGL} = 2.17 \pm 0.30$, $\Gamma_{\rm 3FGL} = 2.22 \pm 0.31$ and $\Gamma_{\rm 4FGL} = 2.23 \pm 0.30$.

Small differences in the power-law index distributions could be related to slightly different systematic uncertainties in the effective area between the IRFs used respectively for 4FGL, 3FGL, 2FGL, and 1FGL (Table \ref{tab:LATcatalogs}).
There is actually no reason why the distribution should remain the same, since the detection threshold depends on the index and the $\log$ N-$\log$ S of flat-spectrum radio quasars, which are soft \Fermilat sources, differs from that of BL Lacs, whose spectra are hard in the LAT band \citep[][Fig.~7]{LAT15_3LAC}. The apparent constancy may largely be the result of competing effects.


\begin{figure}[!ht]
\includegraphics[width=\linewidth]{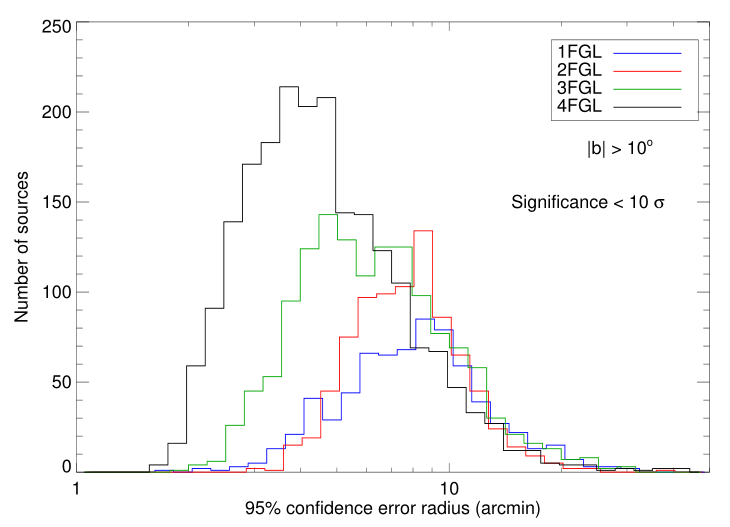}
\caption{Distributions of the 95\% confidence error radii for high-latitude sources with significance $< 10 \sigma$ in 1FGL (713 sources, blue), 2FGL (843 sources, red), 3FGL (1387 sources, green) and 4FGL (2090 sources, black), illustrating the improvement of localizations for sources of equivalent detection significances.
}
\label{fig:conf95_HL_ts100}
\end{figure}

We have compared the distribution of error radii (defined as the geometric mean of the semi-major and semi-minor axes of the 95\% confidence error ellipse) of the 1FGL, 2FGL, 3FGL and 4FGL sources at high Galactic latitude.
Overall the source localization improves with time as more photons are added to previously detected sources. We concentrate instead on what happens specifically for faint sources.
Figure~\ref{fig:conf95_HL_ts100} shows the distribution of 95\% confidence error radii for those sources with $25<TS<100$ in any of the catalogs. The improvement at a given $TS$ level is partly due to the event-level analysis (from Pass 6 to 7 and 8, see Table \ref{tab:LATcatalogs}) and partly to the fact that, at a given significance level and for a given spectrum, fainter sources over longer exposures are detected with more photons. This improvement is key to preserving a high rate of source associations (\S~\ref{sec:assocsum}) even though the source density increases.

\subsubsection{Step-by-step from 3FGL to 4FGL}
\label{compare_stepbystep}


To understand the improvements of the 4FGL analysis with respect to 3FGL, we have considered the effects of changing the analysis and the data set without changing the time range (i.e., leaving it as four years). To that end we started with the same seeds as the 3FGL catalog, changed each element in sequence (in the order of the list below) and compared each intermediate result with the previous one. The effect of introducing energy dispersion was described in \S~\ref{catalog_significance}.
\begin{itemize}
\item We first switched from P7REP to Pass 8 (P8R3), eliminating the Earth limb by cutting zenith angles $> 90\degr$ at 100 to 300~MeV and $> 97.5\degr$ at 300~MeV to 1~GeV for Front, $> 80\degr$ at 100 to 300~MeV and $> 95\degr$ at 300~MeV to 1~GeV for Back. The resulting $TS$ increased by 27\%, in keeping with the effective area increase (the number of sources at $TS > 25$ did not rise, for lack of seeds). Energy flux decreased by 7\% in faint sources. In the Galactic plane, source spectra tended to soften, with power-law indices increasing by 0.04 on average. Both effects appear to be due to the diffuse emission modeling, because they are absent in the bright sources. The isotropic spectrum was recomputed, and even though the Galactic diffuse model was the same, its effects differed because the effective area increase with Pass 8 is stronger at low energy. Those offsets are accompanied by a large scatter: only 72\% of P7REP $\gamma$ rays are still in P8R3, and even for those the reconstructed direction differs.
\item Accounting for energy dispersion increased energy flux on average by 2.4\%. The effect was larger for soft sources (3\% at $\Gamma > 2.1$). The average power-law index did not change, but hard sources got a little softer and soft sources a little harder (with shifts no larger than 0.02), reducing the width of the power-law index distribution. Spectra became more curved as expected (energy dispersion can only broaden the spectra): the curvature $\beta$ increased by 0.014 on average. None of these trends depends on Galactic latitude. The logLikelihood improved, but only by a few tens.
\item Switching from Front/Back to PSF event types increased $TS$ by 10\% (140 more sources). This was the intended effect (not diluting good events with bad ones should increase significance). No systematic effect was noted on energy flux. Soft sources got somewhat softer with PSF event types (power-law indices larger than 2.7 increased by 0.1 on average), but the bias averaged over all sources was only +0.01. The number of curved sources decreased by 50 and the curvature $\beta$ by 0.025 (this is the same effect: low energies moved up, so spectra got closer to a power law).
\item Applying the weights results in a general decrease of $TS$ and increase of errors, as expected. However, because source detection is dominated by energies above 1~GeV even without weights, the effect is modest (the number of sources decreased by only 40). The difference is of course largest for soft sources and in the Galactic plane, where the background is larger and the weights are smaller. There are a few other side effects. The number of curved sources decreased by 30. This is because the lever arm is less as the contributions from low-energy $\gamma$ rays are unweighted. The pivot energy tended to increase for the same reason, and this resulted in a softening of the power-law index of curved sources (not exceeding 0.1). Overall in the Galactic ridge the power-law index increased by 0.025.
\end{itemize}

We evaluated the other two changes on eight years of data:
\begin{itemize}
\item Changing the energy range to start at 50~MeV did not improve $TS$, as expected (the PSF is too broad below 100~MeV to contribute to significance). The energy flux (defined in the same 100~MeV to 100~GeV band) tended to decrease in the Galactic plane (by as much as $-10$\% in the Galactic ridge) and the power-law index tended to become harder (by as much as $-0.05$ in the Galactic ridge). This is because the low-energy information tends to stabilize artificially soft sources. Neither effect was noticeable outside the Galactic plane. The other consequence was to increase the number of significantly curved sources by 80, because the broader energy range made it easier to detect curvature (this was true everywhere in the sky).
\item Changing the Galactic diffuse emission model from \texttt{gll\_iem\_v06} used in 3FGL to that used here (\S~\ref{DiffuseModel}), without changing the analysis or the data, had a noticeable effect. The flags in \S~\ref{catalog_analysis_flags} are based on the comparison to a version of the FL8Y source list (using \texttt{gll\_iem\_v06}) extending the energy range to start at 50~MeV, and using the same extended sources and $TS_{\rm curv}$ threshold as 4FGL. The source significance is lower in 4FGL by 0.1 $\sigma$ on average and the number of sources decreased by 10\%. The energy flux is lower in 4FGL by 2\%, the power-law index is smaller (harder) by 0.02 and there are more curved sources than in FL8Y. This is all because the intensity of the new diffuse model is somewhat higher below 100~MeV. Because this is a background-related effect, it affects primarily the faint sources. The strong overprediction west of Carina in \texttt{gll\_iem\_v06} is gone but overall the residuals are at a similar level.
\end{itemize}

In conclusion, to first order the resulting net changes are not very large, consistent with the general comparison between 4FGL and 3FGL in \S~\ref{fgl_comparison}. Systematic effects are collectively visible but within calibration errors, and within statistical errors of individual sources.

\section{Automated Source Associations}
\label{sec:associations}

\startlongtable
\begin{deluxetable*}{lrr}
\setlength{\tabcolsep}{0.04in}
\tablewidth{0pt}
\tabletypesize{\scriptsize}
\tablecaption{Catalogs Used for the Automatic Source Association Methods
\label{tab:catalogs}
}
\tablehead{
\colhead{Name} & 
\colhead{Objects\tablenotemark{$a$}} & 
\colhead{Ref.}
}

\startdata
High $\dot{E}/d^2$ pulsars & 313 & \citet{ATNFcatalog}\tablenotemark{b} \\
Other normal pulsars & 2248  & \citet{ATNFcatalog}\tablenotemark{b} \\
Millisecond pulsars & 240 & \citet{ATNFcatalog}\tablenotemark{b} \\
Pulsar wind nebulae & 69 & Collaboration internal \\
High-mass X-ray binaries & 137 & \citet{HMXBcatalog_Chaty} \\
Low-mass X-ray binaries & 187  & \citet{LMXBcatalog} \\
Point-like SNR & 158  & \citet{SNRcatalog}\tablenotemark{c}  \\
Extended SNR\tablenotemark{$f$} & 295  & \citet{SNRcatalog}\tablenotemark{c}  \\
Globular clusters & 160  & \citet{GlobClusterCatalog} \\
Dwarf galaxies\tablenotemark{$f$}& 100  & \citet{DwarfGalaxies} \\
Nearby galaxies & 276  & \citet{NearbyGalaxiesCatalog} \\
IRAS bright galaxies & 82  & \citet{IRAScatalog} \\
BZCAT (Blazars) & 3561  & \citet{BZcatalog} \\
BL Lac & 1371  & \citet{AGNcatalog} \\
AGN & 10066  & \citet{AGNcatalog} \\
QSO & 129,853  & \citet{AGNcatalog} \\
Seyfert galaxies & 27651  & \citet{AGNcatalog} \\
Narrow-line Seyfert galaxies & 18  &  \citet{NLSy1catalog_Berton} \\
Narrow-line Seyfert galaxies & 556  & \citet{NLSy1catalog} \\
FRICAT (Radio galaxies)& 233 & \citet{FRICAT}\\ 
FRIICAT (Radio galaxies) & 123 & \citet{FRIICAT}\\ 
Giant Radio Source & 349 & \citet{GRSKcatalog} \\
\hline
2WHSP & $~$1691 & \citet{2WHSP} \\
{\sl WISE} blazar catalog & 12319 & \citet{WISE}\\
Radio Fundamental Catalog (2019a) & 15740 & \url{http://astrogeo.org/rfc} \\
CGRaBS & 1625 & \citet{CGRaBS} \\
CRATES & 11499  & \citet{CRATES} \\
ATCA 20 GHz southern sky survey & 5890  & \citet{AT20G} \\
105-month {\sl Swift}/BAT catalog &1632 & \citet{BATcatalog} \\
4$^{th}$ IBIS catalog & 939 & \citet{IBIScatalog} \\
\hline
2nd {\sl AGILE} catalog\tablenotemark{$e$} & 175   & \citet{AGILEcatalog2} \\
3rd EGRET catalog\tablenotemark{$e$} & 271  & \citet{3EGcatalog} \\
EGR catalog\tablenotemark{$e$} & 189  & \citet{EGRcatalog} \\
0FGL list\tablenotemark{$e$} & 205  & \citet[][0FGL]{LAT09_BSL} \\
1FGL catalog\tablenotemark{$e$} & 1451 & \citet[][1FGL]{LAT10_1FGL} \\
2FGL catalog\tablenotemark{$e$} & 1873 & \citet[][2FGL]{LAT12_2FGL} \\
3FGL catalog\tablenotemark{$e$} & 3033 & \citet[][3FGL]{LAT15_3FGL} \\
1FHL catalog\tablenotemark{$e$} & 514 & \citet[][1FHL]{LAT13_1FHL} \\
2FHL catalog\tablenotemark{$e$} & 360 & \citet[][1FHL]{LAT16_2FHL} \\
3FHL catalog\tablenotemark{$e$} & 1556 & \citet[][1FHL]{LAT17_3FHL} \\
TeV point-like source catalog\tablenotemark{e,f} & 108 & \url{http://tevcat.uchicago.edu/} \\
TeV extended source catalog\tablenotemark{$g$} & 72   & \url{http://tevcat.uchicago.edu/} \\
\hline
LAT pulsars & 234 &  Collaboration internal\tablenotemark{$d$}  \\
LAT identified & 145  & Collaboration internal \\ %
\enddata

\tablenotetext{a}{Number of objects in the catalog.}
\tablenotetext{b}{version 1.56, \url{http://www.atnf.csiro.au/research/pulsar/psrcat}}
\tablenotetext{c}{Green D. A., 2017, `A Catalogue of Galactic Supernova Remnants (2017 June version)', Cavendish Laboratory, Cambridge, United Kingdom (available at \url{http://www.mrao.cam.ac.uk/surveys/snrs/})}
\tablenotetext{d}{\url{https://confluence.slac.stanford.edu/display/GLAMCOG/Public+List+of+LAT-Detected+Gamma-Ray+Pulsars}}
\tablenotetext{e}{For these catalogs, the association is performed according to  Eq.~\ref{eq:compnFGL}.}
\tablenotetext{f}{Version of 2018 November 30.}
\tablenotetext{g}{For these catalogs of extended sources, the association is performed by requiring that the separation from the 4FGL sources is less than the quadratic sum of the 95\% confidence error radii.} 
\end{deluxetable*}


We use two complementary methods in the association task. The Bayesian method is based only on spatial coincidence between the gamma-ray sources and their potential counterparts. This method  does not require any additional information (like an available  log N-log S) for the considered catalogs. It  is of general use and applicable to many counterpart catalogs. However it is inapproppriate when considering large surveys (e.g., in the radio or X-ray bands) because of their high source densities. The Likelihood Ratio method on the other hand can be applied to these surveys, owing to the use of their log N-log S. This method  allows us to retrieve some associations with relatively bright counterparts that were missed with the Bayesian method.  The mitigation of the effect of large effective counterpart densities is not perfect. The resulting association probabilities are typically lower than for the Bayesian method.
     
The Bayesian source association method  \citep{LAT10_1FGL} for the {\it Fermi}-LAT, implemented with the {\sl gtsrcid} tool\footnote{\url{https://fermi.gsfc.nasa.gov/ssc/data/analysis/scitools/overview.html}},  was developed following the prescription devised by \citet{mattox97} for EGRET. It relies on the fact that the angular distance between a LAT source and a candidate counterpart  is driven by i) the position uncertainty in the case of a real association and ii) the counterpart density in the case of a false (random) association. In addition to the angular-distance probability density functions for real and false associations, the posterior probability depends on a prior. This prior is calibrated via Monte Carlo simulations so that the number of false associations, $N_{\rm false}$ is equal to the sum of the association-probability complements. For a given counterpart catalog, the so-obtained prior is found to be close to $N_{\rm assoc}/N_{\rm tot}$, where $N_{\rm assoc}$ is the number of associations from this catalog and $N_{\rm tot}$ is the number of catalog sources. The sum of the association probabilities over all pairs ($\gamma$-ray source, potential counterpart) gives the total number of real associations for a particular catalog, allowing the number of subthreshold associations to be estimated. The total numbers of associations are  reported  in \S~\ref{sec:assocsum}  for the various classes, where the overlap between associations from different catalogs is taken into account.  A uniform threshold of $P \geq 0.8$ is applied to the posterior probability for the association to be retained. The reliability of the Bayesian associations is assessed by verifying that  the distribution of the angular offset between $\gamma$-ray source and counterpart matches well the expected one in the case of a true association, i.e., a Rayleigh function with its width parameter given by the sources' positional uncertainties. 

The counterpart catalogs (Table \ref{tab:catalogs}) include known $\gamma$-ray-emitting source classes: Active Galactic Nuclei \citep[AGNs,][]{LAT15_3LAC}, galaxies \citep{LAT10_starbursts}, pulsars \citep{LAT13_2PC}, pulsar-wind nebulae \citep[PWNe,][]{LAT11_PWNcat}, supernova remnants  \citep[SNRs,][]{LAT2016_SNRCat}, globular clusters \citep[GLCs,][]{LAT10_globular}, low- and high-mass X-ray binaries \citep{LAT10_EtaCarina,LAT09_LSI} or surveys of candidate blazars at other frequencies (radio, IR, X-rays). The reported source classes are derived in the same way as in 3FGL. For non-AGN sources, this classification is based on the nature of the association catalogs. For AGNs, the subclasses  as flat-spectrum radio quasars (FSRQs), BL~Lac-type objects (BLLs), blazar candidates of uncertain type (BCUs), radio galaxies (RDGs), narrow-line Seyfert 1 (NLSY1s), steep spectrum radio quasars (SSRQs), Seyfert galaxies (SEYs) or simply AGNs (if no other particular subclass can be assigned), have been selected according to the counterpart properties at other wavelengths.  Please note that we did not use the blazar classes from the Simbad database\footnote{\url{http://simbad.u-strasbg.fr/simbad/}} since some of them correspond to predictions  based on the {\sl WISE}-strip approach \citep{WISE} and not to assessment with the measured strengths of the emission lines.

In complement to the Bayesian method, the Likelihood-Ratio (LR) method \citep{LAT11_2LAC,LAT15_3LAC}, following \cite{deRuiter77}  provides supplementary associations with blazar candidates based on large radio and X-ray surveys: NVSS \citep{NVSScatalog}, SUMSS \citep{SUMSScatalog}, {\sl ROSAT} \citep{RASSbright,RASSfaint} and AT20G \citep{AT20G}.  This method  is similar in essence to the Bayesian method but the false association rate is derived from the density of objects brighter than the considered candidate, assessed from the survey log N-log S distribution. 
While the LR method is able to handle large surveys, its fraction of false associations is notably larger than for the Bayesian method (typically 10\% vs. 2\% ).  The overlap between the results of the Bayesian and LR methods is about 75\% for blazars. Because the surveys include a large number of Galactic sources at low Galactic latitudes, the class of $|b|<10\arcdeg$ sources associated solely via the LR-method has been  set to UNK (standing for unknown) as opposed to the BCU class used by default for sources at higher latitudes.   
 
Firm identifications are based on periodic variability for LAT-detected pulsars or X-ray binaries, correlated variability at other wavelengths for AGNs or spatial morphology related to that found in another band for extended sources. 

The association and classification procedures greatly benefited from data of recent intensive follow-up programs, motivated by the study of the unidentified/unassociated $\gamma$-ray sources. This study was recognized as one of the major scientific goals of the {\it Fermi} mission. Many groups carried out follow-up observations and/or applied statistical procedures to investigate and discern the nature of the unassociated sources from their gamma-ray properties \citep[see, e.g.,][]{ackermann12,hassan13,doert14}. In particular, follow-up campaigns were carried out at different wavelengths with both ground-based and space telescopes  above GHz frequencies \citep[see, e.g.,][]{kovalev09,petrov11,r:aofus1,hovatta12,hovatta14,r:aofus2,r:aofus3} and below \citep[see, e.g.,][]{pap1,nori14,giroletti16}, or using sub-millimeter \citep[see, e.g.,][]{giommi12,lopez13} and infrared observations \citep[see, e.g.,][]{pap2,pap3,pap4,arsioli15,pap6,WISE} up to the X-rays with {\it Swift} \citep[e.g.,][]{mirabal09,paggi13,takeuchi13,stroh13,acero13,landi15,paiano17} as well as with {\it Chandra} and {\it Suzaku} \citep[e.g.,][]{maeda11,cheung12,kataoka12,takahashi12,takeuchi13}. Over the years, these observations allowed additions to the lists of potential counterparts, which were then used with the methods previously described. In addition,  to assess the real nature and classify all newly associated sources, it has been crucial to perform additional spectroscopic optical observations, which for extragalactic objects were also able to provide estimates of their cosmological distances \citep[see, e.g.,][]{shaw13a,shaw13b,paggi14,pap9,ricci15,pap10,landoni15a,landoni15b,chiaro16,crespo16a,crespo16b,landoni18,paiano17b,paiano17c,paiano17d,pena17,marchesi18,marchesini19}. These campaigns are continuously updated including searches in the optical databases of the major surveys \citep[see, e.g.,][]{cowperthwaite13,pap8,maselli15,bcu,quest16,deM19}.

 The false-association rate is difficult to estimate for the new associations resulting from these follow-up observations, preventing them from being treated on the same footing as those obtained as described above. The most-recent {Radio Fundamental Catalog\footnote{Available at \url{http://astrogeo.org/rfc}} (RFC) includes many new entries that came from dedicated follow-up observations. Applying the Bayesian method to the whole catalog and retaining associations with $P \geq$0.8, the association probability attached to the recent additions (181 sources) are reported as NULL to distinguish them from the others.
   
\section{Association summary}
\label{sec:assocsum}

\begin{deluxetable*}{lcrcr}
\setlength{\tabcolsep}{0.04in}
\tablewidth{0pt}
\tabletypesize{\scriptsize}
\tablecaption{LAT 4FGL Source Classes 
\label{tab:classes}
}
\tablehead{
\colhead{Description} & 
\multicolumn{2}{c}{Identified} &
\multicolumn{2}{c}{Associated} \\
& 
\colhead{Designator} &
\colhead{Number} &
\colhead{Designator} &
\colhead{Number}
}
\startdata
Pulsar, identified by pulsations & PSR & 232 & \nodata & \nodata \\
Pulsar, no pulsations seen in LAT yet & \nodata & \nodata & psr & 7 \\
Pulsar wind nebula & PWN & 11 & pwn & 6 \\
Supernova remnant & SNR & 24 & snr &   16 \\
Supernova remnant / Pulsar wind nebula & SPP &  0  & spp  & 78 \\
Globular cluster & GLC &  0  & glc & 30 \\
Star-forming region & SFR & 3 & sfr &  0  \\
High-mass binary & HMB & 5 & hmb & 3 \\
Low-mass binary & LMB & 1 & lmb & 1 \\
Binary & BIN & 1 & bin & 0 \\
Nova & NOV & 1 & nov & 0  \\
BL Lac type of blazar & BLL & 22 & bll & 1109 \\ 
FSRQ type of blazar &   FSRQ & 43 & fsrq &   651 \\
Radio galaxy & RDG & 6 & rdg & 36 \\
Non-blazar active galaxy & AGN & 1 & agn & 10 \\ 
Steep spectrum radio quasar & SSRQ & 0 & ssrq & 2 \\
Compact Steep Spectrum radio source & CSS &  0  & css & 5 \\
Blazar candidate of uncertain type & BCU & 2 & bcu & 1310 \\ 
Narrow-line Seyfert 1 & NLSY1 & 4 & nlsy1 & 5 \\
Seyfert galaxy & SEY &  0  & sey & 1 \\
Starburst galaxy & SBG &  0  & sbg & 7 \\
Normal galaxy (or part) & GAL & 2 & gal & 1 \\
Unknown & UNK & 0 & unk & 92 \\
Total & \nodata &  358 & \nodata &  3370 \\
\hline
Unassociated & \nodata & \nodata & \nodata &  1336 \ 
\enddata
\tablecomments{The designation `spp' indicates potential association with SNR or PWN.  Designations shown in capital letters are firm identifications; lower case letters indicate associations. 
}
\end{deluxetable*}

\begin{figure}[!ht]
\centering
\includegraphics[width=\linewidth]{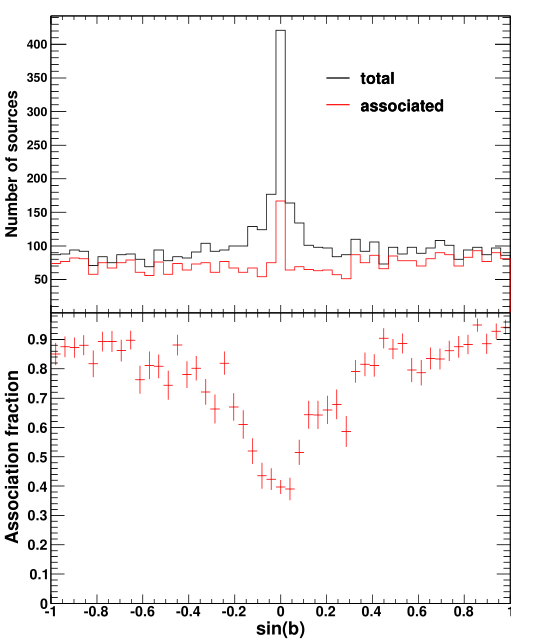}
\caption{Upper panel: Distributions in Galactic latitude $b$ of 4FGL sources (black histogram) and associated sources (red histogram). Lower panel: Association fraction as a function of  Galactic latitude.}
\label{fig:sinb_assoc}
\end{figure}

The association summary is given in Table \ref{tab:classes}. Out of 5064 LAT sources in 4FGL, 1336 are unassociated (26.4\%). Some  92 others are classified  as UNKs, and  78 as SPPs (sources of unknown nature but overlapping with known SNRs or PWNe and thus candidates to these classes), representing 3.3\% in total. Some 3463 sources are associated with the Bayesian method (1069 associations from this method only, overall  $N_{\rm false}$=36.6), 2604 sources with the LR method (210 associations from this method only, $N_{\rm false}$= 22.2 for the latter). The overall association fraction, 70\%, is similar to that obtained in previous LAT catalogs. The association fraction is lower for fainter sources (essentially all $TS>500$ sources are associated), in particular due to their larger error regions.  This fraction also decreases  as sources lie closer to the Galactic plane as illustrated in Figure  \ref{fig:sinb_assoc}. It decreases from about 85\% at high Galactic latitudes to $\simeq$ 40\% close to the Galactic plane. The reason for such an effect is twofold. 
We are not able to associate many of the Galactic sources with high confidence. In addition, the association of background extragalactic sources is impeded by the larger flux limits of some extragalactic-counterpart catalogs due to absorption effects for the X-ray band through the Galactic plane. The properties of the unassociated sources are discussed below.  

\begin{deluxetable*}{cccccc}
\tabletypesize{\scriptsize}
\tablecaption{3FGL sources with different counterparts in 4FGL \label{tab:other_count}}
\tablehead{
\colhead{3FGL name} &
\colhead{3FGL counterpart} &
\colhead{3FGL class} &
\colhead{4FGL name} &
\colhead{4FGL counterpart} &
\colhead{4FGL class}
} 
\startdata
 J0500.3+5237 & \nodata & spp &  J0500.2+5237 & GB6 J0500+5238 & bcu \\ 
 J0618.0+7819 & 1REX J061757+7816.1 & fsrq &  J0618.1+7819 & NGC 2146 & sbg \\ 
 J0647.1$-$4415 & SUMSS J064648$-$441929 & bcu &  J0647.7$-$4418 & RX J0648.0$-$4418 & hmb \\
 J0941.6+2727 & MG2 J094148+2728 & fsrq  & J0941.9+2724 & GB6 J0941+2721 & bll
\\
 J1048.6+2338 & NVSS J104900+233821 & bll &  J1048.6+2340 & PSR J1048+2339 & PSR \\ 
 J1111.9$-$6038 & \nodata & spp &  J1111.8$-$6039 & PSR J1111$-$6039 & PSR \\
 J1132.8+1015 & 4C +10.33 & fsrq &     J1130.8+1016 &  2MASS J11303636+1018245 & bcu \\             
 J1741.1$-$3053 & MSH 17$-$39 & snr &  J1741.4$-$3046 & NVSS J174122$-$304712 & unk \\ 
 J1811.3$-$1927c & \nodata & spp &  J1811.5$-$1925 & PSR J1811$-$1925 & psr \\ 
 J1817.2$-$1739 & \nodata & spp &  J1817.1$-$1742 & PSR J1817$-$1742 & PSR \\ 
 J2022.2+3840 & \nodata & spp &  J2022.3+3840 & PSR J2022+3842 & PSR \\
 J2224.6$-$1122 &  PKS 2221$-$116 & bll &  J2225.5$-$1114 & PKS 2223$-$114 & bll \\
\enddata
\end{deluxetable*}

Sources reported as new below were not in previous FGL catalogs, although their detections may have been reported in other works \citep[e.g.,][]{Zhang2016,Arsioli2018} or in specialized LAT catalogs. Table \ref{tab:other_count} lists the 12 3FGL sources that have different counterparts in 4FGL. Pulsations have been detected for 5 sources previously classified as SPPs. As discussed below, the association of 4FGL  J0647.7$-$4418 with RX J0648.0$-$4418 instead of SUMSS  J064744$-$441946 remains uncertain.

\subsection{Extragalactic sources}
\subsubsection{Active Galactic Nuclei} The largest source population in 4FGL is that of AGNs, with 3137 blazars, 42 radio galaxies and 28 other AGNs. The blazar sample comprises  694  FSRQs, 1131 BLLs  and 1312 BCUs. The detailed properties of the 4FGL AGNs, including redshifts and fitted synchrotron-peak positions, will be the subject of the 4LAC companion catalog. We note here that the separation in $\gamma$-ray spectral hardness between FSRQs and BL~Lacs already reported in previous LAC AGN catalogs is confirmed: 93\% of FSRQs and 81\% of BL~Lacs have power-law photon indices greater and lower than 2.2 respectively. Among the 70 non-blazar AGNs, 35 were present in 3FGL. Note that the location of the $\gamma$-ray source associated with Cen B is not coincident with that of the radio-galaxy core but points to the southern radio jet.
Twenty-three  radio galaxies, listed in Table \ref{tab:new_rdg}, are new.  Four 3FGL sources have changed classes to radio galaxies: three former BCU (IC 1531, TXS 0149+710, PKS 1304$-$215) and one former BLL (B3 1009+427). The 28 other AGNs  include  five compact steep spectrum radio sources (CSS, three are new: 3C 138, 3C 216, 3C 309.1), two steep spectrum radio quasars (SSRQ, new is 3C 212), 9 narrow-line Seyferts 1 (NLSY1), one Seyfert galaxy (the Circinus galaxy, SEY) and 11 AGNs of other types (AGN). Three NLSY1 are new: IERS B1303+515, B3 1441+476, TXS 2116$-$077. 

\begin{deluxetable}{ccc}
\tabletypesize{\scriptsize}
\tablecaption{New radio galaxies in 4FGL \label{tab:new_rdg}}
\tablewidth{0pt}
\tablehead{
\colhead{4FGL name} &
\colhead{4FGL counterpart}
} 
\startdata
J0038.7$-$0204 &  3C 17 \\
J0057.7+3023 & NGC 315 \\
J0237.7+0206 &  PKS 0235+017 \\
J0312.9+4119 & B3 0309+411B \\
J0433.0+0522 & 3C 120  \\
J0708.9+4839 & NGC 2329 \\
J0931.9+6737 &  NGC 2892 \\
J1116.6+2915 & B2 1113+29 \\ 
J1149.0+5924 &  NGC 3894 \\ 
J1236.9$-$7232 & PKS 1234$-$723 \\
J1306.3+1113 & TXS 1303+114 \\
J1449.5+2746 & B2 1447+27 \\
J1516.5+0015 & PKS 1514+00 \\ 
J1518.6+0614 & TXS 1516+064 \\
J1521.1+0421 & PKS B1518+045 \\
J1724.2$-$6501 & NGC 6328 \\
J1843.4$-$4835 & PKS 1839$-$48 \\
J2156.0$-$6942 & PKS 2153$-$69 \\
J2227.9$-$3031 & PKS 2225$-$308 \\
J2302.8$-$1841 & PKS 2300$-$18 \\
J2326.9$-$0201 & PKS 2324$-$02 \\
J2329.7$-$2118 & PKS 2327$-$215 \\
J2341.8$-$2917 & PKS 2338$-$295 \\
\enddata
\end{deluxetable}

\subsubsection{Other galaxies}  No other nearby galaxies, besides the SMC, LMC, and M 31, are detected. Seven  starburst galaxies in the {\sl IRAS} catalog \citep{IRAScatalog} are associated with 4FGL sources. Two sources, Arp 220 \citep{Pen16,Gri16,Yoa17} and NGC 2146 \citep{Tan14}, have been reported as LAT detections since the 3FGL release.  \citet{Yoa17} found an excess of $\gamma$ rays over the expected starburst contribution in Arp 220, similar to the case of the Circinus galaxy \citep{Hay13}. NGC 2146 being close ($0\fdg1$) to the FSRQ 1REX J061757+7816.1, the association is ambiguous. We favor the NGC 2146 association as no evidence for variability is found and the 4FGL photon index (2.17$\pm$0.17) is somewhat low for a FSRQ. Another source, NGC 3424, was not present in 3FGL. The {\sl IRAS} source UGC 11041, which could have been classified as sbg shows significant variability in the LAT band, so the $\gamma$-ray emission most likely arises from an AGN (there is a flat-spectrum radio source, MG2 J175448+3442 at a distance of 2.4$\arcmin$) and it is classified as such.  In addition to these seven associations, the Bayesian method predicts that three more 4FGL sources should be starburst galaxies (corresponding to the subthreshold associations mentioned in \S~\ref{sec:associations}). Some 4FGL sources are positionally consistent with known galaxy clusters, but these clusters host radio galaxies which are the most likely emitters.  No dwarf galaxies have been detected.

\subsection{Galactic sources} 

The Galactic sources include: 
\begin{itemize}
\item 239 pulsars (PSR).  The public list of LAT-detected pulsars is regularly updated\footnote{See \url{https://confluence.slac.stanford.edu/display/GLAMCOG/Public+List+of+LAT-Detected+Gamma-Ray+Pulsars}}. Some 232 pulsars in this list are included in 4FGL (68 would have been missed by the association pipeline using the ATNF catalog), while 6 are absent because they did not pass the $TS> 25$ criterion. These pulsars  represent by far the largest population of identified sources in 4FGL. Another 7 pulsars from the ATNF database are associated with 4FGL sources with high-confidence  via the Bayesian method that we consider $\gamma$-ray pulsar candidates. This method predicts that about 30 other 4FGL sources are ATNF pulsars. Note that out of the 24 pulsar candidates presented in 3FGL, pulsations have now been detected for 19 of them. The other 5 are not associated with pulsars in 4FGL.

\item  40 supernova remnants (SNR). Out of them, 24 are extended and thus firmly identified. The other 16 are not resolved. SNR G150.3+4.5 has a log-normal spectral shape with a very hard  photon index $\Gamma$ of 1.6, which indicates that the emission is most likely leptonic and makes this source an excellent candidate for the Cherenkov Telescope Array (CTA). In contrast, the softer spectrum of the LMC SNR N~132D (photon index=2.07) makes the  hypothesis of a dominant hadronic emission likely.  The significant spectral curvature seen in Puppis A is consistent with its non-detection in the TeV domain. 

\item  17 pulsar wind nebulae (PWN), 15 of them being extended.  New associations are N~157B, PWN G63.7+1.1, HESS J1356$-$645,   FGES J1631.6$-$4756,  FGES J1836.5$-$0651, FGES J1838.9$-$0704, HESS J1857+026. The median photon index of the 4FGL PWNe is 2.31.
N~157B, located in the LMC, has a photon index of 2.0, hinting at an additional contribution from a (yet-undetected) pulsar at low energy on top of the PWN. 

\item 78 unassociated sources overlapping with known PWNe or SNRs (SPP). 
Estimation of missed associations of SNR, PWN and SPP sources is made difficult by the intrinsic spatial extension of the sources;  no attempts have thus been  made along this line.

\item 30 globular clusters (GLC). Missing relative to 3FGL is 2MS$-$GC01.  The 16 new associations are  NGC 362, NGC 1904, NGC 5286, NGC 5904, NGC 6139,  NGC 6218,  NGC 6304,  NGC 6341, NGC 6397, NGC 6402, NGC 6838, NGC 7078,  Terzan 1, Terzan 2, GLIMPSE C01, GLIMPSE C02.  Only two other 4FGL sources are estimated to be GLCs.   
 
\item Six high-mass X-ray binaries (HMB).  The three new sources are HESS J0632+057, which has a reported LAT detection after 3FGL \citep{Li17},  Cyg X-1, an archetypical black-hole binary reported after the  3FGL \citep{Zdz14,Zan16}, and RX J0648.0$-$4418/HD 49798, which is a peculiar X-ray binary \citep{Mer11,Pop18}. The association probability of RX J0648.0$-$4418/HD 49798 is just barely larger (0.85 vs 0.80) than that of the blazar candidate SUMSS J064744$-$441946. Three other  4FGL sources are estimated to be HMBs according to the Bayesian method. 

\item Three star-forming regions; new since 3FHL is the association of the extended source FHES J1626.9$-$2431 (\S~\ref{catalog_extended}) with the $\rho$ Ophiuchi star-forming region. Positional coincidences between 4FGL sources and two of the brightest extended  H II regions present in the catalog of \cite{Pal03} have been found. They are reported here as candidate associations: one region corresponds to NGC 6618 in M17, whose extension of 6$\arcmin$ at 2.7 GHz encompasses  4FGL J1820.4$-$1609; the second one corresponds to NGC 4603, which has a similar extension  of 6$\arcmin$ at 2.7 GHz and encompasses  4FGL J1115.1$-$6118.

\item Two low-mass X-ray binaries (LMB). PSR J1023+0038 is a known binary millisecond pulsar/LMB transition system, with a change in $\gamma$-ray flux detected \citep{Sta14} simultaneously with a state change, and was previously detected as 2FGL J1023.6+0040 (but not detected in 3FGL). 
The LMB 2S 0921$-$630 (V395 Car) is a well-studied binary involving a neutron star and a K0 III star with an orbital period of 9 days \citep{Sha07} and is a new LAT detection. 
 
\item One binary star system (BIN), $\eta$  Carinae \citep{LAT10_EtaCarina,Rei15}.

\item One nova (NOV), V5668 Sagittarii \citep{Che16}. Other novae detected by the LAT are missing.  Novae have short durations, and most are below the significance threshold because their signal is diluted over the eight years of 4FGL data. As discussed in Section \ref{lctwomonths}, Nova V959 Mon 2012 is confused with the SNR Monoceros.
\end{itemize}


\subsection{Low-probability associations}

As a new feature relative to previous catalogs,  the most probable counterpart to a 4FGL unassociated source is given in a separate column of the FITS table, along with the corresponding association probability (applying a threshold of 0.1 on that probability). This additional information, to be used with care given its low confidence,  is meant to foster further investigations regarding the nature of these 4FGL sources and to help clarify why detections claimed in other works are sometimes missing in 4FGL.  We report  124 low-confidence (0.1$<P<$0.8) associations with the Bayesian method. Note that the relative distances between $\gamma$-ray and counterpart sources remain quite small (53 are within $r_{95}$ and  all within 1.85 $r_{95}$). This quite small number of low-association sources illustrates how quickly the Bayesian association probability drops with increasing relative distance in the case of 4FGL.  Except for rare exceptions, the other 1199 4FGL sources (having not even low-confidence associations) will not get associated with any of the  tested sources (i.e., belonging to the  catalogs listed in Table \ref{tab:catalogs}) in a future LAT catalog. We also report 42 matches (classified as UNK) with sources from the {\it Planck} surveys (with 0.1$<P \leq 1$) to guide future investigations.

\subsection{Unassociated sources}

\begin{figure}[!ht]
\centering
\includegraphics[width=\linewidth]{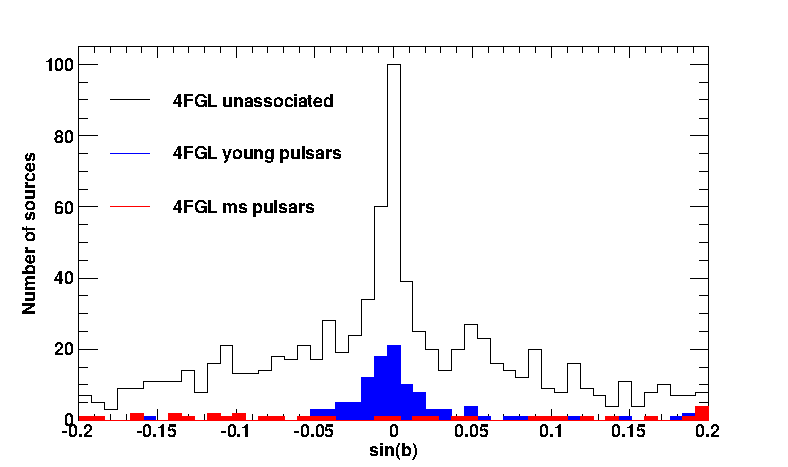}
\caption{Distributions in Galactic latitude $b$ of  4FGL low-latitude,  unassociated sources (black histogram), compared to those of LAT-detected pulsars (young pulsars: blue histogram, millisecond pulsars (MSP): red histogram).}
\label{fig:sinb_unassoc}
\end{figure}

\begin{figure}[!ht]
\centering
\includegraphics[width=\linewidth]{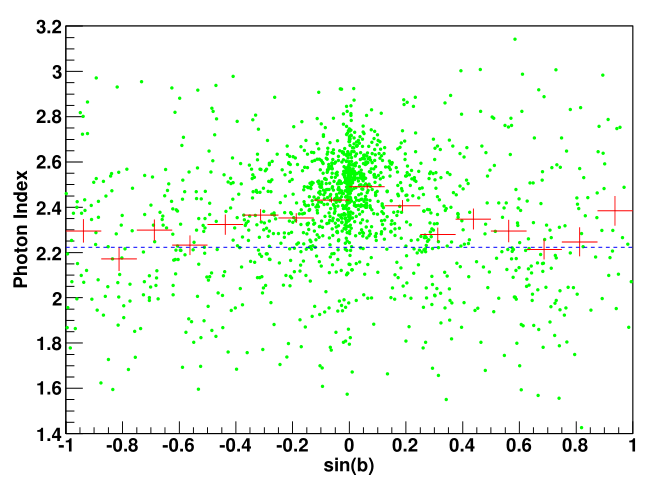}
\caption{Green symbols: Power-law photon index versus Galactic latitude, $b$,  for the unassociated 4FGL sources. Red bars: average photon index for different bins in $b$. Dashed blue line: average photon index of 4FGL BCU blazars.}
\label{fig:index_lat}
\end{figure}

Out of the 1336 sources unassociated in 4FGL, 368 already present in 3FGL had no associations there. Another 27 sources previously associated in 3FGL have now lost their associations  because of a shift in their locations relative to 3FGL.

About half of the unassociated sources are located less than 10$\degr$ away from the Galactic plane. Their wide latitude extension is hard to reconcile with those of known classes of Galactic $\gamma$-ray sources. For instance, Figure \ref{fig:sinb_unassoc} compares this latitude distribution with that of LAT pulsars.  In addition to nearby millisecond pulsars, which have a quasi isotropic distribution, the LAT detects only young isolated pulsars (age $<$10$^6$ y) which are by nature clustered close to the plane. Older pulsars, which have had time to drift further off the plane,  show a wider Galactic-latitude distribution, more compatible with the  observed distribution of the unassociated sources, but these pulsars have crossed the `$\gamma$-ray death line' \citep[see ][]{LAT13_2PC} and are hence undetectable.   Attempts to spatially cross correlate the unassociated population with other potential classes, e.g., LMBs \citep{LMXBcatalog}, O stars\footnote{Galactic O-star catalog (GOSC) \url{https://gosc.cab.inta-csic.es/}}, and Be stars\footnote{Be Star Spectra (BeSS) \url{http://basebe.obspm.fr/basebe/}} have been unsuccessful. The observed clustering of these unassociated sources in high-density `hot spots' may be a clue that they actually correspond to yet-to-be identified, relatively nearby  extended sources. The Galactic latitude distribution near the plane is clearly non-Gaussian as visible in  Figure \ref{fig:sinb_unassoc}, which may indicate the presence of several components.

The spectral properties of these sources can also provide insight into their nature, as illustrated in Figure \ref{fig:index_lat} which shows the latitude distribution of their spectral indices. The change in spectral hardness with sky location demonstrates the composite nature of the unassociated population. The high-latitude sources have an average photon index compatible with that of blazars of unknown type ($\Gamma$=2.24), a hint that these sources could be  predominantly blazars. Unassociated sources lying closer to the Galactic plane have softer spectra, closer to that expected for young pulsars ($\Gamma$=2.42). Another interesting possibility is that some of these unassociated sources actually correspond to WIMP dark matter annihilating in Galactic subhalos \citep{LAT_DMhalo,Cor19}. Indeed, $\Lambda$CDM cosmology predicts the existence of thousands of subhalos below $\sim10^7M_{\odot}$, i.e., not massive enough to retain gas or stars at all. As a result, they are not expected to emit at other wavelengths and therefore they would not possess astrophysical counterparts. Annihilation of particle dark matter may yield a pulsar-like spectrum \citep{Bal07}.

\subsection{Sources missing from previous \Fermi catalogs} 

The correspondence of 4FGL sources with previous \Fermilat catalogs (reported in the \texttt{ASSOC\_FGL} and \texttt{ASSOC\_FHL} columns) was based, as in 3FGL, on error-circle overlap at the 95\% confidence level, amounting to
\begin{equation}
\Delta \le d_{x,a} = \sqrt{\theta_{x,a}^2 +\theta_{x,{\rm 4FGL}}^2}
\label{eq:compnFGL}
\end{equation}
where $\Delta$ is the angular distance between a 4FGL source and a source in catalog $a$, and the $\theta_x$ are derived from the \texttt{Conf\_95\_SemiMajor} columns in the two catalogs at the $x$\% confidence level (assuming a 2-D normal distribution).
We also considered that a previous LAT source corresponds to a 4FGL source whenever they have the same association (the associations can have offsets greater than $\theta_{95}$, depending on the density of sources in the catalogs of counterparts at other wavelengths).

We checked all sources that did not have an obvious counterpart in 4FGL inside $d_{95}$, nor a common association.
The procedure is described in detail in \S~4.2.3 of the 3FGL paper. The result is provided in one FITS file per catalog\footnote{The files are available at \url{https://www-glast.stanford.edu/pub_data/1626/}.}, reporting the same information as Table~11 of the 3FGL paper: counterparts up to $1\degr$, whether they are inside $d_{99.9}$ ( = $1.52 \; d_{95}$) or not, and specific conditions (flagged, \texttt{c} source, close to an extended source, split into several sources).
The number of missed sources and their nature are provided in Table~\ref{tab:lost}.

\begin{deluxetable*}{lcccc|ccc}

\tablecaption{Statistics of previous \Fermi sources missing in 4FGL}
\label{tab:lost}
\tablehead{
 & \colhead{0FGL} & \colhead{1FGL} & \colhead{2FGL} & \colhead{3FGL} & \colhead{1FHL} & \colhead{2FHL} & \colhead{3FHL}
}

\startdata
All & 16 & 283 & 311 & 469 & 23 & 34 & 33 \\
With flags \tablenotemark{(a)} & \nodata & 117 & 229 & 262 & \nodata & \nodata & \nodata \\
Name-FGL \texttt{c} \tablenotemark{(b)} & \nodata & 83 & 97 & 52 & \nodata & \nodata & \nodata \\
Split into several 4FGL sources \tablenotemark{(c)} & 13 & 58 & 68 & 65 & 3 & 3 & 5 \\
Within $1\degr$ of a 4FGL \texttt{e} \tablenotemark{(d)} & 11 & 45 & 65 & 93 & 4 & 6 & 5 \\
\hline
AGN & 1 & 8 & 17 & 55 & 1 & 2 & 10 \\
PSR & 0 & 1 & 2 & 3 & 0 & 0 &  0 \\
spp & 4 & 7 & 19 & 11 & 2 & 0 & 0 \\
Other class-type & 0 & 1 & 2 & 3 & 0 & 1 & 3 \\
Unassociated & 11 & 266 & 271 & 397 & 20 & 31 & 20 \\
\hline
Present in 0FGL & \nodata & 6 & 2 & 6 & 1 & 1 & 0 \\
Present in 1FGL & 8 & \nodata & 56 & 35 & 4 & 3 & 3 \\
Present in 2FGL & 4 & 74 & \nodata & 78 & 4 & 6 & 1 \\
Present in 3FGL & 7 & 52 & 91 & \nodata & 6 & 4 & 4 \\
Present in 1FHL & 0 & 12 & 7 & 2 & \nodata & 8 & 2 \\
Present in 2FHL & 1 & 3 & 0 & 2 & 5 & \nodata & 1 \\
Present in 3FHL & 0 & 8 & 4 & 4 & 2 & 4 & \nodata \\
Not in any other \Fermilat catalog &  4 & 186 & 188 & 369 & 12 & 21 & 27 \\
\enddata
\tablenotetext{a}{Those are flagged as F in the FITS files.}
\tablenotetext{b}{\texttt{c} indicates that based on the region of the sky the source is considered to be potentially confused with Galactic diffuse emission.}
\tablenotetext{c}{Those are flagged as S in the FITS files.}
\tablenotetext{d}{\texttt{e} indicates a source that was modeled as spatially extended. Those are flagged as E in the FITS files.}

\end{deluxetable*}

We have looked at the most-recent catalogs, 3FGL and 3FHL, in more detail. 
Because the first four years are in common, we expect the 3FGL and 4FGL positions to be correlated. That correlation is however less than one might think because the data have changed (from Pass 7 to Pass 8, \S~\ref{LATData}). We found that the distribution of $\Delta/d_{95,{\rm 3FGL}}$ (when it is less than 1) is narrower by a factor 0.83 than the Rayleigh distribution. This means that, by cutting at $d_{95,{\rm 3FGL}}$, we expect only 1.3\% misses by chance (about 40 sources).
With 3FHL the correlation is larger because it used Pass 8 already, the overlap is 7 years, and for the hard sources present in 3FHL the lower-energy photons do not contribute markedly to the localization. The distribution of $\Delta/d_{95,{\rm 3FHL}}$ is narrowed by a factor 0.62, and the number of chance misses by cutting at $d_{95,{\rm 3FHL}}$ should be only 0.04\% (about 1 source). The correlation is similarly large with 2FHL (6 years of Pass 8 data).
That correlation effect is less for earlier catalogs, so for them the fraction of true counterparts that are found outside the combined 95\% error circle is closer to 5\%. Most of those true sources are expected to have a 4FGL counterpart at the 99.9\% level in the FITS files.

Out of 3033 3FGL sources, 469 are missing in 4FGL for various reasons, including the change of diffuse emission model, point sources being absorbed into new extended ones, or variability effects. Most of these missing sources had low significance in 3FGL. Only 72 sources were associated. The majority  are blazars (35 BCUs, 17 FSRQs, one BLL, and one SSRQ) plus one AGN. While BLLs are 36\% more numerous relative to FSRQs in 3FGL, only one has gone away in 4FGL, an effect possibly related to the larger variability of FSRQs relative to BLLs observed in the LAT energy band \citep{LAT15_3LAC}. Other missing sources include 11 SPPs, 3 PSRs, one SNR, and one PWN. The nova V407 Cyg is now missing as it no longer fulfills the average-significance criterion.

Two LAT pulsars are considered lost. PSR J1513$-$5908 (= 3FGL J1513.9$-$5908) inside the PWN MSH 15$-$52 is a pulsar peaking at MeV energies \citep{PSRB1509_99}, very soft in the LAT band \citep{AGILEPSRs_09, LAT10_PSR1509}, which has gone below threshold after applying the weights. PSR J1112$-$6103 (= 3FGL J1111.9$-$6058) was split into two 4FGL sources. One is still associated to the pulsar, but it is not the one closest to the 3FGL position. 
The third missing pulsar association was between 3FGL J1632.4$-$4820 and the non-LAT PSR J1632$-$4818, in a confused region now covered by the extended source 4FGL J1633.0$-$4746e.
Among the five most significant lost 3FGL sources ($> 20 \sigma$), the brightest one (3FGL J1714.5$-$3832 = CTB 37A) was split into two 4FGL sources, the brighter of which is associated instead to the newly discovered pulsar PSR J1714$-$3830 \citep{PSRJ1714_2018} inside the CTB 37A SNR, and hence was not recognized as a common association. Two others (3FGL J1906.6+0720 and 3FGL J0536.4$-$3347) were also split, and now both members of each pair are associated. This is definitely an improvement. The last two (3FGL J1745.3$-$2903c and 3FGL J1747.0$-$2828) were within $0\fdg6$ of the Galactic center, a region of the sky where changing the diffuse model had a strong impact. They have no 4FGL counterpart at all.

Concerning sources missing from 3FHL, established with Pass 8 data as 4FGL, they amount to 33, with 17 unassociated, 9  blazars (4 BLLs and 5 BCUs), one AGN, one SNR,  four UNKs and the transient HMB PSR B1259$-$63 (diluted over 8 years). All these sources had a $TS$ close to the $TS = 25$ significance threshold.  

\subsection{TeV sources} 

\startlongtable
\begin{deluxetable}{ll}
\setlength{\tabcolsep}{0.04in}
\tablewidth{0pt} 
\tabletypesize{\scriptsize}
\tablecaption{Associations of 4FGL with Extended TeV Sources\label{tab:ext_tev_assoc}}
\tablehead{
\colhead{TeVCat Name\tablenotemark{$a$}} & \colhead{4FGL Name}}
\startdata
Boomerang & J2229.0+6114\\
CTA 1 & J0007.0+7303\\
CTB 37A & J1714.4$-$3830\\
CTB 37B & J1714.1$-$3811\\
Crab & J0534.5+2201e\\
G318.2+00.1 & J1453.4$-$5858\\
Geminga & J0633.9+1746\\
HESS J1018$-$589B & J1016.3$-$5857\\
HESS J1026$-$582 & J1028.5$-$5819 \\
HESS J1303$-$631 & J1303.0$-$6312e\\
HESS J1356$-$645 & J1355.2$-$6420e\\
HESS J1420$-$607 & J1420.3$-$6046e\\
HESS J1427$-$608 & J1427.8$-$6051\\
HESS J1458$-$608 & J1456.7$-$6050, J1459.5$-$6053\\
HESS J1507$-$622 & J1507.9$-$6228e\\
HESS J1534$-$571 & J1533.9$-$5712e\\
HESS J1614$-$518 & J1615.3$-$5146e\\
HESS J1616$-$508 & J1616.2$-$5054e\\
HESS J1632$-$478 & J1633.0$-$4746e\\
HESS J1640$-$465 & J1640.6$-$4632\\
HESS J1702$-$420 & J1705.7$-$4124\\
HESS J1718$-$385 & J1718.2$-$3825\\
HESS J1729$-$345 & J1730.1$-$3422\\
HESS J1745$-$303 & J1745.8$-$3028e\\
HESS J1800$-$240A & J1801.8$-$2358\\
HESS J1800$-$240B & J1800.2$-$2403, J1800.7$-$2355, J1800.9$-$2407 \\
HESS J1804$-$216 & J1804.7$-$2144e\\
HESS J1808$-$204 & J1808.2$-$2028e\\
HESS J1809$-$193 & J1810.3$-$1925e\\
HESS J1813$-$126 & J1813.4$-$1246\\
HESS J1813$-$178 & J1813.1$-$1737e\\
HESS J1825$-$137 & J1824.5$-$1351e\\
HESS J1826$-$130 & J1826.1$-$1256\\
HESS J1834$-$087 & J1834.5$-$0846e\\
HESS J1841$-$055 & J1840.9$-$0532e\\
HESS J1848$-$018 & J1847.2$-$0141, J1848.6$-$0202, J1848.7$-$0129\\
HESS J1857+026 & J1857.7+0246e\\
HESS J1858+020 & J1858.3+0209\\
HESS J1912+101 & J1911.7+1014, J1912.7+0957, J1913.3+1019\\
IC 443 & J0617.2+2234e\\
Kookaburra (Rabbit) & J1417.7$-$6057, J1418.7$-$6057\\
Kookaburra PWN & J1420.0$-$6048\\
MGRO J1908+06 & J1906.2+0631, J1907.9+0602\\
MGRO J2031+41 & J2028.6+4110e\\
MSH 15$-$52 & J1514.2$-$5909e\\
RCW 86 & J1443.0$-$6227e\\
RX J0852.0$-$4622 & J0851.9$-$4620e\\
RX J1713.7$-$3946 & J1713.5$-$3945e\\
SNR G292.2$-$00.5 & J1119.1$-$6127\\
TeV J1626$-$490 & J1628.2$-$4848 \\
Terzan 5 & J1748.0$-$2446\\
VER J2019+407 & J2021.0+4031e\\
Vela X & J0833.1$-$4511e\\
W 28 & J1801.3$-$2326e\\
W 51 & J1923.2+1408e\\
Westerlund 1 & J1645.8$-$4533, J1648.4$-$4611, J1649.2$-$4513, \\
 & J1650.3$-$4600, J1652.2$-$4516\\
Westerlund 2 & J1023.3$-$5747e\\
\enddata 
\tablenotetext{a}{From \url{http://tevcat.uchicago.edu}.}
\end{deluxetable}

The synergy between the LAT and the Cherenkov telescopes operating in the TeV energy domain has proven extremely fruitful, in particular by bringing out promising TeV candidates in the LAT catalogs. This approach, further motivated by the upcoming deployment of the Cherenkov Telescope Array, has  justified the release of LAT source catalogs above 10 GeV, like the 3FHL \citep{LAT17_3FHL} based on 7 years of data.  The associations of 4FGL sources with extended sources  listed in TeVCat\footnote{\url{http://tevcat.uchicago.edu/ }} are presented in Table \ref{tab:ext_tev_assoc}. Relative to 3FHL, 9 new extended TeV sources  are associated with 4FGL extended sources (TeV sources: HESS J1534$-$571, HESS J1808$-$204, HESS J1809$-$193, see \S~\ref{catalog_extended}), or (sometimes multiple) 4FGL point sources  (TeV sources: HESS J1718$-$385, HESS J1729$-$345, HESS J1848$-$018, HESS J1858+020, MGRO J1908+06, HESS J1912+101). All TeV blazars have 4FGL counterparts. The median value of $\Gamma$ for  4FGL point sources associated with TeV point sources is 1.95, indicating hard spectra as expected. In associations with extended TeV sources, the median  $\Gamma$ changes from  2.09 to  2.38  depending on whether the 4FGL sources are extended or not.  This fairly  large difference favors the interpretation that most associations between extended TeV sources and non-extended 4FGL sources are accidental.   

\subsection{Counterpart positions}

Whenever a high-confidence association with a point-like counterpart is obtained, we provide the most accurate counterpart position available and its uncertainty. In particular, 2775 4FGL AGNs have Very Long Baseline Interferometry (VLBI) counterparts. VLBI, i.e., radio interferometry with baseline lengths of $>$1000\,km, is sensitive to radio emission from compact regions of AGNs that are smaller than 20  milliarcsecond (mas), which corresponds to parsec scales. Such observations allow the determination of  positions of the AGN jet base with mas level accuracy. We used the RFC catalog based on the
dedicated on-going observing program \citep{r:aofus2,r:aofus3} with the Very
Long Baseline Array \citep{r:vlba}, as well as VLBI data under other programs. The association between $\gamma$-ray source and VLBI counterpart was evaluated along a
similar, but distinct, scheme as that presented in \S~\ref{sec:associations}. This scheme \citep[see ][ for more details]{r:aofus1} is based on the strong connection between the $\gamma$-ray emission
and radio emission at parsec scales and on the sky density of bright compact radio sources being relatively low. The chance to find a bright background,
unrelated compact radio source within the LAT positional error
ellipse is low enough to establish association. The likelihood ratio (with a somewhat different definition from that implemented in the LR-method)  was required to be  greater than 8 to claim an association, with an estimated false association fraction of  1\%. 

For AGNs without VLBI counterparts, the position uncertainties were set to typical values of 20$\arcsec$ for sources associated from the RASS survey and 10$\arcsec$ otherwise. For identified  pulsars, the position uncertainties come from the rotation ephemeris used to find $\gamma$-ray pulsations, many of which were obtained from radio observations \citep{Smith19}.
If the ephemeris does not include the uncertainties and for pulsar candidates, we use the ATNF psrcat values. If neither of those exist, we use the 0.1$\degr$ uncertainties from the list maintained by the WVU Astrophysics group\footnote{\url{http://astro.phys.wvu.edu/GalacticMSPs/GalacticMSPs.txt}}. Ephemeris position uncertainties are often underestimated, so we arbitrarily apply a minimum uncertainty of 1 mas. For GLC from \citet{GlobClusterCatalog}\footnote{\url{https://heasarc.gsfc.nasa.gov/w3browse/all/globclust.html}}, the position uncertainties were assigned a typical value of 2$\arcsec$.


\section{Conclusions}
\label{conclusions}

The fourth {\it Fermi} LAT source catalog is the deepest-yet in the GeV energy range.  The increased sensitivity relative to the 3FGL catalog is due to both the longer time interval (8 years versus 4 years) and the use of Pass 8 data, which provides more acceptance over the entire energy range and a narrower PSF at high energy.  The 4FGL catalog also benefits from higher-level improvements in the analysis, including an improved model for Galactic diffuse emission, a weighted log-likelihood method to mitigate the systematic effects due to that diffuse emission model, and systematic testing of three spectral representations, useful to classify unassociated sources.   

The 4FGL catalog includes 5064 sources.  The sources are detected ($TS > 25$) based on their average fluxes in the 8-year data set; 1327 of the sources are found to be significantly variable on one-year timescales, and 1173 on two-month timescales.  We mark 92 (1.8\%) of the sources as potentially related to imperfections in the model for Galactic diffuse emission; the character \texttt{c} is appended to their names (except those already marked as \texttt{e} for extended).  An additional 1071 (21.1\%) are flagged in the catalog for less serious concerns, e.g., for the spectral model having a poor fit or for being close to a brighter source.  Of the 5064 sources in the catalog, 358 (7.1\%) are considered identified, based on pulsations, correlated variability, or correlated angular sizes with observations at other wavelengths. We find likely lower-energy counterparts for 3370 other sources (66.5\%).  The remaining 1336 sources (26.4\%) are unassociated.

The identified and associated sources in the 4FGL catalog include many Galactic and extragalactic source classes.  The largest Galactic source class continues to be pulsars, with 232 known $\gamma$-ray pulsars and 7 associations to non-LAT pulsars.  Other Galactic source classes have continued to grow; 30 globular clusters, 40 supernova remnants and 17 pulsar wind nebulae are now associated with LAT sources.  Blazars remain the largest class of extragalactic sources, with more than 1800 identified or associated with BL Lac or FSRQ active galaxies.  Non-blazar classes of active galaxies are also found, including 9 narrow-line Seyfert galaxies, 5 compact steep spectrum radio sources and 42 radio galaxies. The populations of active galaxies in 4FGL are considered in more detail in the companion 4LAC catalog.

\acknowledgments
The \textit{Fermi} LAT Collaboration acknowledges generous ongoing support
from a number of agencies and institutes that have supported both the
development and the operation of the LAT as well as scientific data analysis.
These include the National Aeronautics and Space Administration and the
Department of Energy in the United States, the Commissariat \`a l'Energie Atomique
and the Centre National de la Recherche Scientifique / Institut National de Physique
Nucl\'eaire et de Physique des Particules in France, the Agenzia Spaziale Italiana
and the Istituto Nazionale di Fisica Nucleare in Italy, the Ministry of Education,
Culture, Sports, Science and Technology (MEXT), High Energy Accelerator Research
Organization (KEK) and Japan Aerospace Exploration Agency (JAXA) in Japan, and
the K.~A.~Wallenberg Foundation, the Swedish Research Council and the
Swedish National Space Board in Sweden.
 
Additional support for science analysis during the operations phase is gratefully
acknowledged from the Istituto Nazionale di Astrofisica in Italy and the Centre
National d'\'Etudes Spatiales in France. This work performed in part under DOE
Contract DE-AC02-76SF00515.

This work made extensive use of the ATNF pulsar  catalog\footnote{\url{http://www.atnf.csiro.au/research/pulsar/psrcat}}  \citep{ATNFcatalog}.  This research has made use of the NASA/IPAC Extragalactic Database (NED) which is operated by the Jet Propulsion Laboratory, California Institute of Technology, under contract with the National Aeronautics and Space Administration, and of archival data, software and online services provided by the ASI Science Data Center (ASDC) operated by the Italian Space Agency.
We used the Manitoba SNR catalog \citep{Ferrand2012_SNRCat} to check recently published extended sources.

\software{Gardian \citep{Diffuse2}, GALPROP\footnote{\url{http://galprop.stanford.edu}} \citep{GALPROP17}, HEALPix\footnote{\url{http://healpix.jpl.nasa.gov/}} \citep{Gorski2005}, Aladin\footnote{http://aladin.u-strasbg.fr/}, TOPCAT\footnote{\url{http://www.star.bristol.ac.uk/\~mbt/topcat/}} \citep{Tay05}, APLpy, an open-source plotting package for Python\footnote{\url{http://aplpy.github.com}} \citep{Rob12}, XSPEC\footnote{\url{https://heasarc.gsfc.nasa.gov/xanadu/xspec/}}}

\facility{\Fermilat}


\appendix

\section{Description of the FITS version of the 4FGL catalog}
\label{appendix_fits_format}

\startlongtable
\begin{deluxetable*}{lccl}
\setlength{\tabcolsep}{0.04in}
\tablewidth{0pt}
\tabletypesize{\scriptsize}
\tablecaption{LAT 4FGL FITS Format: LAT\_Point\_Source\_Catalog Extension
\label{tab:description}}
\tablehead{
\colhead{Column} &
\colhead{Format} &
\colhead{Unit} &
\colhead{Description}
}
\startdata
Source\_Name & 18A & \nodata & Source name 4FGL JHHMM.m+DDMMa\tablenotemark{a} \\
RAJ2000 & E & deg & Right Ascension \\
DEJ2000 & E & deg & Declination \\
GLON & E & deg & Galactic Longitude \\
GLAT & E & deg & Galactic Latitude \\
Conf\_68\_SemiMajor & E & deg & Long radius of error ellipse at 68\% confidence\tablenotemark{b} \\
Conf\_68\_SemiMinor & E & deg & Short radius of error ellipse at 68\% confidence\tablenotemark{b} \\
Conf\_68\_PosAng & E & deg & Position angle of the 68\% ellipse\tablenotemark{b} \\
Conf\_95\_SemiMajor & E & deg & Long radius of error ellipse at 95\% confidence \\
Conf\_95\_SemiMinor & E & deg & Short radius of error ellipse at 95\% confidence \\
Conf\_95\_PosAng & E & deg & Position angle (eastward) of the long axis from celestial North \\
ROI\_num & I & \nodata & RoI number (cross-reference to ROIs extension) \\
Extended\_Source\_Name & 18A & \nodata & Cross-reference to the ExtendedSources extension \\
Signif\_Avg & E & \nodata & Source significance in $\sigma$ units over the 100~MeV to 1~TeV band \\
Pivot\_Energy & E & MeV & Energy at which error on differential flux is minimal \\
Flux1000 & E & cm$^{-2}$ s$^{-1}$ & Integral photon flux from 1 to 100~GeV \\
Unc\_Flux1000 & E & cm$^{-2}$ s$^{-1}$ & $1\sigma$ error on integral photon flux from 1 to 100~GeV \\
Energy\_Flux100 & E & erg cm$^{-2}$ s$^{-1}$ & Energy flux from 100~MeV to 100~GeV obtained by spectral fitting \\
Unc\_Energy\_Flux100 & E & erg cm$^{-2}$ s$^{-1}$ & $1\sigma$  error on energy flux from 100~MeV to 100~GeV \\
SpectrumType & 18A & \nodata & Spectral type in the global model (PowerLaw, LogParabola, PLSuperExpCutoff) \\
PL\_Flux\_Density & E & cm$^{-2}$ MeV$^{-1}$ s$^{-1}$ & Differential flux at Pivot\_Energy in PowerLaw fit \\
Unc\_PL\_Flux\_Density & E & cm$^{-2}$ MeV$^{-1}$ s$^{-1}$ & $1\sigma$  error on PL\_Flux\_Density \\
PL\_Index & E & \nodata & Photon index when fitting with PowerLaw \\
Unc\_PL\_Index & E & \nodata & $1\sigma$ error on PL\_Index \\
LP\_Flux\_Density & E & cm$^{-2}$ MeV$^{-1}$ s$^{-1}$ & Differential flux at Pivot\_Energy in LogParabola fit \\
Unc\_LP\_Flux\_Density & E & cm$^{-2}$ MeV$^{-1}$ s$^{-1}$ & $1\sigma$  error on LP\_Flux\_Density \\
LP\_Index & E & \nodata & Photon index at Pivot\_Energy ($\alpha$ of Eq.~\ref{eq:logparabola}) when fitting with LogParabola \\
Unc\_LP\_Index & E & \nodata & $1\sigma$ error on LP\_Index \\
LP\_beta & E & \nodata & Curvature parameter ($\beta$ of Eq.~\ref{eq:logparabola}) when fitting with LogParabola \\
Unc\_LP\_beta & E & \nodata & $1\sigma$ error on LP\_beta \\
LP\_SigCurv & E & \nodata & Significance (in $\sigma$ units) of the fit improvement between PowerLaw and \\
& & & LogParabola. A value greater than 4 indicates significant curvature \\
PLEC\_Flux\_Density & E & cm$^{-2}$ MeV$^{-1}$ s$^{-1}$ & Differential flux at Pivot\_Energy in PLSuperExpCutoff fit \\
Unc\_PLEC\_Flux\_Density & E & cm$^{-2}$ MeV$^{-1}$ s$^{-1}$ & $1\sigma$  error on PLEC\_Flux\_Density \\
PLEC\_Index & E & \nodata & Low-energy photon index ($\Gamma$ of Eq.~\ref{eq:expcutoff}) when fitting with PLSuperExpCutoff \\
Unc\_PLEC\_Index & E & \nodata & $1\sigma$ error on PLEC\_Index \\
PLEC\_Expfactor & E & \nodata & Exponential factor ($a$ of Eq.~\ref{eq:expcutoff}) when fitting with PLSuperExpCutoff \\
Unc\_PLEC\_Expfactor & E & \nodata & $1\sigma$ error on PLEC\_Expfactor \\
PLEC\_Exp\_Index & E & \nodata & Exponential index ($b$ of Eq.~\ref{eq:expcutoff}) when fitting with PLSuperExpCutoff \\
Unc\_PLEC\_Exp\_Index & E & \nodata & $1\sigma$ error on PLEC\_Exp\_Index \\
PLEC\_SigCurv & E & \nodata & Same as LP\_SigCurv for PLSuperExpCutoff model \\
Npred & E & \nodata & Predicted number of events in the model \\
Flux\_Band & 7E & cm$^{-2}$ s$^{-1}$ & Integral photon flux in each spectral band \\
Unc\_Flux\_Band & $2 \times 7$E & cm$^{-2}$ s$^{-1}$ & $1\sigma$ lower and upper error on Flux\_Band\tablenotemark{c} \\
nuFnu\_Band & 7E & erg cm$^{-2}$ s$^{-1}$ & Spectral energy distribution over each spectral band \\
Sqrt\_TS\_Band & 7E & \nodata & Square root of the Test Statistic in each spectral band \\
Variability\_Index & E & \nodata & Sum of 2$\times$log(Likelihood) difference between the flux fitted in each time \\
& & & interval and the average flux over the full catalog interval; a value greater \\
& & & than 18.48 over 12 intervals indicates $< $1\% chance of being a steady source \\
Frac\_Variability & E & \nodata & Fractional variability computed from the fluxes in each year \\
Unc\_Frac\_Variability & E & \nodata & $1\sigma$ error on fractional variability \\
Signif\_Peak & E & \nodata & Source significance in peak interval in $\sigma$ units \\
Flux\_Peak & E & cm$^{-2}$ s$^{-1}$ & Peak integral photon flux from 100~MeV to 100~GeV \\
Unc\_Flux\_Peak & E & cm$^{-2}$ s$^{-1}$ &  $1\sigma$ error on peak integral photon flux \\
Time\_Peak & D & s (MET) & Time of center of interval in which peak flux was measured \\
Peak\_Interval & E & s & Length of interval in which peak flux was measured \\
Flux\_History & 12E & cm$^{-2}$ s$^{-1}$ & Integral photon flux from 100~MeV to 100~GeV in each year (best fit from \\
& & &  likelihood analysis with spectral shape fixed to that obtained over full interval)\\
Unc\_Flux\_History & $2 \times 12$E & cm$^{-2}$ s$^{-1}$ &  $1\sigma$ lower and upper error on integral photon flux in each year\tablenotemark{c} \\
Sqrt\_TS\_History & 12E & \nodata & Square root of the Test Statistic in each year \\
Variability2\_Index & E & \nodata & Variability\_Index over two-month intervals; a value greater than 72.44 \\
& & &  over 48 intervals indicates $< $1\% chance of being a steady source \\
Frac2\_Variability & E & \nodata & Fractional variability computed from the fluxes every two months \\
Unc\_Frac2\_Variability & E & \nodata & $1\sigma$ error on Frac2\_Variability \\
Signif2\_Peak & E & \nodata & Source significance in peak interval in $\sigma$ units \\
Flux2\_Peak & E & cm$^{-2}$ s$^{-1}$ & Peak integral photon flux from 100~MeV to 100~GeV \\
Unc\_Flux2\_Peak & E & cm$^{-2}$ s$^{-1}$ &  $1\sigma$ error on peak integral photon flux \\
Time2\_Peak & D & s (MET) & Time of center of interval in which peak flux was measured \\
Peak2\_Interval & E & s & Length of interval in which peak flux was measured \\
Flux2\_History & 48E & cm$^{-2}$ s$^{-1}$ & Integral photon flux from 100~MeV to 100~GeV in each two-month interval \\
Unc\_Flux2\_History & $2 \times 48$E & cm$^{-2}$ s$^{-1}$ &  $1\sigma$ lower and upper error on Flux2\_History\tablenotemark{c} \\
Sqrt\_TS2\_History & 48E & \nodata & Square root of the Test Statistic in each two-month interval \\
ASSOC\_FGL & 18A & \nodata & Most recent correspondence to previous FGL source catalogs, if any \\
ASSOC\_FHL & 18A & \nodata & Most recent correspondence to previous FHL source catalogs, if any \\
ASSOC\_GAM1 & 18A & \nodata & Name of likely corresponding 2AGL source, if any \\
ASSOC\_GAM2 & 18A & \nodata & Name of likely corresponding 3EG source, if any \\
ASSOC\_GAM3 & 18A & \nodata & Name of likely corresponding EGR source, if any \\
TEVCAT\_FLAG & A & \nodata & P if positional association with non-extended source in TeVCat \\
& & & E if associated with an extended source in TeVCat, N if no TeV association \\
ASSOC\_TEV & 24A & \nodata & Name of likely corresponding TeV source from TeVCat, if any \\
CLASS1 & 5A & \nodata & Class designation for associated source; see Table~\ref{tab:classes} \\
CLASS2 & 5A & \nodata & Class designation for low-confidence association \\
ASSOC1 & 28A & \nodata & Name of identified or likely associated source \\
ASSOC2 & 26A & \nodata & Name of low-confidence association or of enclosing extended source \\
ASSOC\_PROB\_BAY & E & \nodata & Probability of association according to the Bayesian method\tablenotemark{d} \\
ASSOC\_PROB\_LR & E & \nodata & Probability of association according to the Likelihood Ratio method\tablenotemark{e} \\
RA\_Counterpart & D & deg & Right Ascension of the counterpart ASSOC1 \\
DEC\_Counterpart & D & deg & Declination of the counterpart ASSOC1 \\
Unc\_Counterpart & E & deg & 95\% precision of the counterpart localization\tablenotemark{f} \\
Flags & I & \nodata & Source flags (binary coding as in Table~\ref{tab:flags})\tablenotemark{g} \\
\enddata
\tablenotetext{a} {The coordinates are rounded, following the IAU convention. The letter at the end can be \texttt{c} (coincident with interstellar clump), 
\texttt{e} (extended source), \texttt{i} (for Crab nebula inverse Compton) or \texttt{s} (for Crab nebula synchrotron).}
\tablenotetext{b} {from the 95\% ellipse, assuming a Gaussian distribution.}
\tablenotetext{c} {Separate $1\sigma$ errors are computed from the likelihood profile toward lower and larger fluxes. The lower error is set equal to NULL and the upper error is derived from a Bayesian upper limit if the $1\sigma$ interval contains 0 ($TS < 1$).}
\tablenotetext{d} {NaN in this column when ASSOC1 is defined means that the probability could not be computed, either because the source is extended or because the counterpart is the result of dedicated follow-up.}
\tablenotetext{e} {Probabilities $< 0.8$ are formally set to 0.}
\tablenotetext{f} {For extended counterparts, this reports their extension radius.}
\tablenotetext{g} {Each condition is indicated by one bit among the 16 bits forming \texttt{Flags}. The bit is raised (set to 1) in the dubious case, so that sources without any warning sign have \texttt{Flags} = 0.}
\end{deluxetable*}

The FITS format version of the second release of the 4FGL catalog has eight binary table extensions.  The extension {\tt LAT\_Point\_Source\_Catalog Extension} has all of the information about the sources. Its format is described in Table~\ref{tab:description}. The table has 5065 rows for 5064 sources because the Crab nebula is described by two entries (the synchrotron component and the inverse Compton component) but counted as only one source. The Crab pulsar is another entry and counted as a separate source.

The extension {\tt GTI} is a standard Good-Time Interval listing the precise time intervals (start and stop in Mission Elapsed Time, MET) included in the data analysis.  The number of intervals is fairly large because on most orbits ($\sim$95~min) \Fermi passes through the SAA, and science data taking is stopped during these times.  In addition, data taking is briefly interrupted on each non-SAA-crossing orbit, as \Fermi crosses the ascending node.  Filtering of time intervals with large rocking angles, gamma-ray bursts, solar flares, data gaps, or operation in non-standard configurations introduces some more entries.  The GTI is provided for reference and is useful, e.g., for reconstructing the precise data set that was used for the analysis.

The extension {\tt ExtendedSources} (format unchanged since 2FGL) contains information about the 75 spatially extended sources that are modeled in the 4FGL source list (\S~\ref{catalog_extended}), including locations and shapes. The extended sources are indicated by an e appended to their names in the main table.

The extension {\tt ROIs} contains information about the 1748 RoIs over which the analysis ran. In particular it reports the best-fit diffuse parameters. Its format is very close to that in 3FGL, with one exception. The {\tt RADIUS} column is replaced by {\tt CoreRadius} which reports the radius of the RoI core (in which the sources which belong to the RoI are located). The RoI radius (half-width in binned mode) depends on the component, and is given by the core radius plus {\tt RingWidth}, where the latter is given in the {\tt Components} extension.

\begin{deluxetable*}{lccl}
\setlength{\tabcolsep}{0.04in}
\tablewidth{0pt}
\tabletypesize{\scriptsize}
\tablecaption{LAT 4FGL FITS Format: Components Extension\label{tab:compdesc}}
\tablehead{
\colhead{Column} &
\colhead{Format} &
\colhead{Unit} &
\colhead{Description}
}
\startdata
LowerEnergy & E & MeV & Lower bound of component's energy interval \\
UpperEnergy & E & MeV & Upper bound of component's energy interval \\
ENumBins & I & \nodata & Number of bins inside energy interval \\
EvType & I & \nodata & Event type selection for this component \\
ZenithCut & E & deg & Maximum zenith angle for this component \\
RingWidth & E & deg & Difference between RoI radius and core radius \\
PixelSize & E & deg & Pixel size for this component (of exposure map in unbinned mode) \\
BinnedMode & I & \nodata & 0=Unbinned, 1=Binned \\
Weighted & I & \nodata & 1 if weights were applied to this component \\
\enddata
\end{deluxetable*}

The extension {\tt Components} is new to 4FGL. It reports the settings of each individual component (15 in all) whose sum forms the entire data set for the SummedLikelihood approach, as described in Table~\ref{tab:components}. Its format is given by Table~\ref{tab:compdesc}.

The extension {\tt EnergyBounds} is new to 4FGL. It contains the definitions of the bands in which the fluxes reported in the {\tt xx\_Band} columns of the main extension were computed, and the settings of the analysis. Its format is the same as that of the {\tt Components} extension, plus one more column ({\tt SysRel}) reporting the systematic uncertainty on effective area used to flag the sources with Flag 10 (Table~\ref{tab:flags}). When several components were used in one band, several lines appear with the same {\tt LowerEnergy} and {\tt UpperEnergy}.

The extension {\tt Hist\_Start} (format unchanged since 1FGL) contains the definitions of the time intervals used to build the light curves.
The new extension {\tt Hist2\_Start} (same format) describes the time intervals used to build the second series of light curves.

\section{Weighted log-likelihood}
\label{appendix_weights}

\begin{figure}
\centering
\includegraphics[width=\linewidth]{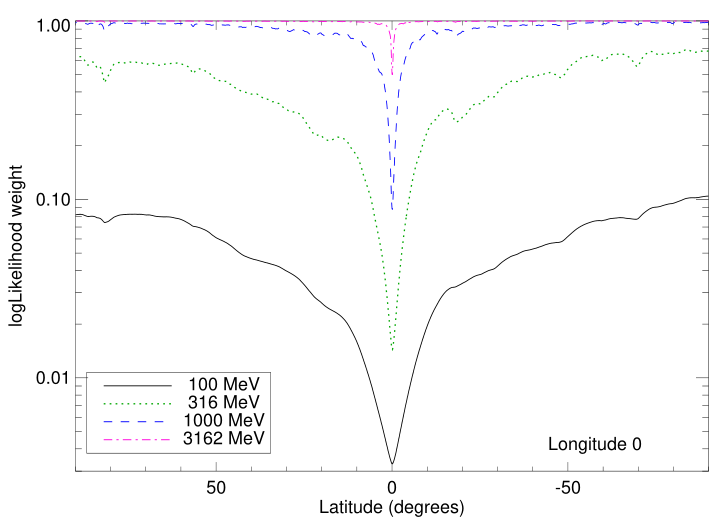}
\caption{Data-based log-likelihood weights as a function of latitude across the Galactic Center, at 100~MeV, 300~MeV, 1~GeV and 3~GeV, assuming all events are used throughout, and the same zenith cut at 105$\degr$. The dips at some latitudes are point sources, which are included in the data-based weights. Those weights were not used in 4FGL (which uses separate event types), they are shown here only for illustration.
}
\label{fig:weights_energy}
\end{figure}

\begin{figure}
\centering
\includegraphics[width=\linewidth]{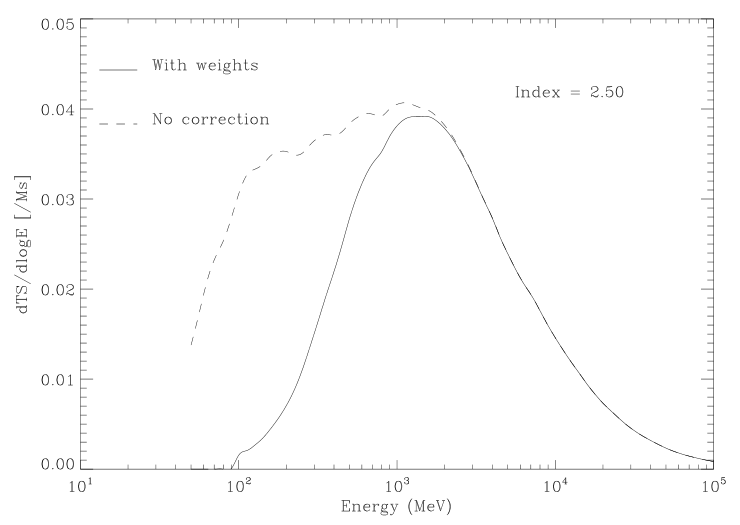}
\caption{Contribution to $TS$ as a function of energy for a power-law source with $\Gamma$ = 2.5 at high latitude, with and without weights.
This assumes all events are used throughout (and with the same zenith cut at 105$\degr$), as in Figure~\ref{fig:weights_energy}.
}
\label{fig:weights_TS}
\end{figure}

In 3FGL we introduced a first attempt at accounting for systematic errors in the maximum likelihood process itself, at the source detection level. It was not used in the source characterization, however, for lack of a suitable framework.
The standard way to account for systematic errors (for example in XSPEC\footnote{\url{https://heasarc.gsfc.nasa.gov/xanadu/xspec/}}) is to define them as a fraction $\epsilon$ of the signal and add them to the statistical errors in quadrature, in a $\chi^2$ formalism.
This can be adapted to the maximum likelihood framework by introducing weights $w_i < 1$ \citep{HuZidek02} as
\be
\log\mathcal{L} = \sum_i w_i (n_i \log M_i - M_i)
\label{eq:wlogL}
\ee
where $M_i$ and $n_i$ are the model and observed counts in each bin, and the sum runs over all bins in space and energy.
The source significance can then be quantified in the same way, via the Test Statistic $TS=2 \log(\mathcal{L}/\mathcal{L}_0)$ in which $\mathcal{L}$ and $\mathcal{L}_0$ are the (weighted) log-likelihood with and without the source of interest, respectively.

Since the statistical variance in Poisson statistics is the signal itself, a first guess for the weights could be 
\be
w_i = \frac{M_i}{M_i + (\epsilon M_i)^2} = \frac{1}{1 + \epsilon^2 M_i}
\ee
However, that formulation has a serious flaw, which is that it is not stable to rebinning. If one splits the bins in half, then $M_i$ is split in half while $\epsilon$ stays the same (it is defined externally). In the limit of very small bins, obviously the weights will all tend to 1 and the $\log\mathcal{L}$ formula will tend to the unweighted one, even though nothing has changed in the underlying data or the model.

The solution we propose, originally presented in \citet{LAT15_Cat}, is to define a suitable integral over energy ($E$) and space ($\bf r$) $N({\bf r},E)$ which does not depend on binning. $M_i$ in the weight formula is then replaced by $N({\bf r}_i,E_i)$ taken at the event's coordinates. For the integral over space, since the catalog mostly deals with point sources, the logical solution is to integrate the background under the PSF, i.e., to convolve the model with the PSF $P({\bf r},E)$, normalized to 1 at the peak (this is equivalent, for a flat diffuse emission, to multiplying by the PSF solid angle). Note that the model already contains the PSF, so this amounts to applying a double convolution to the sky model.

For the energy integral the choice is less obvious. The source spectrum is not a narrow line, so convolving with the energy dispersion (similar to what is done for space) is not justified. An integral over the full energy range would give the same weight to all energies, which is clearly not what we want (there is no reason to downplay the few high-energy events). The option we adopt here is to start the integration at the current energy.
\begin{eqnarray}
\label{wPSF}
w_i & = & \frac{1}{1 + \epsilon^2 N({\bf r}_i, E_i)} \\
N({\bf r}_i, E_i) & = & \int_{E_i}^{E_{\rm max}} S({\bf r}_i,E) \, dE \label{eq:Ntot} \\
S({\bf r},E) & = & \frac{dM}{dE}({\bf r},E) \ast P({\bf r},E)
\end{eqnarray}
where $dM/dE$ is the differential model.
As energy increases, the spectra (in counts) decrease and the LAT PSF gets narrower so the convolution makes $S$ even steeper than $dM/dE$.
As a result, the integral giving $N$ is nearly always dominated by the lowest energies, so the exact upper bound $E_{\rm max}$ is not critical. The only spectral region where it is important is the very lowest energies ($<$ 100~MeV) where the effective area rises steeply. In order not to penalize the lowest energies too much, we set $E_{\rm max} = 2 E_i$ in Eq.~\ref{eq:Ntot}.

There are two possibilities to define $dM/dE$. Since the main origin of the systematic error is the diffuse emission, we can restrict $dM/dE$ to the diffuse emission model only (we call the result model-based weights). On the other hand there are also systematic uncertainties on sources due to PSF calibration and our imperfect spectral representation, so another option is to enter the full model (or the data themselves) into $dM/dE$ (we call the result data-based weights). That second choice limits spurious sources next to bright sources.
There is of course no reason why the level of systematics $\epsilon$ should be the same for the diffuse emission model and the sources, but in practice it is a reasonable approximation.

Another important point, for the procedure to be stable, is that the weights should not change with the model parameters. So $dM/dE$ must be defined beforehand (for example from a previous fit).
In this work we use data-based weights computed from the data themselves, with a common $\epsilon$.
The data are not as smooth as the model, but this is not a problem in the regime of large counts where weights play a role.

We assume here that $\epsilon$ is a true constant (it depends neither on space nor on energy). For a given $\epsilon$ the weights are close to 1 at high energy and decrease toward low energy. At a given energy the weights are smallest where the data is largest (in the Galactic ridge).
We illustrate that behavior in Figure~\ref{fig:weights_energy}, merging all event types together (not what we do in 4FGL), for 8 years and $\epsilon$ = 3\%. The width of the trough in the Galactic Ridge gets narrower at high energy, as the PSF improves. At 100~MeV the weights are everywhere less than 12\%. They reach 50\% at high latitude at 250~MeV, and 90\% at 500~MeV.
This justifies our choice of discarding 75\% of the events below 100~MeV and 50\% below 300~MeV (Table~\ref{tab:components}). The entire sky is limited by systematic effects below 300~MeV.
On average in the Galactic ridge (a little better than the very center shown in Figure~\ref{fig:weights_energy}), the weights are 0.5\% at 100~MeV, 1.5\% at 250~MeV, 5\% at 500~MeV, 20\% at 1~GeV, 60\% at 2~GeV and reach 90\% at 4.5~GeV.

Another way to illustrate the effect of the weights is Figure~\ref{fig:weights_TS} (similar to Figure 18 of the 1FGL paper). It shows the contribution to $TS$ of all energies, for a rather soft source at high latitude (the background and exposure are averaged over all latitudes larger than 10$\degr$), with and without weights. Energies below 300~MeV contribute very little when the weights are applied. This remains true with the actual data selection used in 4FGL.

\begin{figure}
\centering
\includegraphics[width=\linewidth]{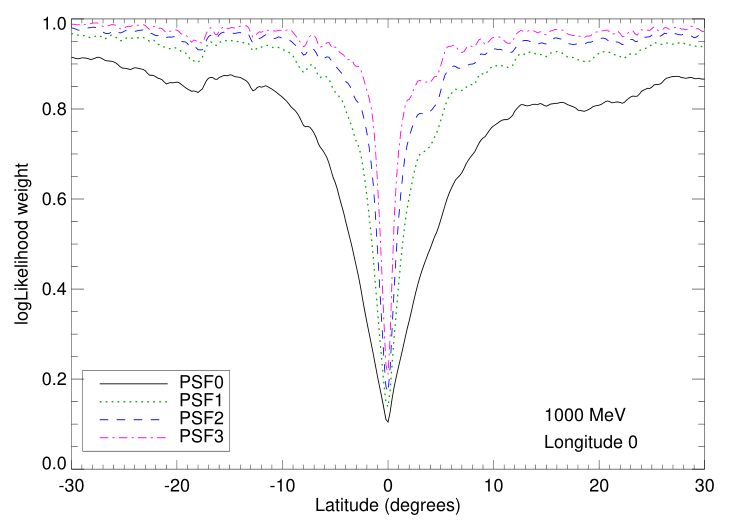}
\caption{Data-based weights at 1~GeV for ZMax = 105$\degr$ as a function of latitude (for the interesting [$-30\degr$, 30$\degr$] region) across the Galactic Center, for different PSF event types, computed according to Eq.~\ref{eq:wtype}. These weights were actually used in 4FGL. The average (over event types) weight is larger than the weight using all events together at the same 1~GeV energy (blue dashed line in Figure~\ref{fig:weights_energy}). This is because keeping event types separate is more favorable than merging them and losing the event type information.
}
\label{fig:weights_evtype}
\end{figure}

A specific difficulty remains because at a given energy we split the data into several components, each corresponding to a particular event type (with a different PSF). Since the systematics act in the same way on all components, the weights must be computed globally (i.e., weights must be lower when using PSF2 and PSF3 events than when using PSF3 alone). On the other hand, the resulting uncertainties with two components should be smaller than those with a single component (adding a second one adds information). In this work, we started by computing weights $w_k$ individually for each component $k$ (the dependence on $E$ and $\bf r$ is left implicit). Then we assumed that the final weights are simply proportional to the original ones, with a factor $\alpha < 1$ ($\alpha$ depends on $E$ and $\bf r$ as well). A reasonable solution is then
\begin{eqnarray}
N_{\rm min} & = & \min_k N_k \\
K_{\rm tot} & = & \sum_k \left( \frac{N_{\rm min}}{N_k} \right)^2 \\
\alpha & = & \frac{1 + \epsilon^2 N_{\rm min}}{1 + \epsilon^2 N_{\rm min} K_{\rm tot}}
\label{eq:alpha} \\
w_k & = & \frac{\alpha}{1 + \epsilon^2 N_k}
\label{eq:wtype}
\end{eqnarray}
$K_{\rm tot}$ and $\alpha$ are 1 if one component dominates over the others, and $K_{\rm tot}$ is the number of components if they are all similar.
The effect of this procedure is depicted in Figure~\ref{fig:weights_evtype} at 1~GeV, the lowest energy at which we use all event types. It illustrates quantitatively how the PSF0 events are unweighted at low latitudes, compared to better event types.
 
\end{document}